\documentclass[aps,prb,twocolumn,superscriptaddress,
groupedaddress,10pt]{revtex4}
\usepackage{amsmath}
\usepackage{amssymb}
\usepackage{graphicx}
\usepackage{epsfig}
\usepackage{dcolumn}
\usepackage{bm}
\usepackage{bm}
\usepackage{blindtext}
\usepackage{natbib}
\usepackage{siunitx} 

\usepackage[titletoc]{appendix}

\usepackage{simplewick}

\usepackage{bbm}
\makeatletter
\newcommand{\mathleft}{\@fleqntrue\@mathmargin0pt}
\newcommand{\mathcenter}{\@fleqnfalse}
\makeatother

\usepackage[usenames,dvipsnames]{xcolor}
\usepackage{amsthm}
\usepackage{booktabs}
\usepackage{hyperref}
\usepackage{tikz}
\usetikzlibrary{calc}
\usepackage[printwatermark]{xwatermark}
\newcommand{\tikzmark}[1]{\tikz[overlay,remember picture] \node (#1) {};}
\newcommand{\DrawBox}[3][]{%
    \tikz[overlay,remember picture]{
    \draw[black,#1]
      ($(#2)+(-0.1em,2.0ex)$) rectangle
      ($(#3)+(0.1em,-0.5ex)$);}
}

\usepackage{mathtools}

\def\be{\begin{equation}} \def\ee{\end{equation}}
\def\bea{\begin{eqnarray}} \def\eea{\end{eqnarray}}

\def\nn{\nonumber}

\begin{document}
\title{
Spin wave theory of one-dimensional  generalized Kitaev model
}

\author{Wang Yang}
\affiliation{Department of Physics and Astronomy and Stewart Blusson Quantum Matter Institute,
University of British Columbia, Vancouver, B.C., Canada, V6T 1Z1}

\author{Alberto Nocera}
\affiliation{Department of Physics and Astronomy and Stewart Blusson Quantum Matter Institute, 
University of British Columbia, Vancouver, B.C., Canada, V6T 1Z1}


\author{Ian Affleck}
\affiliation{Department of Physics and Astronomy and Stewart Blusson Quantum Matter Institute, 
University of British Columbia, Vancouver, B.C., Canada, V6T 1Z1}

\begin{abstract}
In this work, we perform a combination of classical and spin wave analysis on the one-dimensional spin-$S$ Kitaev-Heisenberg-Gamma model in the region of an antiferromagnetic Kitaev coupling.
Four phases are found, including a N\'eel ordered phase, a phase with $O_h\rightarrow D_3$ symmetry breaking, and ``$D_3$-breaking I, II" phases which both break $D_3$ symmetries albeit in different ways,
where $O_h$ is the full octahedral group and $D_3$ is the dihedral group of order six.
The lowest-lying spin wave mass is calculated perturbatively in the vicinity of the hidden SU(2) symmetric ferromagnetic point.

\end{abstract}
\maketitle

\section{Introduction}

Frustration in low dimensional strongly correlated magnetic systems leads to a plethora of fascinating  behaviors \cite{Fazekas1999,Lauchli2006,Balents2010,Witczak-Krempa2014,Rau2016,Savary2017,Winter2017,Zhou2017}.
An unusual way of introducing magnetic frustrations is by strong spin-orbit couplings, which induce bond- and direction-dependent magnetic interactions \cite{Jackeli2009,Chaloupka2010,Rau2014}. 
A famous example of frustrated magnetic system of this type is the two-dimensional (2D) Kitaev model on the honeycomb lattice \cite{Kitaev2006}.
The model was proposed to host exotic fractionalized excitations including Majorana fermions and anyons \cite{Kitaev2006},
and  has triggered tremendous research interests in recent years \cite{Singh2010,Reuther2011,Jiang2011,Price2012,Choi2012,Singh2012,Chaloupka2013,Modic2014,Plumb2014,Kim2015,Johnson2015,Sandilands2015,Sears2015,Banerjee2016,Yadav2016,Baek2017,Banerjee2017,Zheng2017,Ran2017,Wang2017,Janssen2017,Liu2018,Catuneanu2018,Gohlke2018,Jansa2018,Yu2018,Hentrich2018,Kasahara2018,Gordon2019,Motome2020}.
The 2D Kitaev model can be realized in Mott insulating A$_2$IrO$_4$ (A=Li, Na) compounds and $\alpha$-RuCl$_3$ systems.
In real materials, additional symmetry allowed couplings also appear. 
The generalized Kitaev model has been proposed to describe the real systems \cite{Jackeli2009,Rau2014,Wang2017}, which includes Heisenberg and Gamma interactions in addition to the Kitaev coupling.

Since quantum fluctuations are enhanced by reducing the spatial dimension,
exotic behaviors are expected to emerge also in one-dimensional (1D)  strongly spin-orbit coupled quantum magnetic systems.
A series of recent works have performed both analytical and numerical studies on the phase diagram of 1D spin-1/2 generalized Kitaev models \cite{Agrapidis2018,Yang2020a,Yang2020,Yang2020b}.
The two-leg ladder case has also been analyzed \cite{Agrapidis2019,Catuneanu2019}, which already shows a similar phase diagram with the 2D case \cite{Catuneanu2019}. 
In particular, in Ref. \onlinecite{Yang2020b}, the phase diagram of the 1D spin-1/2 Kiteav-Heisenberg-Gamma chain has been studied in detail, which reveals a rich phase diagram with eleven distinct phases.

\begin{figure}[h]
\includegraphics[width=8.5cm]{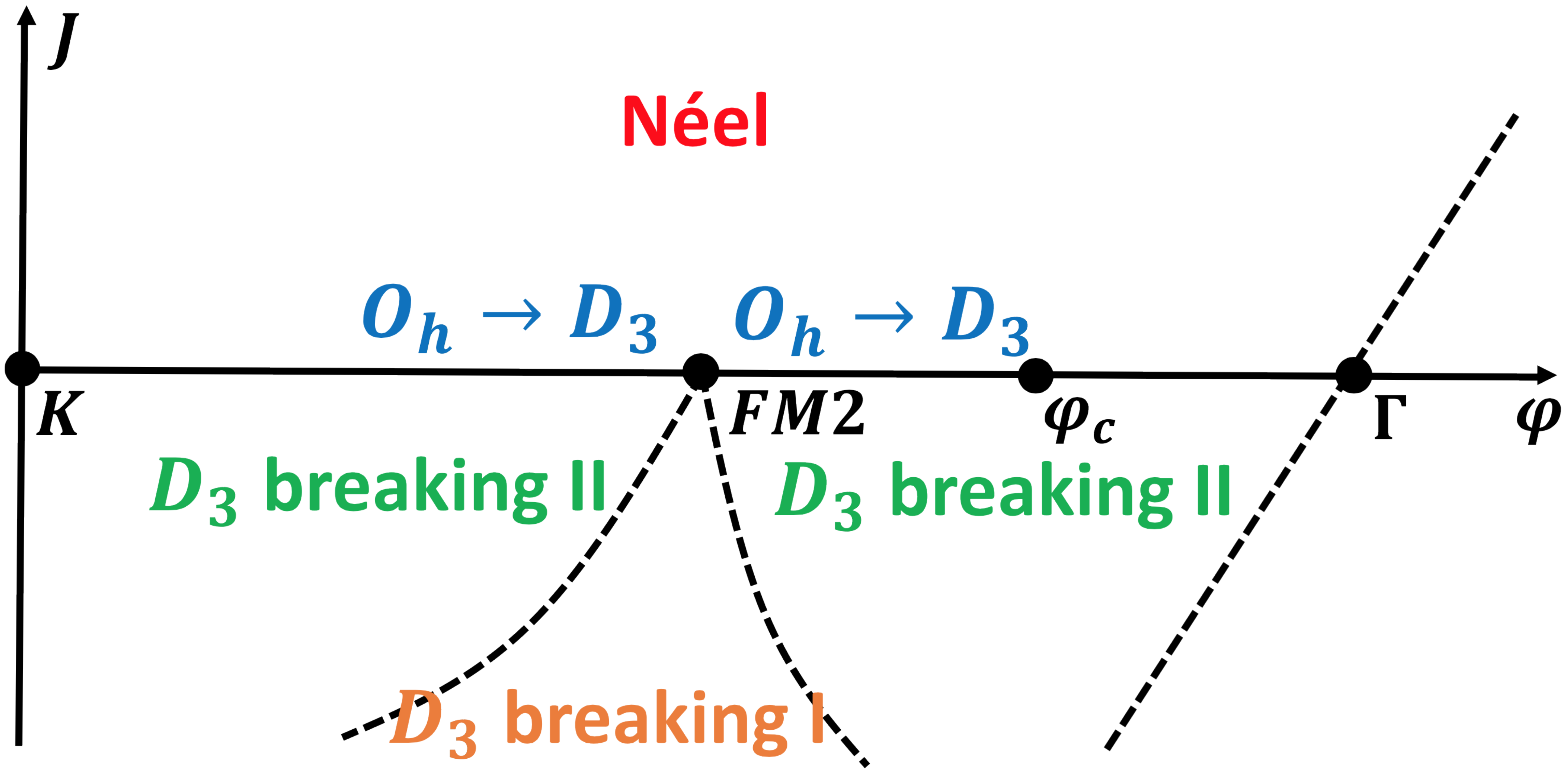}
\caption{Classical phase diagram in the vicinity of the FM2 point.
The horizontal coordinate $\varphi$ is defined through $K=\cos(\varphi)$, $\Gamma=\sin(\varphi)$.
The $\varphi$-coordinates of $K$, FM2 and $\Gamma$ points when $J=0$ are $0$, $\pi/4$ and $\pi/2$, respectively.
The classical phase transition at $\Gamma$ is shifted to $\varphi_c$ by quantum fluctuations.
} \label{fig:phase}
\end{figure}

In this work, we perform a combination of  classical  and spin wave analysis on the  1D spin-$S$ Kitaev-Heisenberg-Gamma model with an antiferromagnetic (AFM) Kitaev coupling. 
The phase diagram is shown in Fig. \ref{fig:phase}.
The N\'eel and ``$D_3$-breaking I, II" phases for the spin-1/2 case found in Ref. \onlinecite{Yang2020b} are also confirmed for higher spins.
On the other hand, the classical analysis predicts an $O_h\rightarrow D_3$ symmetry breaking for $J=0$, which is in contrast with the $O_h\rightarrow D_4$ symmetry breaking for the spin-1/2 case as discussed in Ref. \onlinecite{Yang2020a}.
Our DMRG numerics provide evidence for the $O_h\rightarrow D_3$  symmetry breaking for $S=1$ and $3/2$, based on which we conjecture that the spin-1/2 case is the only exception where strong quantum fluctuations invalidate the classical analysis. 

We have also constructed the spin wave theory which captures the small fluctuations around the classical configurations.
The lowest-lying spin wave mass $m_1$ is calculated perturbatively in the ``N\'eel", ``$O_h\rightarrow D_3$" and ``$D_3$-breaking I" phases close to the hidden SU(2) symmetric ferromagnetic (FM) FM2 point in Fig. \ref{fig:phase}. 
Interestingly, although $m_1\propto (K-\Gamma)^2$  in the ``$O_h\rightarrow D_3$" phase (where $J=0$)  and $m_1\propto J^2$ in the ``$D_3$-breaking I" phase for $K=\Gamma$, the former requires a second order symplectic perturbation calculation, whereas to obtain the latter, one has to go to third order perturbation,
where $K$, $\Gamma$  and $J$ represent the Kitaev, Gamma and Heisenberg couplings, respectively.
In the ``$D_3$-breaking II" phase, we encounter intrinsic difficulties in the perturbative calculation of the spin wave mass, and $m_1$ is studied numerically.
The origin of such difficulty is worth further explorations. 
Finally, we emphasize that the phase diagram in Fig. \ref{fig:phase} possibly can only be trusted in a neighborhood of the FM2 point.
When approaching the origin of Fig. \ref{fig:phase} (i.e., the AFM Kitaev point), enhanced quantum fluctuations arising from frustrations may destroy the classical order.

\section{Model Hamiltonian}

\subsection{The Hamiltonian}

\begin{figure}
\includegraphics[width=0.48\textwidth]{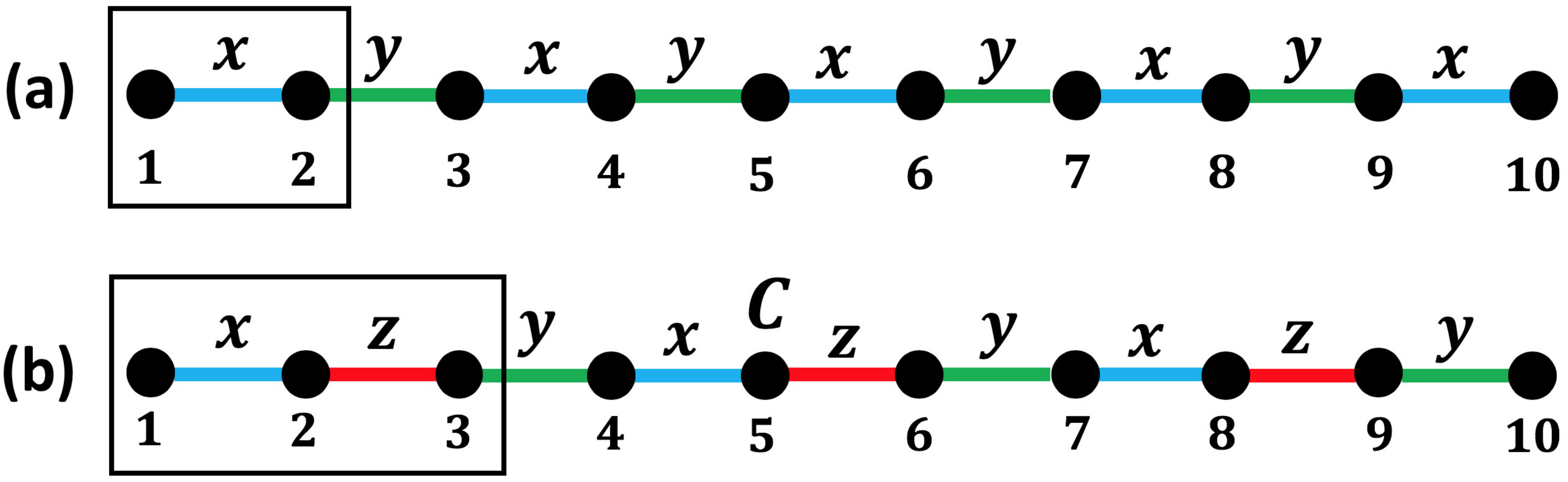}
\caption{Bond structures (a) before and (b) after the six-sublattice rotation.
The rectangular boxes denote unit cells.
}
\label{fig:bonds}
\end{figure}

The spin-$S$ Kitaev-Heisenberg-Gamma ($KH\Gamma$) chain \cite{Rau2014} is defined as
\begin{flalign}
&H=\sum_{<ij>\in\gamma\,\text{bond}}\big[ KS_i^\gamma S_j^\gamma+ J\vec{S}_i\cdot \vec{S}_j+\Gamma (S_i^\alpha S_j^\beta+S_i^\beta S_j^\alpha)\big],
\label{eq:Ham}
\end{flalign}
in which $<ij>$ is used to denote that $i,j$ are nearest neighboring lattice sites;
$\gamma=x,y$ is the spin direction associated with the $\gamma$ bond shown in Fig. \ref{fig:bonds} (a); 
$\alpha\neq\beta$ are the two remaining spin directions other than $\gamma$; 
$K$, $J$, and $\Gamma$ 
are the Kitaev, Heisenberg, and Gamma couplings, respectively;
and the spin operators satisfy $\sum_{\alpha=x,y,z}(S_i^\alpha)^2=S(S+1)$.
Since $R(\hat{z},\pi)$ changes the sign of $\Gamma$ but leaves $K$ and $J$ invariant, there is the equivalence \cite{Yang2020b}
\bea
(K,J,-\Gamma)\simeq (K,J,\Gamma),
\label{eq:equivalence}
\eea
where the notation $R(\hat{n},\alpha)$ is used to represent a global spin rotation around the $\hat{n}$-direction by an angle $\alpha$.
Parametrizing $K$ and $\Gamma$ as 
\bea
K=\cos(\varphi), ~\Gamma=\sin(\varphi),
\eea
it is enough to consider $\varphi\in[0,\pi]$ due to the equivalence in Eq. (\ref{eq:equivalence}).
Occasionally, we also use the following parametrization
\bea
K=\sin(\theta)\cos(\varphi), ~\Gamma=\sin(\theta)\sin(\varphi),~J=\cos(\theta).
\label{eq:parametize_theta_phi}
\eea
In this work, we will be interested in the region with an antiferromagnetic Kitaev coupling, i.e., $\varphi\in[0,\pi/2]$.
In particular, we mainly study the region in the vicinity of the FM2 point in Fig. \ref{fig:phase} where the coordinates of FM2 are $\varphi=\pi/4$, $J=0$ (i.e., $\theta=\pi/2$).
Here we note that the notation ``FM2" is chosen in accordance with Ref. \onlinecite{Yang2020b}.

A particularly useful six-sublattice rotation $U_6$ is defined as 
\cite{Stavropoulos2018,Yang2020a}
\bea
\text{Sublattice $1$}: & (x,y,z) & \rightarrow (x^{\prime},y^{\prime},z^{\prime}),\nn\\ 
\text{Sublattice $2$}: & (x,y,z) & \rightarrow (-x^{\prime},-z^{\prime},-y^{\prime}),\nn\\
\text{Sublattice $3$}: & (x,y,z) & \rightarrow (y^{\prime},z^{\prime},x^{\prime}),\nn\\
\text{Sublattice $4$}: & (x,y,z) & \rightarrow (-y^{\prime},-x^{\prime},-z^{\prime}),\nn\\
\text{Sublattice $5$}: & (x,y,z) & \rightarrow (z^{\prime},x^{\prime},y^{\prime}),\nn\\
\text{Sublattice $6$}: & (x,y,z) & \rightarrow (-z^{\prime},-y^{\prime},-x^{\prime}),
\label{eq:6rotation}
\eea
in which "Sublattice $i$" ($1\leq i \leq 6$) represents the collection of the sites $\{ i+6n\}_{n\in \mathbb{Z}}$, and $S^\alpha$ ($S^{\prime \alpha}$) is abbreviated as $\alpha$ ($\alpha^\prime$) for short, where $\alpha=x,y,z$.
The transformed Hamiltonian $H^\prime=U_6 H U_6^{-1}$ acquires the form
\begin{eqnarray}
H^\prime&=\sum_{<ij>\in \gamma\,\text{bond}}\big[ -KS_i^\gamma S_j^\gamma-\Gamma (S_i^\alpha S_j^\alpha+S_i^\beta S_j^\beta) \nn\\
& -J(S_i^\gamma S_j^\gamma+S_i^\alpha S_j^\beta+S_i^\beta S_j^\alpha)\big],
\label{eq:6rotated}
\end{eqnarray}
in which the bond $\gamma=x,z,y$ is periodic under translation by three sites as shown in Fig. \ref{fig:bonds} (b),
and the  prime has been dropped in $\vec{S}_i^\prime$ for simplicity.
The explicit form of $H^\prime$ is included in Appendix \ref{app:Ham}.
It is clear from Eq. (\ref{eq:6rotated}) that the FM2 point is SU(2) invariant in the six-sublattice rotated frame with an FM coupling. 

In the remaining parts of the paper, we will stick to the six-sublattice rotated frame from here on  unless otherwise stated.

\subsection{Review of the  symmetries}
\label{sec:review_Oh}

In this section, we give a quick review of the symmetries of the model within the six-sublattice rotated frame.

\begin{figure}[h]
\includegraphics[width=7.3cm]{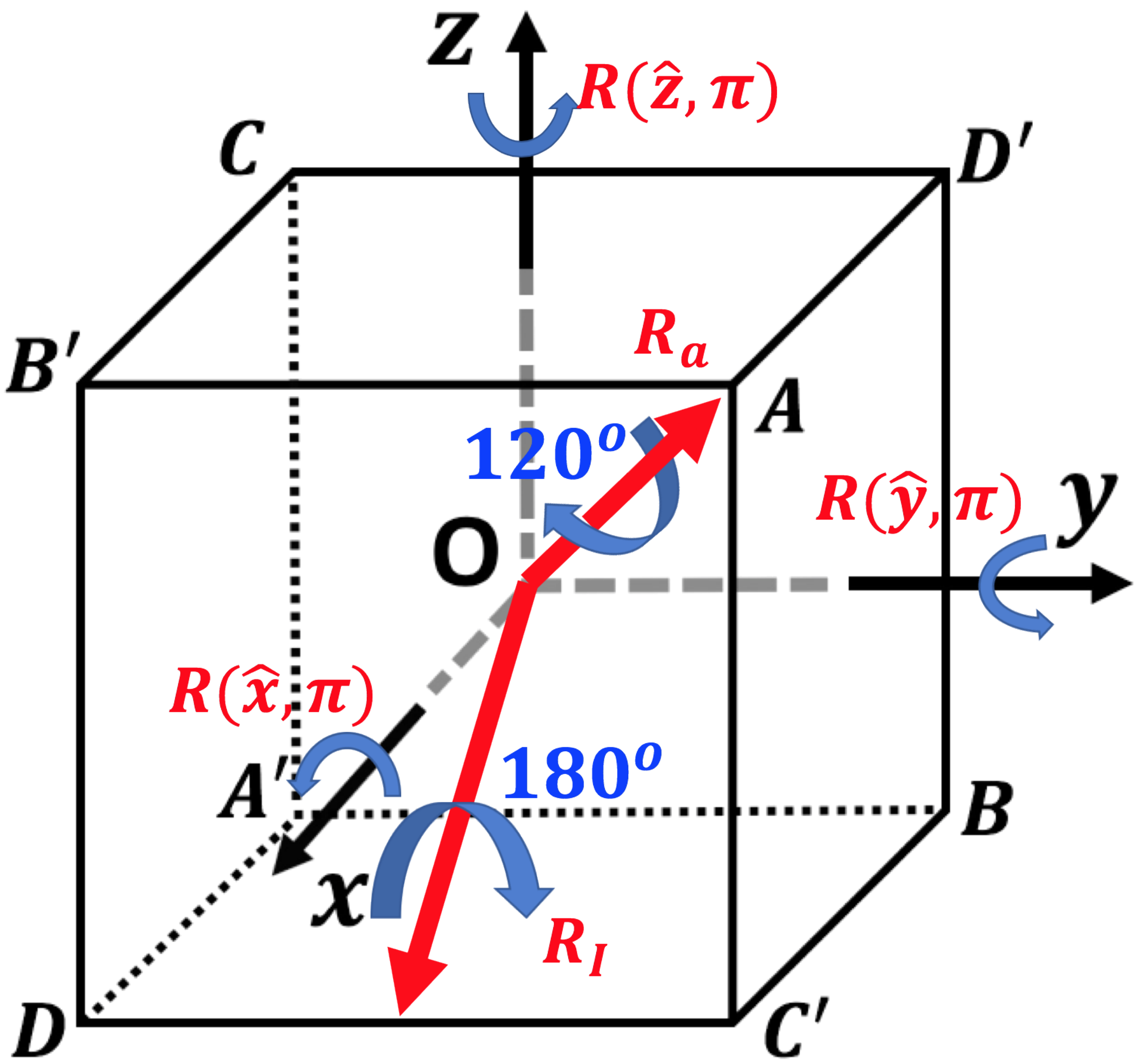}
\caption{Actions of the elements in $G/\mathopen{<}T_{3a}\mathclose{>}$ in  spin space as symmetry operations of a cube.
} \label{fig:Oh_ops}
\end{figure}

We first consider the $J=0$ case, i.e., the Kitaev-Gamma chain.
The symmetry group has been discussed in detail in Ref. \onlinecite{Yang2020a}.
The symmetry transformations include:
\begin{eqnarray}
1.&T &:  (S_i^x,S_i^y,S_i^z)\rightarrow (-S_{i}^x,-S_{i}^y,-S_{i}^z)\nn\\
2.& R_aT_a&:  (S_i^x,S_i^y,S_i^z)\rightarrow (S_{i+1}^z,S_{i+1}^x,S_{i+1}^y)\nn\\
3.&R_I I&: (S_i^x,S_i^y,S_i^z)\rightarrow (-S_{10-i}^z,-S_{10-i}^y,-S_{10-i}^x)\nn\\
4.&R(\hat{x},\pi)&: (S_i^x,S_i^y,S_i^z)\rightarrow (S_{i}^x,-S_{i}^y,-S_{i}^z)\nn\\
5.&R(\hat{y},\pi)&: (S_i^x,S_i^y,S_i^z)\rightarrow (-S_{i}^x,S_{i}^y,-S_{i}^z)\nn\\
6.&R(\hat{z},\pi)&: (S_i^x,S_i^y,S_i^z)\rightarrow (-S_{i}^x,-S_{i}^y,S_{i}^z),
\label{eq:syms_KG}
\end{eqnarray}
in which $T$ is time reversal; $T_a$ is translation by one lattice site;
$I$ is the spatial inversion around the point $C$ in Fig. \ref{fig:bonds} (b); 
and $R_a=R(\hat{n}_a,-2\pi/3)$, $R_I=R(\hat{n}_I,\pi)$ where 
\bea
\hat{n}_a=\frac{1}{\sqrt{3}}(1,1,1)^T,~\hat{n}_I=\frac{1}{\sqrt{2}}(1,0,-1)^T.
\label{eq:na_nI}
\eea
We note that the inversion center $C$ can be chosen modulo three.
The symmetry group $G$ is generated by the above transformations as
\bea
G=\mathopen{<}  T,R_aT_a,R_I I, R(\hat{x},\pi),R(\hat{y},\pi),R(\hat{z},\pi) \mathclose{>}.
\eea

Since $T_{3a}=(R_aT_a)^3$ is an abelian normal subgroup of $G$, we can consider the quotient group $G/\mathopen{<}T_{3a}\mathclose{>}$.
It has been worked out in Ref. \onlinecite{Yang2020a} that the quotient group is isomorphic to $O_h$, 
where $O_h$ is the full octahedral group which is the symmetry group of a cube.
There is an intuitive understanding of this isomorphism.
Neglecting the spatial components in the operations,
the actions in spin space are all symmetries of a spin cube as shown in Fig. \ref{fig:Oh_ops}.
It is proved in Ref. \onlinecite{Yang2020a} that the isomorphism still holds even if the spatial components are also included.
Hence we conclude that
\bea
G\cong O_h \ltimes 3\mathbb{Z},
\eea
where $3\mathbb{Z}=\mathopen{<}T_{3a}\mathclose{>}$ and $\ltimes$ is the semi-direct product.

Next we consider the $J\neq 0$ case, i.e., a general Kitaev-Heisenberg-Gamma chain.
In this case, the system is no longer invariant under the operations $R(\hat{\alpha},\pi)$ ($\alpha=x,y,z$).
Thus the symmetry group $G_1$ is
\bea
G_1=\mathopen{<}  T,R_aT_a,R_I I \mathclose{>}.
\eea
It has been shown in Ref. \onlinecite{Yang2020} that the group structure of $G_1$ is
\bea
G_1\cong D_{3d}\ltimes 3\mathbb{Z},
\eea
in which $\mathopen{<}T,R_aT_a,R_II\mathclose{>}/\mathopen{<}T_{3a}\mathclose{>}\cong D_{3d}$ is used. 

\subsection{Summary of the classical phase diagram}
\label{sec:cl_phase}

\begin{figure}
\includegraphics[width=7cm]{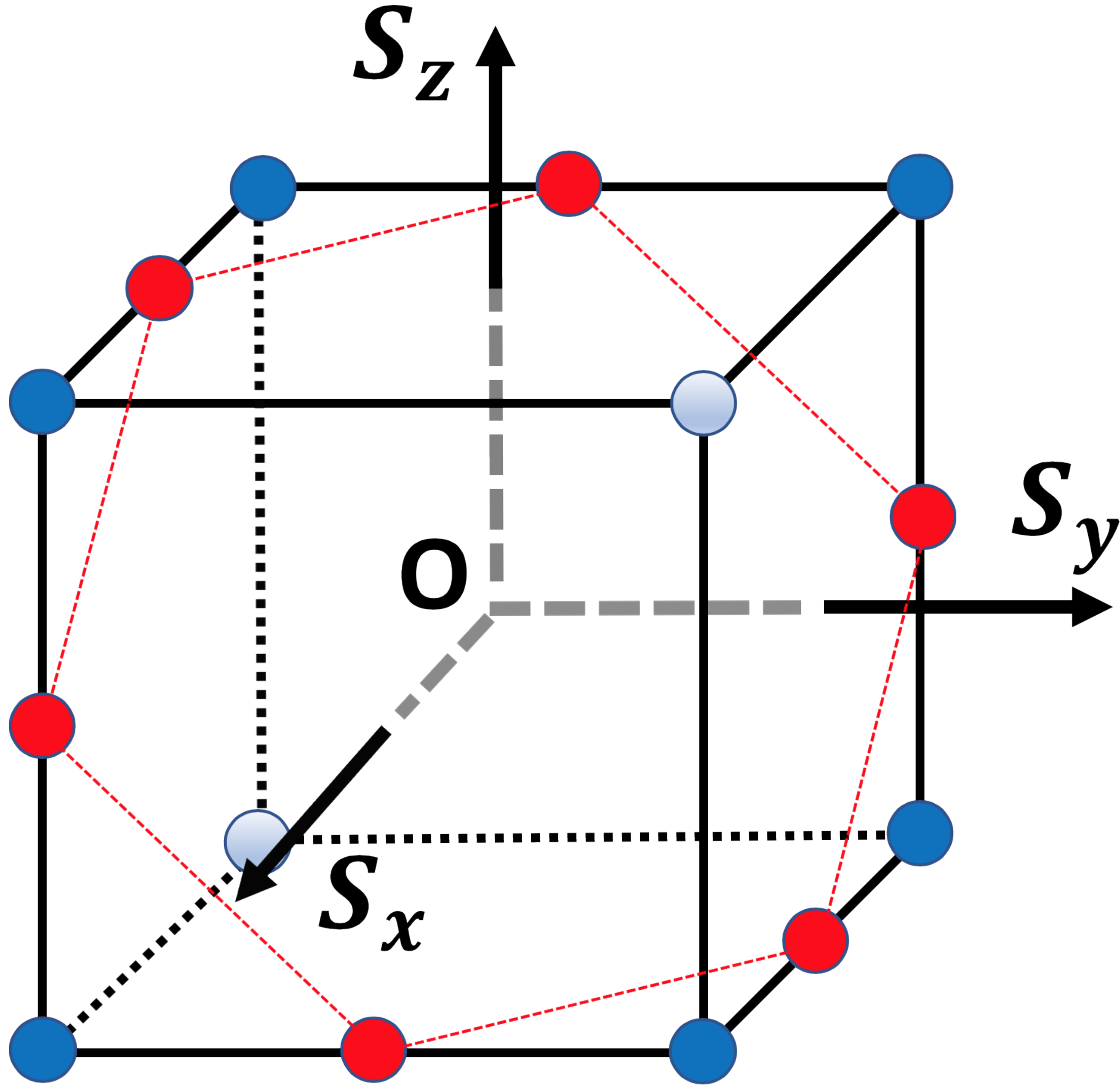}
\caption{``Center of mass" directions of the three spins within a unit cell 
in the six-sublattice rotated frame as represented by: 
the eight solid blue circles  for the eight degenerate ground states in the ``$O_h\rightarrow D_3$" phase;
the two solid light blue circles (along the $\pm\hat{n}_a$-directions) for the two degenerate ground states in the N\'eel phase; 
the six solid red circles  for the six degenerate ground states in the ``$D_3$-breaking I" phase;
the six solid dark blue circles (removing the two light blue ones among the eight) for the six degenerate ground states in the ``$D_3$-breaking II" phase.
 In the ``$D_3$-breaking II" phase, the plots are for $J\rightarrow 0$ according to the classical analysis.
In this paper, the convention  of the coordinates is taken such that  the eight vertices of the cube are  located at $(\pm 1,\pm1,\pm1)$.
} 
\label{fig:spin_orders}
\end{figure}

Here we give a brief summary  on the classical phase diagram as shown in Fig. \ref{fig:phase}.

The system has a long-range N\'eel order for $J>0$ where N\'eel refers to the original frame \cite{Yang2020b}. 
In the six-sublattice rotated frame, the ``center of mass" directions of the three spins in a unit cell are along $\pm\hat{n}_a$-directions as shown by the two solid light blue circles in Fig. \ref{fig:spin_orders},
where $\hat{n}_a$ is defined in Eq. (\ref{eq:na_nI}).
For $|\Delta|, |J|\ll1$, the lowest-lying spin wave mass is calculated to be $(\frac{4}{81}\Gamma \Delta^2+\frac{2}{3}J)S$,
where $\Delta=(K-\Gamma)/\Gamma$.

When $J=0$, the ground states are eight-fold degenerate with an $O_h\rightarrow D_3$ symmetry breaking. 
Our DMRG numerics provide evidence for the $O_h\rightarrow D_3$ symmetry breaking for $S=1$ and $3/2$,
though the spin-1/2 case is different which has  an $O_h\rightarrow D_4$ symmetry breaking as discussed in Ref. \onlinecite{Yang2020a}. 
The ``center of mass" spin directions of a unit cell in the eight degenerate $O_h\rightarrow D_3$ ground states are shown by the eight solid blue circles in Fig. \ref{fig:spin_orders}.
The classical phase transition point for $J=0$ is located at the $\Gamma$-point (i.e., $\varphi=\pi/2$), which is shifted to a different point $\varphi_c$ due to quantum fluctuations.
For $|\Delta|\ll1$, the lowest-lying spin wave mass is calculated to be $\frac{4}{81}S \Gamma  \Delta^2$.

When $J<0$, there are two phases, namely ``$D_3$-breaking I, II", both having six-fold degenerate ground states.
The symmetry breaking patterns of the two phases are $D_{3d}\rightarrow \mathbb{Z}_2^{\text{(I)}}$ and $D_{3d}\rightarrow \mathbb{Z}_2^{\text{(II)}}$, respectively, 
where $\mathbb{Z}_2^{\text{(I)}}$ and $\mathbb{Z}_2^{\text{(II)}}$ are two different symmetry groups albeit both isomorphic to $\mathbb{Z}_2$.
In the ``$D_3$-breaking I" phase, the ``center of mass" spin directions of a unit cell 
in the six degenerate ground states
within the six-sublattice rotated frame are plotted as the six solid red circles in Fig. \ref{fig:spin_orders}. 
We have calculated the lowest-lying spin wave mass $m_1$ for $\Delta=0$, $|J|\ll 1$
and the result is $S J^2/\Gamma $.
Although $m_1$ is proportional to $J^2$, it requires a third order symplectic perturbation calculation as discussed in Sec. \ref{sec:sw_D3I}.
In the ``$D_3$-breaking II" phase, the ``center of mass" spin directions
in the six degenerate ground states
 in the limit $J\rightarrow 0$ are plotted as the six solid dark blue circles in Fig. \ref{fig:spin_orders}.
For larger $|J|$, the ``center of mass" directions are distorted away from the vertices of the cube.
Due to intrinsic difficulties in doing perturbation in the ``$D_3$-breaking II" phase, we are not able to obtain a perturbative expression for the spin wave mass.
On the other hand, the lowest-lying spin wave mass has been studied numerically as shown in Fig. \ref{fig:sw_mass_D3}.
We note that our DMRG numerics provide evidence for  the spin ordering patterns in both ``$D_3$-breaking I, II" phases for $S=1,3/2$. 

Finally we make a comment on the numerical methods that we employ in this work.
The DMRG method\cite{White1992} was used on chains with length of $L=18$ sites and periodic boundary conditions within the six-sublattice rotated frame. The calculation of the first ten eigenstates was performed using standard DMRG multi-targeting approaches\cite{White1993}.
Even though it is known that DMRG convergence is hard for periodic boundary conditions, we have checked that for the system size considered our results are converged using up to m= 1000 states with a truncation error below $10^{-6}$ as in previous investigations\cite{Yang2020,Yang2020a,Yang2020b}.

\section{The ``$O_h\rightarrow D_3$" phase for $J=0$}

In this section, we perform a combination of classical and spin wave analysis for $J=0$ in the vicinity of the FM2 point in Fig. \ref{fig:phase}.
In Sec. \ref{sec:classical_equator}, the trial classical solution is demonstrated to be a minimum of the classical free energy by showing that the eigenvalues of the Hessian matrix are all positive. 
In Sec. \ref{sec:sym_break_equator}, the symmetry breaking pattern of the classical solution is shown to be $O_h\rightarrow D_3$, exhibiting an eight-fold degeneracy.
Then in Sec. \ref{sec:SW_equator}, we derive the spin wave theory by quantizing the Gaussian fluctuations around the classical minima in the long wavelength limit.
The smallest spin wave mass is shown to be $\frac{4}{81}S\Gamma \Delta^2$ up to the leading nonvanishing order in $\Delta$.
Finally in Sec. \ref{sec:numerics_OhD3}, we provide numerical evidence for the ``$O_h\rightarrow D_3$" symmetry breaking for $S=1$  and $S=3/2$.
We work in the six-sublattice rotated frame throughout this section unless otherwise stated.

\subsection{The classical solutions}
\label{sec:classical_equator}

The classical analysis is the saddle point approximation in the spin path integral formalism which is valid in the large-$S$ limit. 
In what follows, we neglect  quantum fluctuations of the spins and approximate them as classical three-vectors, i.e., 
\bea
\vec{S}_i = S\hat{n}_i,
\eea
in which $S$ is the spin magnitude, and $\hat{n}_i=(x_i,y_i,z_i)^T$ is a unit vector.
The classical free energy of a general $KH\Gamma$ chain is
\bea
f&=&\sum_n (f_{1+3n}+f_{2+3n}+f_{3+3n}),
\label{eq:f}
\eea
in which
\begin{flalign}
&f_{1+3n}=-(K+J)S^2 x_{1+3n}x_{2+3n}-\Gamma S^2[y_{1+3n}y_{2+3n}\nn\\
&~~+z_{1+3n}z_{2+3n}]-JS^2[y_{1+3n}z_{2+3n}+z_{1+3n}y_{2+3n}],\nn\\
&f_{2+3n}=-(K+J)S^2 z_{2+3n}z_{3+3n}-\Gamma S^2[x_{2+3n}x_{3+3n}\nn\\
&~~+y_{2+3n}y_{3+3n}]-JS^2[x_{2+3n}y_{3+3n}+y_{2+3n}x_{3+3n}],\nn\\
&f_{3+3n}=-(K+J)S^2 y_{3+3n}y_{4+3n}-\Gamma S^2[z_{3+3n}z_{4+3n}\nn\\
&~~+x_{3+3n}x_{4+3n}]-JS^2[z_{3+3n}x_{4+3n}+x_{3+3n}z_{4+3n}].
\end{flalign}
The constraints
\bea
x_j^2+y_j^2+z_j^2=1
\label{eq:constraints}
\eea 
can be introduced via Lagrange multipliers $\{\lambda_j\}_{j\in \mathbb{Z}}$
so that the free energy becomes
\bea
f^\prime=f-\frac{1}{2}\sum_j \lambda_j(x_j^2+y_j^2+z_j^2-1).
\label{eq:fprime}
\eea
We will first write down the saddle point equations for a general $J$, and later take $J=0$ in this section.

Seeking classical minima that are invariant under $T_{3a}$, i.e.,
\bea
x_{i+3n}\equiv x_i,y_{i+3n}\equiv y_i, z_{i+3n}\equiv z_i,~ (1\leq i\leq 3),
\eea
the energy per unit cell $F=3f^\prime/L$ becomes
\bea
F&=& -(K^\prime+J^\prime) x_1x_2-\Gamma^\prime(y_1y_2+z_1z_2)-J^\prime(y_1z_2+z_1y_2)\nn\\
&&-(K^\prime+J^\prime) z_2z_3-\Gamma^\prime(x_2x_3+y_2y_3)-J^\prime(x_2y_3+y_2x_3)\nn\\
&&-(K^\prime+J^\prime) y_3y_1-\Gamma^\prime(z_3z_1+x_3x_1)-J^\prime(z_3x_1+x_3z_1)\nn\\
&&-\sum_{i=1}^3 \frac{1}{2}\lambda_i (x_i^2+y_i^2+z_i^2-1),
\label{eq:energy_KHG}
\eea
in which $\Gamma^\prime,K^\prime,J^\prime$ are defined as
\bea
\Gamma^\prime=\Gamma S^2, ~K^\prime=K S^2, ~J^\prime=J S^2.
\eea
From Eq. (\ref{eq:energy_KHG}), the saddle point equations  can be derived as
\bea
\frac{\partial F}{\partial x_1}&=&-(K^\prime+J^\prime)x_2-\Gamma^\prime x_3-J^\prime z_3-\lambda_1 x_1=0\nn\\
\frac{\partial F}{\partial y_1}&=&-\Gamma^\prime y_2-(K^\prime+J^\prime)y_3-J^\prime z_2-\lambda_1 y_1=0\nn\\
\frac{\partial F}{\partial z_1}&=&-\Gamma^\prime z_2-\Gamma^\prime z_3-J^\prime x_3-J^\prime y_2-\lambda_1 z_1=0\nn\\
\frac{\partial F}{\partial \lambda_1}&=&-(x_1^2+y_1^2+z_1^2-1)=0
\label{eq:partial_F1}
\eea
\bea
\frac{\partial F}{\partial x_2}&=&-(K^\prime+J^\prime)x_1-\Gamma^\prime x_3-J^\prime y_3-\lambda_2 x_2=0\nn\\
\frac{\partial F}{\partial y_2}&=&-\Gamma^\prime y_1-\Gamma^\prime y_3-J^\prime z_1-J^\prime x_3-\lambda_2 y_2=0\nn\\
\frac{\partial F}{\partial z_2}&=&-\Gamma^\prime z_1-(K^\prime+J^\prime) z_3-J^\prime y_1-\lambda_2 z_2=0\nn\\
\frac{\partial F}{\partial \lambda_2}&=&-(x_2^2+y_2^2+z_2^2-1)=0
\label{eq:partial_F2}
\eea
\bea
\frac{\partial F}{\partial x_3}&=&-\Gamma^\prime x_2-\Gamma^\prime x_1-J^\prime y_2-J^\prime z_1-\lambda_3 x_3=0\nn\\
\frac{\partial F}{\partial y_3}&=&-\Gamma^\prime y_2-(K^\prime+J^\prime) y_1-J^\prime x_2-\lambda_3 y_3=0\nn\\
\frac{\partial F}{\partial z_3}&=&-(K^\prime+J^\prime) z_2-\Gamma^\prime z_1-J^\prime x_1-\lambda_3 z_3=0\nn\\
\frac{\partial F}{\partial \lambda_3}&=&-(x_3^2+y_3^2+z_3^2-1)=0.
\label{eq:partial_F3}
\eea

For the purpose of discussing the Kitaev-Gamma chain in this section, $J$ should be taken as zero.
Taking $J=0$, and plugging the following trial solutions 
\begin{flalign}
&\hat{n}^{(0)}_1=(x_1,y_1,z_1)^T=(a,a,b)^T,\nn\\
&\hat{n}^{(0)}_2=(x_2,y_2,z_2)^T=(a,b,a)^T,\nn\\
&\hat{n}^{(0)}_3=(x_3,y_3,z_3)^T=(b,a,a)^T,\nn\\
&\lambda_1=\lambda_2=\lambda_3=\lambda,
\label{eq:spins_OhD4}
\end{flalign}
into Eqs. (\ref{eq:partial_F1},\ref{eq:partial_F2},\ref{eq:partial_F3}), 
where the superscript ``$(0)$" is used to indicate that these are saddle point solutions,
we find that Eqs. (\ref{eq:partial_F1},\ref{eq:partial_F2},\ref{eq:partial_F3}) are reduced to
\bea
-(\lambda+K^\prime)a -\Gamma^\prime b&=&0 \nn\\
-2\Gamma^\prime a -\lambda b&=& 0\nn\\
2a^2+b^2-1&=&0.
\label{eq:partial_red_equator}
\eea
Since there are three variables $a,b,\lambda$ and three equations,
the solution of Eq. (\ref{eq:partial_red_equator}) exists.
In particular, $\lambda$ can be determined from the secular equation
\bea
\det \left(\begin{array}{cc}
-(\lambda+K^\prime) & -\Gamma^\prime \\
-2\Gamma^\prime & -\lambda
\end{array}\right)=0.
\label{eq:lambda_eq_equator}
\eea

When $K=\Gamma$, there are two solutions of $\lambda$ solved from Eq. (\ref{eq:lambda_eq_equator}), i.e., $\lambda^{(1)}=\Gamma^\prime$ and $\lambda^{(2)}=-2\Gamma^\prime$.
The solution $\lambda^{(2)}$ should be kept, since the free energy $F$ in Eq. (\ref{eq:energy_KHG}) acquires a larger value for $\lambda^{(1)}$ than for $\lambda^{(2)}$.
When $K\neq \Gamma$, Eq. (\ref{eq:partial_red_equator}) can be solved perturbatively in an expansion over $\Delta$, where the parameter $\Delta$ is defined as
\bea
\Delta=(K-\Gamma)/\Gamma.
\label{eq:define_Delta}
\eea 
The results up to $O(\Delta^2)$ are
\bea
\lambda(\Delta)&=& (-2-\frac{2}{3}\Delta-\frac{2}{27} \Delta^2)\Gamma^\prime+O(\Delta^3),\nn\\
a(\Delta)&=& \frac{1}{\sqrt{3}} (1+\frac{1}{9} \Delta-\frac{2}{81} \Delta^2)+O(\Delta^3),\nn\\
b(\Delta)&=& \frac{1}{\sqrt{3}} (1-\frac{2}{9} \Delta+\frac{1}{81} \Delta^2)+O(\Delta^3).
\label{eq:saddle_Oh_D4}
\eea
We note that among the two solutions of $\lambda$, the one which reduces to $-2\Gamma^\prime$ for $\Delta=0$ is kept in Eq. (\ref{eq:saddle_Oh_D4}).

On the other hand, Eq. (\ref{eq:saddle_Oh_D4}) only represents a saddle point solution, not necessarily a global minimum of the free energy.
Next we perturbatively show that the eigenvalues of the Hessian matrix of the free energy $F$ are all positive at least for $|\Delta|\ll 1$,
thereby confirming that Eq. (\ref{eq:saddle_Oh_D4}) constitutes a minimal solution.
Numerics of the classical analysis provide evidence for Eq. (\ref{eq:saddle_Oh_D4}) to be a  global  minimum of the free energy as discussed in Appendix \ref{app:numerics_classical}.

Because of the constraints in Eq. (\ref{eq:constraints}), the $T_{3a}$-invariant spin configurations form a six-dimensional manifold in the nine-dimensional Euclidean space spanned by the nine coordinates $\{x_i,y_i,z_i\}_{1\leq i\leq 3}$.
Since the $\lambda_i$ terms in Eq. (\ref{eq:energy_KHG}) vanish as a consequence of the constraints in Eq. (\ref{eq:constraints}), $f^\prime$ in Eq. (\ref{eq:fprime}) acquires the same value as $f$ in Eq. (\ref{eq:f}) on the six-dimensional manifold, where $L\in 3\mathbb{Z}$ is the number of lattice sites.  
Therefore, we will equivalently consider $F=3f^\prime/L$ instead of $3f/L$ in what follows to calculate the Hessian matrix.
The advantage of using $F$ is that its gradient vanishes at the saddle point,
 unlike the case of $3f/L$, where the gradient is perpendicular to the tangent space at the saddle point.

Consider the six eigenvalues of the Hessian matrix of the free energy $F$ restricted to the six-dimensional manifold.
Right at the FM2 point, two of the eigenvalues are zero, 
which is reasonable since there are two gapless spin waves for an FM Heisenberg chain.
Based on this, we expect that for $|\Delta|\ll 1$, 
the Hessian matrix contains two low-lying eigenvalues. 
Since the other four high-lying eigenvalues remain to be gapped with a small correction  dependent on $\Delta$,
it is enough to check that the two-lying eigenvalues are positive.
In what follows, we demonstrate this by perturbatively calculating the two smallest eigenvalues of the Hessian matrix in an expansion in $\Delta$.

Before proceeding on, we first set up some notations.
Denote $R(\Delta)=(\hat{n}^{(0),T}_1(\Delta),\hat{n}^{(0),T}_2(\Delta),\hat{n}^{(0),T}_3(\Delta))^T$  to be the saddle point solution for a fixed value of $\Delta$ within the nine-dimensional space where $\{\hat{n}^{(0)}_i(\Delta)\}_{1\leq i\leq 3}$ are given by Eqs. (\ref{eq:spins_OhD4},\ref{eq:saddle_Oh_D4}).
In what follows, we will ignore the transpose operation on the superscripts,
bearing in mind that we are  always considering a nine-component column vector.
Denote $T_R(\Delta)$ to be the tangent space of the six-dimensional manifold at the point $R(\Delta)$,
and $P(\Delta)$ to be the projection to the tangent space $T_R(\Delta)$.
Explicitly, the expression of $P(\Delta)$ is
\bea
P=\mathbbm{1}_{9\times 9}-r_1r_1^T-r_2r_2^T-r_3r_3^T,
\label{eq:projection_classical}
\eea
in which $r_i$ is $r_i(\Delta)$ for short, 
where 
\bea
r_1&=&(\hat{n}^{(0)}_1,\vec{0},\vec{0}),\nn\\
r_2&=&(\vec{0},\hat{n}^{(0)}_2,\vec{0}),\nn\\
r_3&=&(\vec{0},\vec{0},\hat{n}^{(0)}_3).
\label{eq:define_ri}
\eea 
Now let $H_F(\Delta)$ be the $9\times 9$ Hessian matrix of $F$,
in which the derivatives are taken with respect to the unconstrained coordinates $\{x_i,y_i,z_i\}_{1\leq i\leq3}$,
i.e.,
\bea
(H_F)_{\alpha_i,\beta_j}(\Delta)=\frac{\partial^2 F}{\partial \alpha_i \partial \beta_j}(\Delta), 
\label{eq:HF_equator}
\eea
where $1 \leq i,j\leq 3$ are the site indices in a unit cell
and $\alpha,\beta=x,y,z$.
Notice that if $P(\Delta)H_F(\Delta)P(\Delta)$ is viewed as a $9\times 9$ matrix, 
then there are always three zero eigenvalues, and the three  corresponding null vectors are given by Eq. (\ref{eq:define_ri}), 
since $r_i(\Delta)$ ($1\leq i \leq 3$) are always annihilated by $P(\Delta)$.
Denote $v_1(\Delta),v_2(\Delta)$ to be the eigenvectors of the two low-lying eigenvalues,
and $w_1(\Delta),w_2(\Delta),w_3(\Delta),w_4(\Delta)$ the other four eigenvectors of the high-lying eigenvalues.
We will be only interested in $v_1(\Delta),v_2(\Delta)$.

Consider an FM configuration with all spins aligning along $\hat{n}$-direction.
Let $\hat{e}_{\theta},\hat{e}_\phi$ be the two unit vectors perpendicular to $\hat{n}$
which are along tangent directions of the $\theta$ and $\phi$ coodinates, respectively,
where $\theta,\phi$ are the polar and azimuthal angles of a unit sphere.
When $\Delta=0$, $v_1=v_1(\Delta=0)$ and  $v_2=v_2(\Delta=0)$ are the two acoustic eigenvectors given by
\bea
v_1&=& (\frac{1}{\sqrt{3}} \hat{e}_\theta, \frac{1}{\sqrt{3}} \hat{e}_\theta,\frac{1}{\sqrt{3}} \hat{e}_\theta),\nn\\
v_2&=& (\frac{1}{\sqrt{3}} \hat{e}_\phi, \frac{1}{\sqrt{3}} \hat{e}_\phi,\frac{1}{\sqrt{3}} \hat{e}_\phi),
\label{eq:v1v2}
\eea
whereas $\{w_i=w_i(\Delta=0)\}_{1\leq i\leq4} $ are the optical ones:
\bea
w_1&=&(\frac{1}{\sqrt{2}}\hat{e}_\theta,-\frac{1}{\sqrt{2}}\hat{e}_\theta,\vec{0}),\nn\\
w_2&=&(\frac{1}{\sqrt{2}}\hat{e}_\phi,-\frac{1}{\sqrt{2}}\hat{e}_\phi,\vec{0}),\nn\\
w_3&=&(\frac{1}{\sqrt{2}}\hat{e}_\theta,\frac{1}{\sqrt{6}}\hat{e}_\theta,-\sqrt{\frac{2}{3}}\hat{e}_\theta),\nn\\
w_4&=&(\frac{1}{\sqrt{6}}\hat{e}_\phi,\frac{1}{\sqrt{6}}\hat{e}_\phi,-\sqrt{\frac{2}{3}}\hat{e}_\phi).
\label{eq:wi_s}
\eea
We note that the eigenvalues of the Hessian matrix for $\Delta=0$ corresponding to $v_i$-eigenvectors ($i=1,2$) are both $0$, and those corresponding to $w_i$'s ($i=1,2,3,4$) are all $3$.
Since when $\Delta\rightarrow 0$, the solution reduces to $a=b$ as can be seen from Eq. (\ref{eq:saddle_Oh_D4}),
$\hat{n}$ should be chosen as $\hat{n}_a=\frac{1}{3}(1,1,1)^T$ to determine the zeroth order terms in $v_1(\Delta)$ and $v_2(\Delta)$ in a perturbative expansion over $\Delta$.
As a result, $\hat{e}_\theta$ and $\hat{e}_\phi$ in Eqs. (\ref{eq:v1v2},\ref{eq:wi_s}) are given by
\bea
\hat{e}_\theta&=&(-\frac{1}{\sqrt{6}},-\frac{1}{\sqrt{6}},\sqrt{\frac{2}{3}})^T,\nn\\
\hat{e}_\phi&=&(-\frac{1}{\sqrt{2}},\frac{1}{\sqrt{2}},0)^T.
\eea

The projected Hessian matrix
\bea
\mathcal{H}_F(\Delta)=P(\Delta) H_F(\Delta) P(\Delta)
\label{eq:proj_HF_equator}
\eea
can be expanded in a power series of $\Delta$
\bea
\mathcal{H}_F(\Delta)=\mathcal{H}_F^{(0)}+ \mathcal{H}_F^{(1)}+\mathcal{H}_F^{(2)} +....,
\eea
in which $\mathcal{H}_F^{(n)}$ is proportional to $\Delta^n$.
Since both $v_1$ and $v_2$ have zero eigenvalues of $\mathcal{H}_F^{(0)}$,
a degenerate first order perturbation theory should be considered,
and the first order perturbation Hamiltonian is 
\bea
h^{(1)}=\left(\begin{array}{cc}
v_1^T \mathcal{H}_F^{(1)} v_1& v_1^T \mathcal{H}_F^{(1)} v_2\\
v_2^T \mathcal{H}_F^{(1)} v_1 & v_2^T \mathcal{H}_F^{(1)} v_2
\end{array}
\right).
\label{eq:h1}
\eea
However, straightforward calculation shows that $h^{(1)}$ vanishes
and we have to go to second order perturbation.

The second order perturbation Hamiltonian can be obtained as
\bea
h^{(2)}=\left(\begin{array}{c}
v_1^T\\
v_2^T
\end{array}
\right)
\big(\mathcal{H}^{(2)}_F+\mathcal{H}^{(1)}_F\sum_{i=1}^4 \frac{w_iw_i^T}{E_0-E_i} \mathcal{H}^{(1)}_F\big)(v_1\,v_2),
\eea
in which $E_0=0,E_i=3$ are the eigenvalues of $\mathcal{H}_F^{(0)}$
corresponding to the acoustic and optical eigenvectors, respectively. 
Calculations show that 
\bea
h^{(2)}=\frac{4}{27}\Gamma^\prime \Delta^2 \sigma_0,
\label{eq:h2_equator}
\eea
where $\sigma_0$ is the $2\times 2$ identity matrix.
Since the eigenvalue $\frac{4}{27}\Gamma S^2  \Delta^2$ is positive,
we arrive at the conclusion  that the solution in Eq. (\ref{eq:spins_OhD4})
is indeed a minimum of the classical free energy $F$ regardless of the sign of $\Delta$
at least when $|\Delta|$ is small.

Notice that up to $O(\Delta^2)$, $v_1(\Delta)$ and $v_2(\Delta)$ are degenerate according to Eq. (\ref{eq:h2_equator}).
In fact, this degeneracy holds to all orders in $\Delta$.
This is explained in detail in Appendix \ref{app:proof_degeneracy} based on a group theory analysis.

\subsection{The symmetry breaking pattern}
\label{sec:sym_break_equator}

To identify the symmetry breaking pattern, we work out the unbroken symmetry group of the spin alignments in Eq. (\ref{eq:spins_OhD4}) in the six-sublattice rotated frame.

It is straightforward to verify that the spin orientations in Eq. (\ref{eq:spins_OhD4}) 
are invariant under the symmetry operations $R_aT_a$ and $TR_I I$.
Therefore the unbroken symmetry group $N$ is 
\bea
N=\mathopen{<}R_aT_a,TR_I I \mathclose{>}.
\eea
Since $T_{3a}$ is unbroken, in what follows within this subsection,
we will consider the quotient group $N/\mathopen{<}T_{3a}\mathclose{>}$. 
As proved in Ref. \onlinecite{Yang2020}, $N/\mathopen{<}T_{3a}\mathclose{>}$ is isomorphic to $D_3$.
Here we give a quick demonstration of this isomorphism.
The group $D_n$ (i.e, the dihedral group of order $2n$)  has the following generator-relation representation
\begin{eqnarray}
D_n=\mathopen{<} \alpha,\beta| \alpha^n=\beta^2=(\alpha\beta)^2=e \mathclose{>}.
\label{eq:generator_Dn}
\end{eqnarray}
Define $\alpha=R_aT_a$, and $\beta=TR_I I$.
It is straightforward to verify that the relations in Eq. (\ref{eq:generator_Dn}) are satisfied for $\alpha,\beta$ modulo $T_{3a}$.
Furthermore, it can be checked that $N/\mathopen{<}T_{3a}\mathclose{>}$ contains at least six elements.
Since $|D_3|=6$, we conclude that  $N/\mathopen{<}T_{3a}\mathclose{>}\cong D_3$.
This analysis shows that the symmetry breaking pattern predicted by the classical theory is
\bea
O_h\rightarrow D_3.
\eea
We note that the classical prediction is  different from the symmetry breaking pattern for the spin-1/2 case \cite{Yang2020a} which is numerically identified as $O_h\rightarrow D_4$.
This  indicates strong quantum fluctuations in the spin-1/2 case.
On the other hand, numerical calculations provide evidence for the $O_h\rightarrow D_3$ symmetry breaking for $S=1$  and $S=3/2$ as will be discussed in Sec. \ref{sec:numerics_OhD3}.
Based on this, we conjecture that  spin-1/2 is the only exception and all other spins exhibit an $O_h\rightarrow D_3$ symmetry breaking as predicted by the classical analysis.

The classical solutions are degenerate, and Eq. (\ref{eq:spins_OhD4}) only gives one of the possibilities.
Since $|O_h|=48$ and $|D_3|=6$,
the number of degenerate classical minima is $|O_h/D_3|=8$.
The other minima are related to Eq. (\ref{eq:spins_OhD4})
by $O_h$ operations. 
Note that only operations in different equivalent classes of $O_h/D_3$
give distinct classical spin configurations. 
In fact, the eight degenerate spin orientations of $\vec{S}_1$
are $(\pm a,\pm a, \pm b)$,
and the orientations for $\vec{S}_2$ and $\vec{S}_3$
can be obtained by permuting $a$ and $b$
in accordance with Eq. (\ref{eq:spins_OhD4}).
For a pictorial illustration, the ``center of mass" directions of the three spins within a unit cell
corresponding to the eight classical minima 
are represented as solid blue circles  located at the vertices of a cube as shown in Fig. \ref{fig:spin_orders}.


\subsection{Spin wave theory}
\label{sec:SW_equator}

In this section, we derive the spin wave theory in the path integral formalism which characterizes the small fluctuations around the classical spin configurations.
We focus on the $|\Delta|\ll 1$ region, and only the lowest-lying spin wave will be considered.

\subsubsection{The spin wave Lagrangian}
\label{sec:sw_lag_equator}

The Lagrangin of the spin coherent state path integral is
\bea
L=S\sum_j \vec{A}_j\cdot \partial_t\hat{n}_j-f^\prime[\{\hat{n}_j\}],
\label{eq:spin_path_integral}
\eea
in which the first term is the Berry phase term; the Berry connection $\vec{A}_j(\theta_j,\phi_j)$ can be chosen as $\frac{1-\cos \theta_j}{\sin \theta_j}\hat{e}_{j\phi}$ where  $\theta_j$ and $\phi_j$ are the polar and azimuthal angles of $\hat{n}_j$, respectively,
and $\hat{e}_{j\phi}$ is the unit vector along the azimuthal direction at $\hat{n}_j$;
the functional $f^\prime$ is given by Eq. (\ref{eq:fprime}).
Notice that again by virtue of the constraints in Eq. (\ref{eq:constraints}), 
there is no difference between $f^\prime$ and $f$ in Eq. (\ref{eq:f}).
Therefore, it would be legitimate to write $f^\prime$ instead of $f$ in Eq. (\ref{eq:spin_path_integral}).

Next we expand the Lagrangian around the classical solution in Eq. (\ref{eq:spins_OhD4}).
In the spin wave approximation, only the Gaussian fluctuations will be kept.
For small fluctuations, $\hat{n}_j$ moves in the tangent space of the unit sphere at the point $\hat{n}_j^{(0)}$, in which $\hat{n}_j^{(0)}=\hat{n}_{[j]}^{(0)}$, $[j]\equiv j\mod 3$ and $1\leq [j]\leq 3$,
where $\hat{n}_{[j]}^{(0)}$ is given by Eqs. (\ref{eq:spins_OhD4},\ref{eq:saddle_Oh_D4}).
The local coordinate frame of the tangent space at site $j$ can be set up as $\{\hat{e}^{(0)}_\theta(j),\hat{e}^{(0)}_\phi(j)\}$, where $\hat{e}^{(0)}_\theta(j)$  and $\hat{e}^{(0)}_\phi(j)$ are the unit vectors along the polar and azimuthal directions at $\hat{n}^{(0)}_{[j]}$, respectively.
Then the deviations away from the equilibrium position are characterized by $\{\chi_\theta(j), \chi_\phi(j)\}$ which are the displacements along the $\hat{e}^{(0)}_\theta(j)$ and  $\hat{e}^{(0)}_\phi(j)$ directions.

With the above setup, the Berry phase term becomes
\bea
\frac{1}{2}S\sum_j [\chi_\theta(j)\partial_t\chi_\phi(j)-\chi_\phi(j)\partial_t\chi_\theta(j)].
\label{eq:Berry_chi}
\eea
As can be easily checked, the integration of Eq. (\ref{eq:Berry_chi}) over  time gives the area swept out by the trajectory of $\hat{n}_j$ within the tangent space, which coincides with the geometric meaning of the Berry phase term. 
We note that $\chi_\theta(j)$ and $\chi_\phi(j)$ form a pair of canonical conjugates which can be clearly seen from Eq. (\ref{eq:Berry_chi}).
Alternatively, choosing the quantization axis along $\hat{n}_j^{(0)}$, the angular momentum commutation relation becomes
\bea
[S\chi_\theta(j),S\chi_\phi(j)]=i\hat{n}^{(0)}_j\cdot \vec{S}_j.
\label{eq:commutation}
\eea
Replacing $\hat{n}^{(0)}_j\cdot \vec{S}_j$ with its classical value $S$,
Eq. (\ref{eq:commutation}) becomes
\bea
[\chi_\theta(j),\chi_\phi(j)]=i\frac{1}{S},
\eea
which is the canonical commutation relation where $1/S$ plays the role of $\hbar$.
This also indicates that the classical and spin wave analysis only applies in the large-$S$ (i.e., small $\hbar$) limit.

For later convenience, we rewrite Eq. (\ref{eq:Berry_chi}) in the Cartesian coordinates $\{x_i,y_i,z_i\}$ in the spin space.
The expression under the summation in Eq. (\ref{eq:Berry_chi}) can be written as
\bea
\hat{n}_j^T [\hat{e}^{(0)}_\theta(j)\hat{e}^{(0),T}_\phi(j)-\hat{e}^{(0)}_\phi(j)\hat{e}^{(0),T}_\theta(j)] \partial_t \hat{n}_j.
\label{eq:Berry_chi2}
\eea
Notice that the matrix kernel in Eq. (\ref{eq:Berry_chi2}) is simply the $\pi/2$-rotation matrix around the $\hat{n}^{(0)}_j$-direction.
Since such rotation can be implemented by a cross product with $\hat{n}^{(0)}_j$,
the matrix kernel in Eq. (\ref{eq:Berry_chi2}) is  equal to
\bea
M_j=\left(\begin{array}{ccc}
0& -n^{(0)}_{jz}&n^{(0)}_{jy}\\
n^{(0)}_{jz}&0&-n^{(0)}_{jx}\\
-n^{(0)}_{jy} &n^{(0)}_{jx}&0
\end{array}\right),
\label{eq:Mj}
\eea
in which $n^{(0)}_{j\alpha}$ is the $\alpha$-component of $n^{(0)}_{j}$, where $\alpha=x,y,z$ and $j=1,2,3$.
Therefore, for small fluctuations, Eq. (\ref{eq:spin_path_integral}) becomes
\bea
L=S\sum_j\frac{1}{2} \delta\hat{n}_j^T M_j \partial_t \delta\hat{n}_j -f^\prime[\{\hat{n}_j\}],
\label{eq:spin_path_integral3}
\eea
in which $\delta \hat{n}_j=\hat{n}_j-\hat{n}_j^{(0)}$, and $M_j$ is given by Eq. (\ref{eq:Mj}).

To discuss the spin wave dispersion, it is convenient to transform into the Fourier space. 
In what follows, the Fourier transform of the Cartesian coordinates  $\alpha_{i+3n}$ ($\alpha=x,y,z$; $i=1,2,3$; $n\in \mathbb{Z}$) will be defined as
\bea
\alpha_{i+3n}=\frac{1}{\sqrt{L/3}} \sum_n e^{i kn}\alpha_i(k),
\label{eq:Fourier_sw}
\eea
in which $L$ is the system size.
Plugging Eq. (\ref{eq:Fourier_sw}) into Eq. (\ref{eq:spin_path_integral3}) (setting $J=0$), we obtain
\bea
L&=&-f_0+S\sum_k\frac{1}{2} N^T(k) M \partial_t N(-k)\nn\\
&& -\frac{1}{2}\sum_k N^T(k)[\mathcal{H}_F+\delta H (k)] N(-k),
\label{eq:f_fourier}
\eea
in which $f_0$ is the classical free energy at the saddle points given by
\bea
f_0=-L\Gamma S^2 (1+\frac{1}{3}\Delta+\frac{1}{27}\Delta^2)+O(\Delta^3);
\eea
$M$ is a $9\times 9$ matrix
\bea
M=\left(\begin{array}{ccc}
M_1&0&0\\
0&M_2&0\\
0&0&M_3
\end{array}
\right),
\label{eq:M}
\eea
where $M_j$ ($j=1,2,3$) is defined in Eq. (\ref{eq:Mj});
 $N^T(k)$ defined as
 \bea
 N^T(k)=(\hat{n}^T_1(k),\hat{n}^T_2(k),\hat{n}^T_3(k))
 \label{eq:Nk}
 \eea 
 is a nine-component row vector where $\hat{n}^T_i(k)=(x_i(k),y_i(k),z_i(k))$ ($i=1,2,3$); $\mathcal{H}_F$ is given by Eq. (\ref{eq:proj_HF_equator}); and the $9\times 9$ matrix $\delta H(k)$ can be derived as
\bea
\delta H^{(1,3)}(k)&=&\Gamma S^2 (1-e^{-ik}) \text{diag}(1,1+\Delta,1),\nn\\
\delta H^{(3,1)}(k)&=&[\delta H^{(1,3)}(k)]^\dagger,\nn\\
\delta H^{(\alpha,\beta)}(k)&=&0, ~\text{for}~\{\alpha,\beta\}\neq\{1,3\},
\label{eq:deltaHk}
\eea
where $\text{diag}(\cdot\cdot\cdot)$ denotes the diagonal matrix, and $\delta H^{(\alpha,\beta)}(k)$ is the $3\times 3$  matrix at the $(1,3)$-block of $\delta H(k)$.

\subsubsection{Zero wavevector}
\label{sec:sw_equator_zero_momentum}

Let's first consider the zero wavevector spin waves. 
The Hamiltonian in Eq. (\ref{eq:f_fourier}) for $k=0$ is
\bea
\frac{1}{2} N^T(k=0)H_F N(k=0).
\label{eq:Ham_sw_0}
\eea
To get the spin wave masses, the matrix kernel in Eq. (\ref{eq:Ham_sw_0}) needs to be diagonalized. 
Naively, the matrix $\mathcal{H}_F$ has already been diagonalized in Sec. \ref{sec:classical_equator}.
However, in Eq. (\ref{eq:Ham_sw_0}), $\mathcal{H}_F$ must be diagonalized by symplectic transformation which leaves the symplectic form $M$ in Eq. (\ref{eq:M}) invariant,
unlike the case in Sec. \ref{sec:classical_equator} where $\mathcal{H}_F$ is  diagonalized by an orthogonal transformation.
Recall that we have already proved $\mathcal{H}_F$ to be positive definite which is the restriction of $H_F$ in Eq. (\ref{eq:HF_equator})  to  the six-dimensional tangent space.
Then by the symplectic theory \cite{DelaCruz2016}, $\mathcal{H}_F$ (which is viewed as a $6\times 6$ matrix) can be diagonalized by a symplectic transformation $U$, i.e., 
\bea
\mathcal{H}_F=U^T \Lambda U;
\eea
in which $U$ satisfies
\bea
U^T MU=M
\eea
and the diagonal matrix $\Lambda$ is of the form
\bea
\Lambda=\left(\begin{array}{ccc}
m_1\sigma_0 & 0&0\\
0&m_2\sigma_0&0\\
0&0&m_3\sigma_0
\end{array}
\right)
\eea
where $m_3>m_2>m_1>0$ ($i=1,2,3$); $\sigma_0$ is the $2\times 2$ identity matrix;
and $M$ is viewed as a $6\times 6$ matrix acting in  the six-dimensional tangent space.
We will be interested in $m_1$ which is related to the smallest spin wave mass.
Notice that in general, $m_i$'s do not coincide with the eigenvalues of $\mathcal{H}_F$. 
In what follows,  $m_1$ will be calcualted to the leading nonvanishing order in $\Delta$ by perturbation theory.
The result happens to be the same as the two lowest eigenvalues of $\mathcal{H}_F$ (i.e., $\frac{4}{27}\Gamma S^2  \Delta^2$) derived in Sec. \ref{sec:classical_equator}.

The calculations of $m_i$'s can be converted into an eigenvalue problem by considering the matrix $M\mathcal{H}_F$ \cite{DelaCruz2016}.
In fact, according to the symplectic linear algebra, 
the eigenvalues of $M\mathcal{H}_F$ are $\pm i m_j$ ($j=1,2,3$).
The basics of the symplectic transformations for our purpose are collected in Appendix \ref{app:symplectic}.
In what follows, we will view both $M$ and $\mathcal{H}_F$ as $9\times 9$ matrices.
Since $M_j$ is defined as a cross product operation in Eq. (\ref{eq:Mj}),
$\hat{n}_j^{(0)}$ must be a null vector of $M_j$.
As a result, $r_i(\Delta)$ in Eq. (\ref{eq:define_ri}) are always annihilated by $M$.
Hence the $9\times 9$ matrix $M\mathcal{H}_F$ always has three zero eigenvalues, and we will be interested in the other six eigenvalues.
When $\Delta=0$, among the other six eigenvalues, $M(\Delta=0)\mathcal{H}_F(\Delta=0)$ contains two zero eigenvalues with eigenvectors given by 
\bea
v_\pm=v_1\pm iv_2,
\label{eq:v_pm}
\eea
 where $v_i=v_i(\Delta=0)$ ($i=1,2$) are given by Eq. (\ref{eq:v1v2}).
For $\Delta\neq 0$, $v_\pm$ evolve into $v_\pm(\Delta)$ which have eigenvalues $\pm im_1(\Delta)$.

Let's first consider $\pm im_1(\Delta)$ to the linear order of $\Delta$.
Define $ \mathcal{H}_F^{M,(n)}$ in terms of the power expansion as
\bea
M(\Delta)\mathcal{H}_F(\Delta)
&=& \mathcal{H}_F^{M,(0)}+\mathcal{H}_F^{M,(1)}+ \mathcal{H}_F^{M,(2)}+...,
\eea
where $ \mathcal{H}_F^{M,(n)}$ is proportional to $\Delta^n$.
Define $P_1(\Delta)$ to be projection to the subspace spanned by $v_\pm(\Delta)$.
At a nonzero $\Delta$, the first order degenerate perturbation theory is  given by 
\bea
h^{M,(1)}=P_1^{(0)} \mathcal{H}_F^{M,(1)} P_1^{(0)},
\eea
where $P_1^{(0)}=P_1(\Delta=0)$.
Calculations show that $h^{M,(1)}$ vanishes,
hence  second order degenerate perturbation has to be considered.
We note that there is a quick way to see $h^{M,(1)}=0$.
In fact, there is the relation
\bea
h^{M,(1)}=M^{(0)}P_1^{(0)}h^{(1)}P_1^{(0)},
\label{eq:hM1_equator_relation}
\eea
in which $h^{(1)}$ is defined in Eq. (\ref{eq:h1}).
A proof of Eq. (\ref{eq:hM1_equator_relation}) is given in Appendix \ref{app:proof_symplectic_first}.
Since $P_1^{(0)}h^{(1)}P_1^{(0)}$ vanishes according to the discussion below Eq. (\ref{eq:h1}), $h^{M,(1)}$ has to vanish as a result.

\begin{figure}[h]
\includegraphics[width=8.0cm]{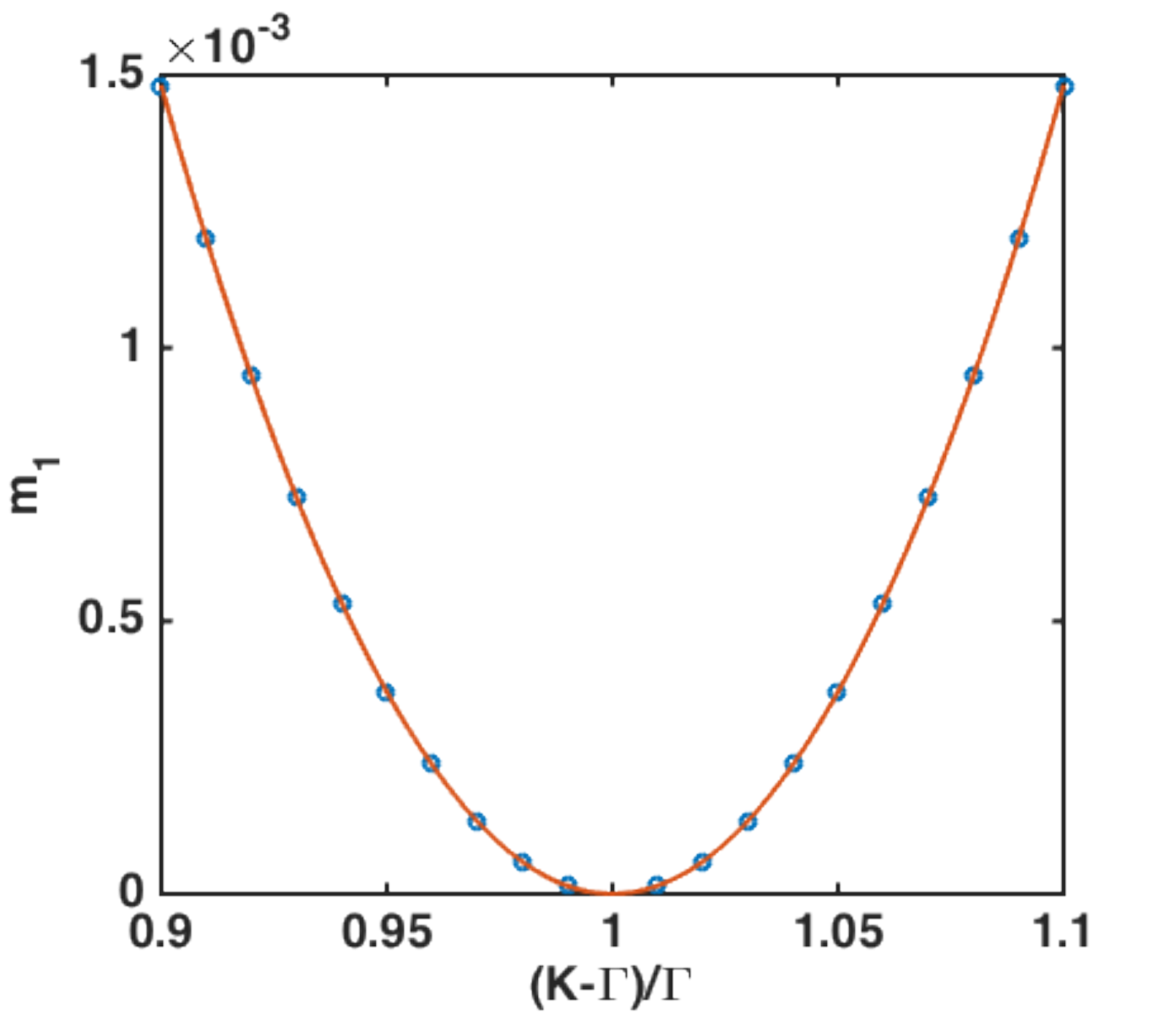}
\caption{$m_1$ vs. $\Delta$ for $J=0$ represented by the hollow circles as obtained by numerical diagonalization of $M(\Delta)\mathcal{H}_F(\Delta)$,
where $\Delta$ is defined in Eq. (\ref{eq:define_Delta}).
The solid line represents $\frac{4}{27}\Delta^2$.
The vertical axis is in unit of $\Gamma^\prime=\Gamma S^2$.
} \label{fig:m1_Delta}
\end{figure}

Next we proceed to second order perturbation.
Define $u_i$ ($i=1,2,3,4$) as
\bea
u_1&=&w_1+iw_2,\nn\\
u_2&=&w_1-iw_2,\nn\\
u_3&=&w_3+iw_4,\nn\\
u_4&=&w_3-iw_4,
\label{eq:Def_us}
\eea
in which $w_i$ are given in Eq. (\ref{eq:wi_s}).
Then $u_i$ are eigenvectors of $\mathcal{H}^{M,(0)}_F$ with eigenvalues equal to $\epsilon_i$, where 
\bea
\epsilon_1=3i,~ \epsilon_2=-3i,~ \epsilon_3=3i, ~\epsilon_4=-3i.
\eea
The second order degenerate perturbation theory is captured by the following matrix,
\bea
h^{M,(2)}&=&
\left(\begin{array}{c}
v_+^\dagger\\
v_-^\dagger
\end{array}
\right)
\big[\mathcal{H}^{M,(2)}_F+\mathcal{H}^{M,(1)}_F\sum_{i=1}^4 \frac{u_iu_i^\dagger}{E_0-\epsilon_i} \mathcal{H}^{M,(1)}_F\big]\nn\\
&&~~~~~~~~~~~~~~~~~~~~~~~~~~~~~~~~~~~~~\times(v_+\,v_-),
\eea
in which $E_0=0$ are the eigenvalues of $\mathcal{H}^{M,(0)}_F$ for $v_{\pm}$.
Calculations show that 
\bea
h^{M,(2)}=i\frac{4}{27}\Gamma^\prime \Delta^2 \sigma_3,
\eea
in which $\sigma_3$ is the third Pauli matrix.
This shows that to the leading nonvanishing order,
\bea
m_1=\frac{4}{27}\Gamma^\prime \Delta^2.
\label{eq:m1_perturbative}
\eea
We have numerically calculated the eigenvalues of $M(\Delta)\mathcal{H}_F(\Delta)$ and the result for $m_1$ is displayed in Fig. \ref{fig:m1_Delta}.
As can be seen from Fig. \ref{fig:m1_Delta}, the numerical results agree well with Eq. (\ref{eq:m1_perturbative}).

\subsubsection{Nonzero wavevectors and the spin wave dispersions}

Next, we consider nonzero wavevectors and diagonalize the matrix kernel $H_F+\delta H(k)$ in Eq. (\ref{eq:f_fourier}).
We will consider the long wavelength limit $k\ll 1$ where the lattice constant has been taken as $1$.
As can be seen from Eq. (\ref{eq:deltaHk}), the matrix elements of $\delta H(k)$ are very small in the long wavelength limit,
hence $\delta H(k)$ can be treated as a perturbation of $\mathcal{H}_F$.

Multiplying with the symplectic matrix, the first order degenerate perturbation is  implemented by the following $2\times 2$ matrix
\bea
\delta h^{(1)}(k)=P_1^{(0)}M^{(0)} \delta H(\Delta=0,k)P_1^{(0)},
\eea
in which we have taken $\Delta=0$ since we are only interested in the leading nonvanishing order terms in $\Delta$.
Straightforward calculations show that the eigenvalues of $\delta h^{(1)}(k)$ are $\pm i\frac{1}{6} \Gamma S^2 k^2$.
Thus, by keeping only the lowest-lying spin wave, the spin wave Lagrangian in Eq. (\ref{eq:f_fourier}) becomes
\begin{flalign}
&L=-f_0+S\sum_k\frac{1}{2} (\xi_\theta(k) \partial_t \xi_\phi(-k) -\xi_\phi(k) \partial_t \xi_\theta(-k) )\nn\\
& -\frac{1}{2}\Gamma S^2 \sum_k (\frac{4}{27} \Delta^2+\frac{1}{3}k^2)[\xi_\theta(k)\xi_\theta(-k)+\xi_\phi(k)\xi_\phi(-k)],
\label{eq:f_fourier_low}
\end{flalign}
in which 
\bea
\xi_\theta(k)=N^T(k)v_1,~\xi_\phi(k)=N^T(k)v_2,
\eea
where $N(k)$ is defined in Eq. (\ref{eq:Nk}).

Finally, we rewrite the spin wave Hamiltonian (i.e., the second line in Eq. (\ref{eq:f_fourier_low})) in real space in the continuum limit.
The summation over $k$ can be converted to $\sum_n=\frac{1}{3}\int dx$ where $x$ is the real space coordinate in the continuum limit.
The momentum $k$ can be converted to $i\partial_n=3i\partial_x$.
Using these, we see that the spin wave Hamiltonian $H_{sw}$ in the real space is
\begin{flalign}
H_{sw}=\Gamma S^2 \int dx[ \frac{1}{2} (\partial_x \xi_\theta)^2+\frac{1}{2} (\partial_x \xi_\phi)^2+ \frac{2}{81}\Delta^2 (\xi_\theta^2+\xi_\phi^2)],
\label{eq:Hsw_equator}
\end{flalign}
in which $\xi_\theta(j),\xi_\phi(j)$ is a pair of canonical conjugates satisfying $[\xi_\theta(j),\xi_\phi(j^\prime)]=i\delta_{jj^\prime}\frac{1}{S}$.
From Eq. (\ref{eq:Hsw_equator}) and the fact that $\hbar=1/S$,
the dispersion of the spin wave can be obtained as
\bea
E(k)=\Gamma S (k^2+\frac{4}{81}\Delta^2).
\eea
Since the spin wave mass $\frac{4}{81}\Gamma S \Delta^2$ is very small, it would be very difficult to determine numerically (for example, in DMRG numerics). 
We note that the path integral calculations to derive the spin wave Hamiltonian in Eq. (\ref{eq:Hsw_equator}) is equivalent with the Bogoliubov transformation based on the Holstein-Primakoff transformation as explained in detail in Appendix \ref{app:HP_equiv}.


\subsection{DMRG numerics}
\label{sec:numerics_OhD3}

\begin{table}
                \begin{tabular}[t]{|c|c|c|c|c|}
		  \multicolumn{5}{c}{}\\ \hline
$E(S=1)$ & No field & $h_{\hat{x}}=10^{-4}$ & $h_{\hat{n}_{a}}=10^{-4}$ & $h_{\hat{n}_{I}}=10^{-4}$ \\ \hline
$E_1$ &\tikzmark{top left 1a} -12.01911&\tikzmark{top left 1b} -12.02010&\tikzmark{top left 1c}  -12.02067\tikzmark{bottom right 1c}&\tikzmark{top left 1d} -12.02106\\
$E_2-E_1$ & $1.57\cdot 10^{-4}$ & $1.47\cdot 10^{-4}$ & $1.17\cdot 10^{-3}$ & $2.3\cdot 10^{-4}$\tikzmark{bottom right 1d}\\
$E_3-E_1$ & $1.57\cdot 10^{-4}$ & $2.21\cdot 10^{-4}$ & $1.17\cdot 10^{-3}$ & $2.06\cdot 10^{-3}$\\
$E_4-E_1$ & $1.57\cdot 10^{-4}$ & $4.07\cdot 10^{-4}$\tikzmark{bottom right 1b}& $1.21\cdot 10^{-3}$ & $2.06\cdot 10^{-3}$\\
$E_5-E_1$ & $3.38\cdot 10^{-4}$ & $2.09\cdot 10^{-3}$ & $2.36\cdot 10^{-3}$ & $2.22\cdot 10^{-3}$\\
$E_6-E_1$ & $3.38\cdot 10^{-4}$ & $2.21\cdot 10^{-3}$ & $2.40\cdot 10^{-3}$ & $2.25\cdot 10^{-3}$\\
$E_7-E_1$ & $3.38\cdot 10^{-4}$ & $2.28\cdot 10^{-3}$ & $2.40\cdot 10^{-3}$ & $4.16\cdot 10^{-3}$\\
$E_8-E_1$ & $5.55\cdot 10^{-4}$\tikzmark{bottom right 1a}& $2.41\cdot 10^{-3}$ & $3.60\cdot 10^{-3}$ & $4.30\cdot 10^{-3}$\\
$E_9-E_1$ & $4.33\cdot 10^{-3}$ & $4.34\cdot 10^{-3}$ & $4.60\cdot 10^{-3}$ & $4.50\cdot 10^{-3}$\\
$E_{10}-E_1$ & $4.33\cdot 10^{-3}$ & $4.48\cdot 10^{-3}$ & $5.56\cdot 10^{-3}$ & $5.32\cdot 10^{-3}$\\ \hline
                \end{tabular}
			 	\DrawBox[ultra thick, red]{top left 1a}{bottom right 1a}
                \DrawBox[ultra thick, blue]{top left 1b}{bottom right 1b}
                \DrawBox[ultra thick, green]{top left 1c}{bottom right 1c}
                \DrawBox[ultra thick, orange]{top left 1d}{bottom right 1d}
                \hfill
                \begin{tabular}[t]{|c|c|c|c|c|}
		  \multicolumn{5}{c}{}\\ \hline
$E(S=3/2)$ & No field & $h_{\hat{x}}=10^{-4}$ & $h_{\hat{n}_{a}}=10^{-4}$ & $h_{\hat{n}_{I}}=10^{-4}$ \\ \hline
$E_1$ &\tikzmark{top left 2a} -26.99084&\tikzmark{top left 2b} -26.99237&\tikzmark{top left 2c} -26.99342 \tikzmark{bottom right 2c}&\tikzmark{top left 2d} -26.99389\\
$E_2-E_1$ & $6.6\cdot 10^{-5}$ & $6.3\cdot 10^{-5}$ & $1.78\cdot 10^{-3}$ & $8.2\cdot 10^{-5}$\tikzmark{bottom right 2d}\\
$E_3-E_1$ & $6.6\cdot 10^{-5}$ & $8.2\cdot 10^{-5}$ & $1.78\cdot 10^{-3}$ & $3.09\cdot 10^{-3}$\\
$E_4-E_1$ & $6.6\cdot 10^{-5}$ & $1.48\cdot 10^{-4}$\tikzmark{bottom right 2b}& $1.78\cdot 10^{-3}$& $3.09\cdot 10^{-3}$\\
$E_5-E_1$ & $1.34\cdot 10^{-4}$ & $3.11\cdot 10^{-3}$ & $3.57\cdot 10^{-3}$ & $3.15\cdot 10^{-3}$\\
$E_6-E_1$ & $1.34\cdot 10^{-4}$ & $3.16\cdot 10^{-3}$ & $3.57\cdot 10^{-3}$ & $3.15\cdot 10^{-3}$\\
$E_7-E_1$ & $1.34\cdot 10^{-4}$ & $3.18\cdot 10^{-3}$ & $3.57\cdot 10^{-3}$ & $6.21\cdot 10^{-3}$\\
$E_8-E_1$ & $2.04\cdot 10^{-4}$\tikzmark{bottom right 2a}& $3.23\cdot 10^{-3}$ & $1.27\cdot 10^{-3}$ & $6.26\cdot 10^{-3}$\\
$E_9-E_1$ & $1.00\cdot 10^{-2}$ & $1.01\cdot 10^{-2}$ & $1.21\cdot 10^{-2}$ & $1.02\cdot 10^{-2}$\\
$E_{10}-E_1$ & $1.00\cdot 10^{-2}$ & $1.01\cdot 10^{-2}$ & $5.37\cdot 10^{-2}$ & $1.10\cdot 10^{-2}$\\ \hline
                \end{tabular}
			 	\DrawBox[ultra thick, red]{top left 2a}{bottom right 2a}
                \DrawBox[ultra thick, blue]{top left 2b}{bottom right 2b}
                \DrawBox[ultra thick, green]{top left 2c}{bottom right 2c}
                \DrawBox[ultra thick, orange]{top left 2d}{bottom right 2d}				\hfill
\caption{Energies of the first ten lowest lying states computed with DMRG. The data refer to $L=18$ sites, $J=0$, and $\phi=0.2\pi$.
The energies enclosed by the colored squares are approximately degenerate.
}
\label{table:OhD3}
\end{table}

\begin{figure*}[htbp]
\includegraphics[width=12.0cm]{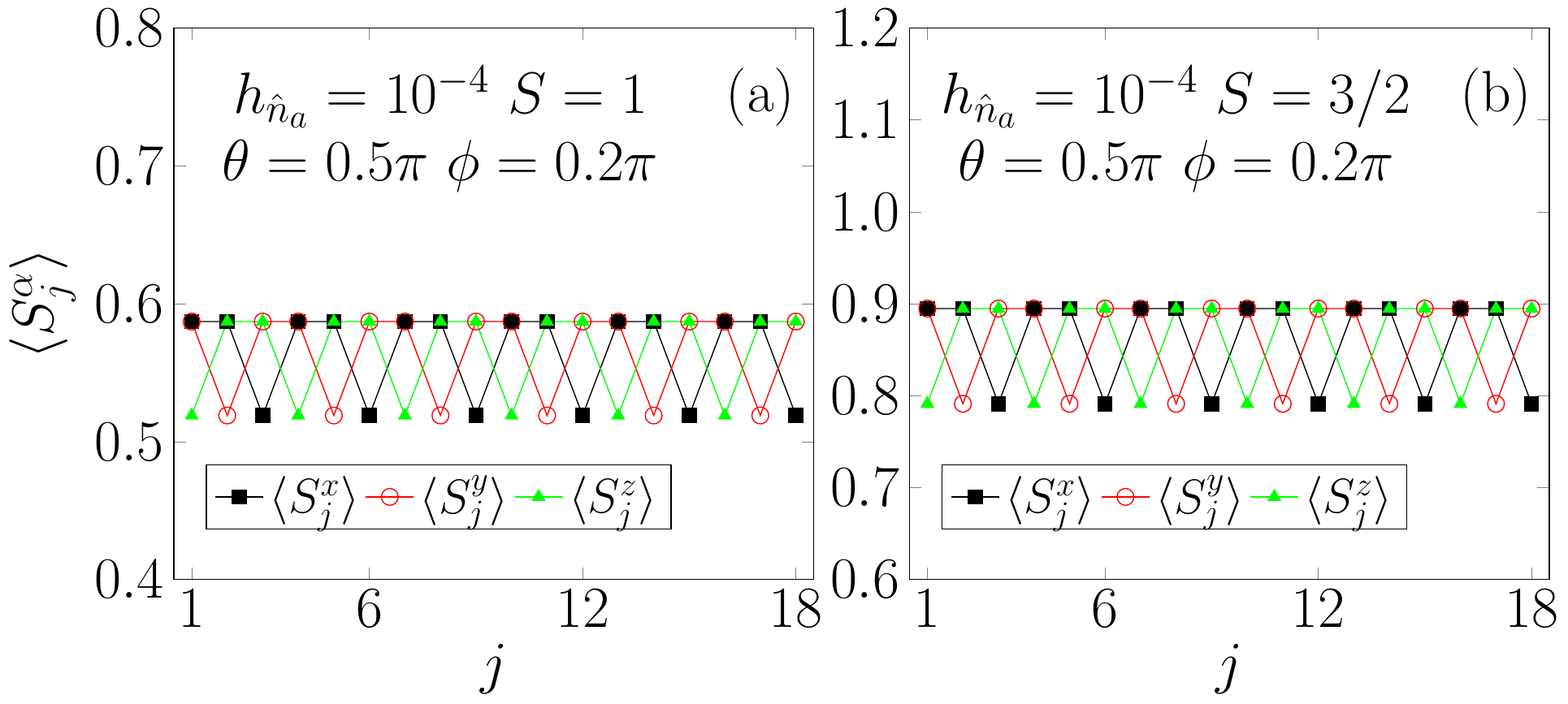}
\caption{Spin expectation values $\left<S_i^\alpha\right>$ ($\alpha=x,y,z$) under a small field $h_{\hat{n}_a}=10^{-4}$ along $(1,1,1)$-direction 
at a representative point $(\theta=\pi/2,\phi=0.2\pi)$ in the $O_h\rightarrow D_3$ phase
 for (a) $S=1$, and (b) $S=3/2$.
 The parametrization $(\theta,\phi)$ is defined in Eq. (\ref{eq:parametize_theta_phi}).
DMRG simulations are performed on a system of  $L=18$ sites.
} \label{fig:OhD3_spin_aligns_111}
\end{figure*}

\begin{figure*}[htbp]
\includegraphics[width=12.0cm]{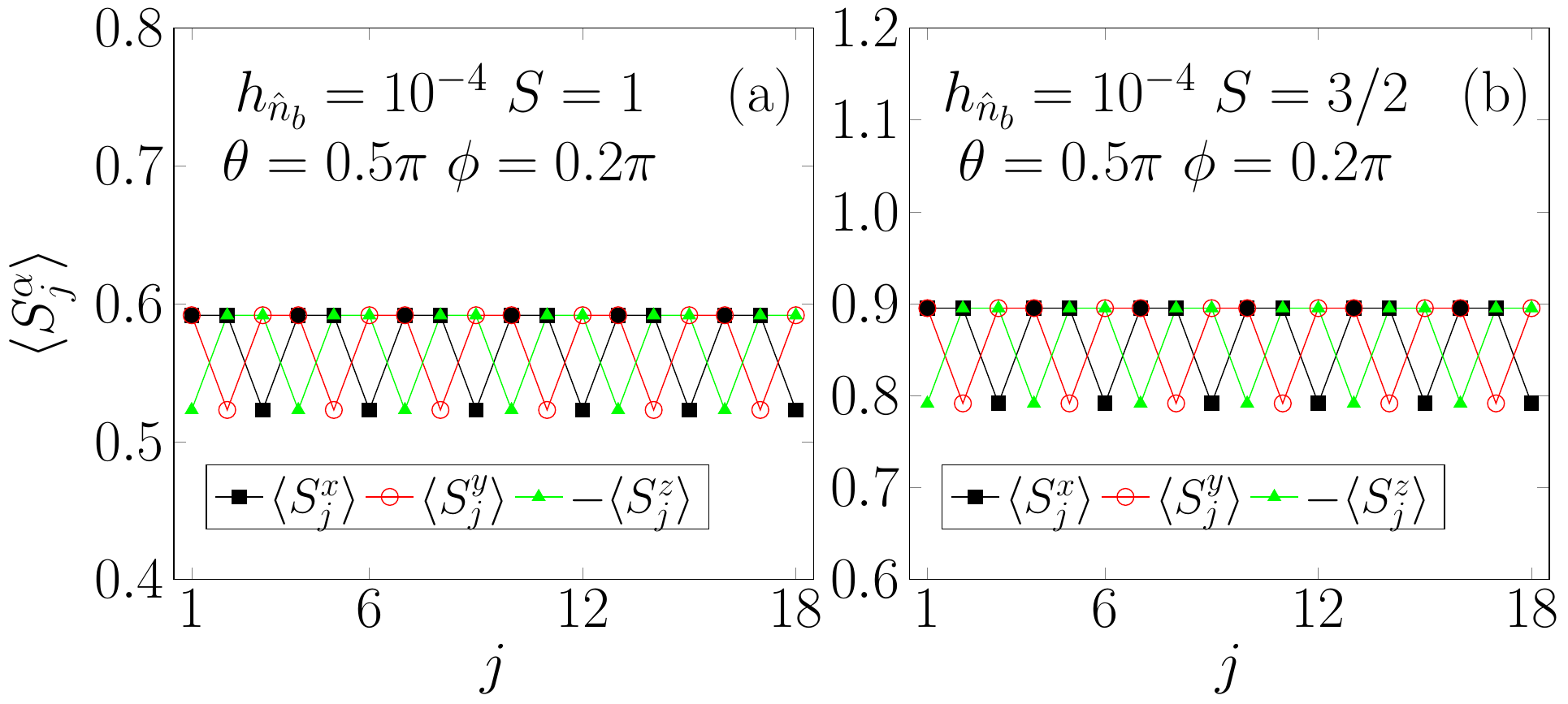}
\caption{Spin expectation values $\left<S_i^\alpha\right>$ ($\alpha=x,y,z$) under a small field $h_{\hat{n}_b}=10^{-4}$ along $(1,1,-1)$-direction 
at a representative point $(\theta=\pi/2,\phi=0.2\pi)$ in the $O_h\rightarrow D_3$ phase
 for (a) $S=1$, and (b) $S=3/2$.
 The parametrization $(\theta,\phi)$ is defined in Eq. (\ref{eq:parametize_theta_phi}).
DMRG simulations are performed on a system of  $L=18$ sites.
} \label{fig:OhD3_spin_aligns_11m1}
\end{figure*}

In this section, we present DMRG numerical results for $S=1,3/2$, which provide evidence for the revealed $O_h\rightarrow D_3$ symmetry breaking based on a classical analysis.

Table \ref{table:OhD3} displays the results for the energies of the ten lowest eigenstates under different magnetic fields at a representative point $(J=0,\phi=0.2\pi)$ in the $O_h\rightarrow D_3$ phase,
in which the first and second tables are for $S=1$ and $S=3/2$, respectively.
DMRG is performed on a system of $L=18$ sites in obtaining the data. 
As can be clearly seen from Table \ref{table:OhD3}, the system is approximately eight-fold degenerate at zero field,
with a ground state energy splitting (characterized by $E_8-E_1$) about one order of magnitude smaller than the excitation gap $E_9-E_1$,
which is consistent with the eight-fold degeneracy predicted by the $O_h\rightarrow D_3$ symmetry breaking as discussed in Sec. \ref{sec:sym_break_equator}.

To test the pattern of the spin alignments as shown in Fig. \ref{fig:spin_orders}, we apply  small magnetic fields $h_{\hat{x}}$, $h_{\hat{n}_a}$ and $h_{\hat{n}_I}$  along  $\hat{x}$, $\hat{n}_a$, and $\hat{n}_I$-directions (within the six-sublattice rotated frame), respectively, where $\hat{n}_a$ and $\hat{n}_I$ are defined in Eq. (\ref{eq:na_nI}).
The magnitude of the field $h=10^{-4}$ is chosen to satisfy
$\Delta E\ll  S|h|L \ll E_g$, in which $L$ is the system size, $E_g$ is the excitation gap, and $\Delta E$ is the finite size splitting of the ground state octet at zero field.  
Such choice of $h$ ensures a degenerate perturbation within the eight-dimensional ground state subspace, and at the same time, no mixing between the ground states and the excited states is induced. 
Hence, it is a thermodynamically small field which only perturbs the ground state subspace. 

As can be read from Fig. \ref{fig:spin_orders}, 
the field $h_{\hat{x}}$ is predicted to lower the energies of the four states located at vertices $(1,\pm1,\pm1)$;
$h_{\hat{n}_a}$ lowers the energy of state at $(1,1,1)$;
and $h_{\hat{n}_I}$ lowers the energies of the two states at $(1,\pm1,-1)$.
Indeed, as can be seen from Table \ref{table:OhD3}, 
the ground state degeneracy becomes $4$-, $1$- and $2$-fold under the fields $h_{\hat{x}}$, $h_{\hat{n}_a}$, and $h_{\hat{n}_I}$, respectively, 
which are consistent with the above analysis.
This provides further evidence for the predicted $O_h\rightarrow D_3$ symmetry breaking.

In addition, we have also directly measured the expectation values of the spin operators under the fields $h_{\hat{n}_a}$ and $h_{\hat{n}_b}$ where $\hat{n}_b$ is along the $(1,1,-1)$-direction.
The results are displayed in Figs. (\ref{fig:OhD3_spin_aligns_111}, \ref{fig:OhD3_spin_aligns_11m1}). 
According to the discussions in Secs. (\ref{sec:classical_equator}, \ref{sec:sym_break_equator}), since the vertices located at $(1,1,1)$ and $(1,1,-1)$ are picked out by $h_{\hat{n}_a}$ and $h_{\hat{n}_b}$, respectively,
the spin alignments are predicted to be:
\bea
\left<\vec{S}_{1+3n}\right>&=&(a,a,b)^T,\nn\\
\left<\vec{S}_{3+3n}\right>&=&(a,b,a)^T,\nn\\
\left<\vec{S}_{3+3n}\right>&=&(b,a,a)^T,
\label{eq:aab}
\eea
for $h_{\hat{n}_a}$;
and 
\bea
\left<\vec{S}_{1+3n}\right>&=&(a,a,-b)^T,\nn\\
\left<\vec{S}_{2+3n}\right>&=&(a,b,-a)^T,\nn\\
\left<\vec{S}_{3+3n}\right>&=&(b,a,-a)^T,
\label{eq:aamb}
\eea
for $h_{\hat{n}_b}$.
Indeed, Fig. \ref{fig:OhD3_spin_aligns_111} (Fig. \ref{fig:OhD3_spin_aligns_11m1}) is consistent with the pattern in Eq. (\ref{eq:aab}) (Eq. (\ref{eq:aamb})).

\section{The N\'eel phase for $J>0$}

In this section, we perform a combination of classical and spin wave analysis for $J>0$ in the vicinity of the FM2 point in Fig. \ref{fig:phase}.
Since the spin alignments exhibit an antiferromagnetic pattern in the original frame,
the region corresponds to a N\'eel phase.
The mass of the lowest spin wave is calculated to the leading nonvanishing order in an expansion over $J$ and $\Delta$.
Throughout this section, we work in the six-sublattice rotated frame unless otherwise stated.

\subsection{Classical analysis and spin wave theory}

The saddle point equations have been derived in Eqs. (\ref{eq:partial_F1},\ref{eq:partial_F2},\ref{eq:partial_F3}).
Assuming the same pattern of spin alignments and relations between $\lambda_i$'s ($1\leq i\leq 3$) as those in Eq. (\ref{eq:spins_OhD4}), the saddle point equations reduce to
\bea
-(\lambda+K^\prime+2J^\prime) a-\Gamma^\prime b&=&0,\nn\\
-2\Gamma^\prime a-(\lambda+2J^\prime)b&=&0,\nn\\
2a^2+b^2-1&=&0.
\label{eq:Saddle_Neel}
\eea
Since there are three variables and three equations, a solution in general exists.
On the other hand, to confirm that this is a minimum of the free energy,
we still need to show that the eigenvalues of the Hessian matrix are all positive.
We will do a perturbative analysis and demonstrate that this is true at least in the vicinity of the FM2 point.

For simplicity, let's first take $\Delta=0$ and turn on a small $J>0$.
The solution of Eq. (\ref{eq:Saddle_Neel}) is given by
\bea
\lambda=-2\Gamma^\prime-2J^\prime,~
a=b=\frac{1}{\sqrt{3}}.
\label{eq:classical_Neel}
\eea
Following the same logic in Sec. \ref{sec:classical_equator}, 
we define the matrix
\bea
\mathcal{H}_F(J)=P(J) H_F(J) P(J),
\label{eq:HFJ}
\eea
in which $H_F(J)$ is the $9\times 9$ Hessian matrix of the free energy in Eq. (\ref{eq:energy_KHG}),
and $P(J)\equiv P(J=0)$ is given by Eq. (\ref{eq:projection_classical})
where $\vec{r}_i=\hat{n}_a$ ($i=1,2,3$) with $\hat{n}_a=\frac{1}{\sqrt{3}}(1,1,1)^T$.
Taking the two gapless acoustic eigenvectors $v_1$ and $v_2$ (defined in Eq. (\ref{eq:v1v2})) as the zeroth order  vectors,
the first order degenerate perturbation  matrix is given by
\bea
h^{(1)}(J)=\left(\begin{array}{c}
v_1^T\\
v_2^T
\end{array}\right)
\Delta \mathcal{H}_F(J)
(v_1~v_2),
\eea
in which 
\bea
\Delta \mathcal{H}_F(J)=\mathcal{H}_F(\Delta=0,J)-\mathcal{H}_F(\Delta=0,J=0).
\eea
Straightforward calculations show that 
\bea
h^{(1)}(J)=2JS^2 \sigma_0,
\eea
where $\sigma_0$ is the $2\times 2$ identity matrix. 
Thus the eigenvalues of the Hessian matrix are positive when $J>0$,
thereby confirming the solution in  Eq. (\ref{eq:classical_Neel})
to be at least a local minimum.
In fact, numerical minimization of the free energy shows that it is also a global minimum
as discussed in Appendix \ref{app:numerics_classical}. 

We note that the above  analysis can be extended to the case where both $J$ and $\Delta$ are nonzero but small (i.e.,  $J,|\Delta|\ll 1$).
To the lowest nonvanishing order in perturbation, the wavefunction is unchanged. 
Hence the eigenvalues are additive for $J$ and $\Delta$.
Therefore, two lowest eigenvalues of the Hessian matrix are both $2JS^2+\frac{4}{27}\Delta^2\Gamma S^2$.

We also briefly discuss the symmetry breaking in the N\'eel phase.
The unbroken symmetry group is the same as the $O_h\rightarrow D_3$ phase, since the spins have the same pattern of alignments.
As discussed in Sec. \ref{sec:review_Oh}, the full symmetry group for a nonzero $J$ is $D_{3d}$ (modulo $T_{3a}$),
therefore, the symmetry breaking in the N\'eel phase is
\bea
D_{3d}\rightarrow D_3.
\label{eq:D3d_D3}
\eea
Since $|D_{3d}/D_3|=2$, there are two degenerate ground states.
The ``center of mass" directions for the three spins within a unit cell in the two degenerate states are plotted as the two solid light blue circles in Fig. \ref{fig:spin_orders}.

We make a comment on the spin ordering in the original frame.
Rotating the spin orientations in Eq. (\ref{eq:spins_OhD4}) back to the original frame using Eq. (\ref{eq:6rotation}),
it is straightforward to verify that the spins align in a N\'eel pattern with a two-site periodicity, i.e.,
\bea
\vec{S}_{1+2n}=S(a,a,b)^T,~\vec{S}_{2+2n}=S(-a,-a,-b)^T.
\label{eq:Neel_orig}
\eea
Thus this phase is termed as ``N\'eel" in the phase diagram in Fig. \ref{fig:phase}.

Finally we build up a spin wave theory for the small fluctuations around the classical configurations.
To obtain the spin wave mass, we need to calculate the eigenvalues of the matrix $M(\Delta,J)\mathcal{H}_F(\Delta,J)$.
The contribution from the $\Delta$-part is the same as Sec. \ref{sec:SW_equator}.
For the $J$-part, within first order perturbation theory, the contribution is the same as the eigenvalues of $\mathcal{H}_F$ as can be seen from Eq. (\ref{eq:hM1_equator_relation}).
Therefore, the spin wave Hamiltonian for the lowest spin wave is
\bea
H_{sw}&=& \frac{1}{2}\Gamma S^2\int dx[ (\partial_x \xi_\theta)^2+(\partial_x \xi_\phi)^2]\nn\\
&&+(\frac{2}{81}\Gamma \Delta^2+\frac{1}{3}J)S^2\int dx (\xi_\theta^2+\xi_\phi^2),
\label{eq:Hsw_Neel}
\eea
in which $\xi_\theta(j),\xi_\phi(j)$ is a pair of canonical conjugates satisfying $[\xi_\theta(j),\xi_\phi(j^\prime)]=i\delta_{jj^\prime}\frac{1}{S}$.

\subsection{DMRG numerics}
\label{sec:Neel_numerics}

\begin{table}
                \begin{tabular}[t]{|c|c|c|}
		  \multicolumn{3}{c}{}\\ \hline
$E(S=1)$ & No field & $h_{\hat{n}_{a}}=10^{-4}$\\ \hline
$E_1$ & \tikzmark{top left 1a}-17.96054 &\tikzmark{top left 1b} -17.96348\tikzmark{bottom right 1b}\\
$E_2-E_1$ & $4.57\cdot 10^{-6}$\tikzmark{bottom right 1a} & $5.887\cdot 10^{-3}$\\
$E_3-E_1$ & $6.054\cdot 10^{-2}$ & $6.056\cdot 10^{-2}$\\
$E_4-E_1$ & $6.054\cdot 10^{-2}$ & $6.110\cdot 10^{-2}$\\
$E_5-E_1$ & $6.515\cdot 10^{-2}$ & $6.518\cdot 10^{-2}$\\
\hline
                \end{tabular}
                \DrawBox[ultra thick, red]{top left 1a}{bottom right 1a}
                \DrawBox[ultra thick, blue]{top left 1b}{bottom right 1b}
                \hfill
                \begin{tabular}[t]{|c|c|c|}
		  \multicolumn{3}{c}{}\\ \hline
$E(S=3/2)$ & No field & $h_{\hat{n}_{a}}=10^{-4}$\\ \hline
$E_1$ & \tikzmark{top left 1a}-39.50646& \tikzmark{top left 1b}-39.51099\tikzmark{bottom right 1b}\\
$E_2-E_1$ & $8.9\cdot 10^{-12}$\tikzmark{bottom right 1a} & $9.060\cdot 10^{-3}$\\
$E_3-E_1$ & $8.908\cdot 10^{-2}$ & $8.910\cdot 10^{-2}$\\
$E_4-E_1$ & $8.908\cdot 10^{-2}$ & $8.997\cdot 10^{-2}$\\
$E_5-E_1$ & $9.487\cdot 10^{-2}$ & $9.489\cdot 10^{-2}$\\
\hline
                \end{tabular}
                \DrawBox[ultra thick, red]{top left 1a}{bottom right 1a}
                \DrawBox[ultra thick, blue]{top left 1b}{bottom right 1b}
\caption{Energies of the five lowest lying states computed with
DMRG simulations. The data refer to $L=18$ sites, $\theta=0.4\pi$, and $\phi=0.2\pi$.
The energies enclosed by the colored squares are approximately degenerate.
\label{table:energies_Neel}
}
\end{table}

\begin{figure*}[htbp]
\includegraphics[width=12.0cm]{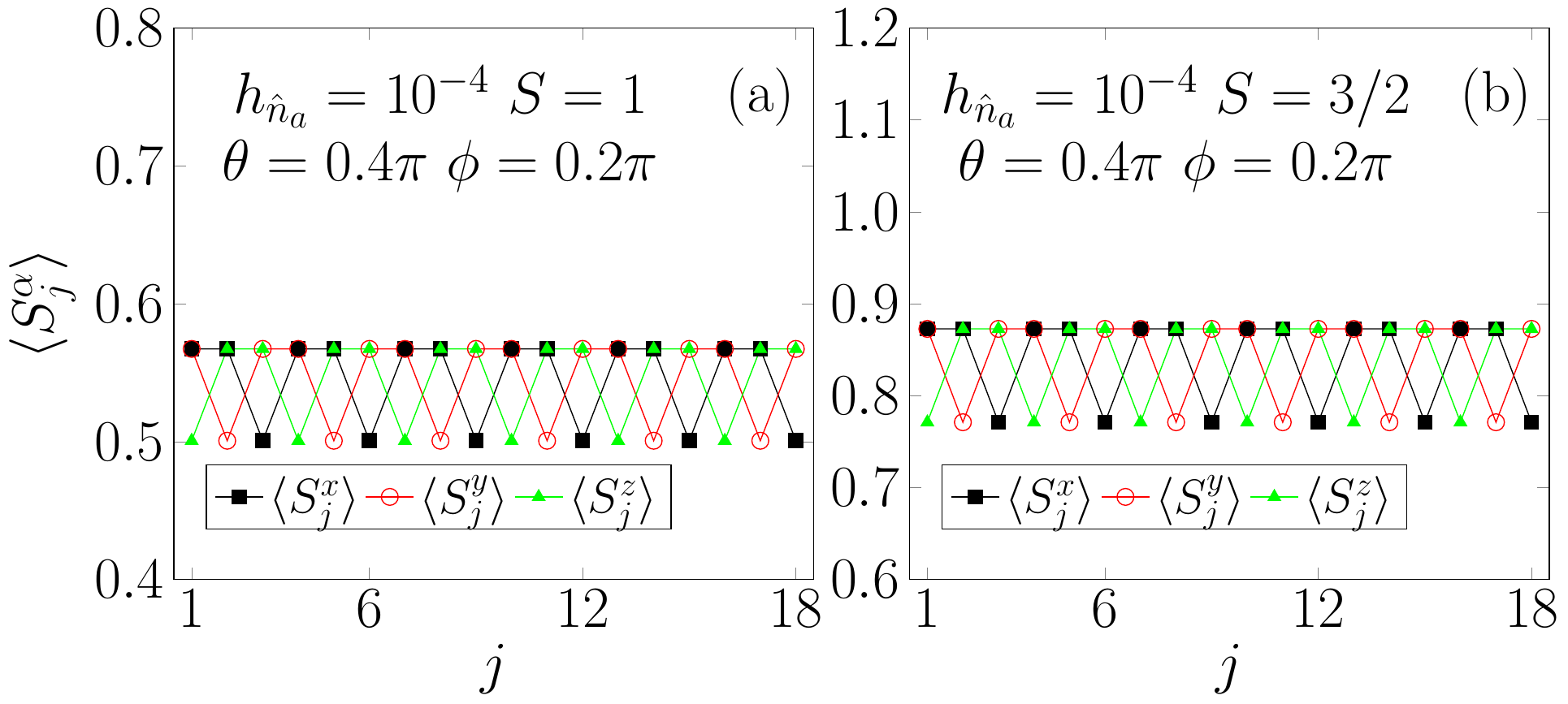}
\caption{Spin expectation values $\left<S_i^\alpha\right>$ ($\alpha=x,y,z$) under a small field $h_{\hat{n}_a}=10^{-4}$ along $(1,1,1)$-direction 
at a representative point $(\theta=0.4\pi,\phi=0.2\pi)$ in the N\'eel phase
 for (a) $S=1$, and (b) $S=3/2$.
 The parametrization $(\theta,\phi)$ is defined in Eq. (\ref{eq:parametize_theta_phi}).
DMRG is performed on a system of  $L=18$ sites.
} \label{Neel_spin_align}
\end{figure*}

In this section, we present DMRG numerical results for $S=1,3/2$, which provide evidence for the revealed $D_{3d}\rightarrow D_3$ symmetry breaking based on a classical analysis.
We proceed similarly as Sec. \ref{sec:numerics_OhD3}.

Table \ref{table:energies_Neel} displays the results for the energies of the five lowest eigenstates under different magnetic fields at a representative point $(\theta=0.4\pi,\phi=0.2\pi)$ in the N\'eel phase,
in which the first and second tables are for $S=1$ and $S=3/2$, respectively,
and $\theta,\phi$ are defined in Eq. (\ref{eq:parametize_theta_phi}).
DMRG is performed for a system of $L=18$ sites in obtaining the data.
On a $L=12$ size system, we have checked that the DMRG results are in agreement with Lanczos Exact Diagonalization.
As can be clearly seen from Table \ref{table:energies_Neel}, the system is approximately two-fold degenerate at zero field,
with a ground state energy splitting (characterized by $E_2-E_1$)  orders of magnitude smaller than the excitation gap $E_3-E_1$,
which is consistent with the two-fold degeneracy predicted by the $D_{3d}\rightarrow D_3$ symmetry breaking as discussed in Eq. (\ref{eq:D3d_D3}).
We have also applied a small magnetic field along the $\hat{n}_a$-direction,
which should be able to pick out the state located at the $(111)$-vertex as shown in Fig. \ref{fig:spin_orders}.
Indeed, as can be seen from Table \ref{table:energies_Neel},
the system becomes nondegenerate when $h_{\hat{n}_a}$ is applied.

In addition, we have also directly measured the expectation values of the spin operators under the fields $h_{\hat{n}_a}$.
The results are displayed in Fig. \ref{Neel_spin_align} (a) for $S=1$ and (b) for $3/2$.
Clearly, the spin alignments revealed in Fig. \ref{Neel_spin_align}
are consistent with the pattern in Eq. (\ref{eq:aab}).

\section{The ``$D_3$-breaking I, II" phases for $J<0$}

In this section, we discuss the ``$D_3$-breaking I, II" phases in the negative $J$ region.
We work within the six-sublattice rotated frame unless otherwise stated.

\subsection{Classical phase diagram}

We first briefly describe the classical phase diagram in the negative $J$ region as shown in Fig.  \ref{fig:phase},
with calculations included in the next two subsections. 
There are two phases denoted as ``$D_3$ breaking I" and ``$D_3$ breaking II".
Both phases break the $D_3$ symmetry albeit in different ways,
hence the ground states are six-fold degenerate.
However, the symmetry breaking patterns are not the same.

To clarify this point, recall that the symmetry group is $G_1\simeq D_{3d}\ltimes 3\mathbb{Z}$  as discussed in Sec. \ref{sec:review_Oh}.
Since $T_{3a}$ is not broken, we consider $G_1^\prime=G_1/\mathopen{<}T_{3a}\mathclose{>}\simeq D_{3d}$ and the spins within a unit cell in what follows.
In the ``$D_3$ breaking I" phase,
the spin orientations in one of the six degenerate ground states are
\bea
\vec{S}_1=S\left(\begin{array}{c}
x\\
y\\
z
\end{array}\right),~
\vec{S}_2=S\left(\begin{array}{c}
-\frac{1}{\sqrt{2}}\\
0\\
\frac{1}{\sqrt{2}}
\end{array}\right),~
\vec{S}_3=S\left(\begin{array}{c}
-z\\
-y\\
-x
\end{array}\right),
\label{eq:order_D6I}
\eea
in which $x^2+y^2+z^2=1$.
As can be checked, the little group of Eq. (\ref{eq:order_D6I}) is generated by $R_I I$.
Hence the symmetry breaking is $D_{3d}\rightarrow \mathopen{<} R_I I \mathclose{>} $.
On the other hand, in the ``$D_3$ breaking II" phase,
the spin orientations in one of the six degenerate ground states are
\bea
\vec{S}_1=S\left(\begin{array}{c}
x\\
y\\
z
\end{array}\right),~
\vec{S}_2=S\left(\begin{array}{c}
m\\
n\\
m
\end{array}\right),~
\vec{S}_3=S\left(\begin{array}{c}
z\\
y\\
x
\end{array}\right),
\label{eq:order_D6II}
\eea
in which $x^2+y^2+z^2=2m^2+n^2=1$.
The little group of Eq. (\ref{eq:order_D6II}) is generated by $TR_I I$,
and the symmetry breaking is $D_{3d}\rightarrow \mathopen{<} TR_I I \mathclose{>} $.
Thus we see that although the symmetry breaking in the two phases are both 
$D_{3d}\rightarrow \mathbb{Z}_2$, the group $\mathbb{Z}_2$ represents different little groups.
We also note that since $D_{3d}/\mathbb{Z}_2 \simeq D_3$,
the two phases both exhibit $D_3$-breaking
which is the origin of the names of the two phases.
The ``center of mass" directions of the three spins within a unit cell in the six degenerate ground states are shown in Fig. \ref{fig:spin_orders},
where the red (dark blue) solid circles correspond to the ``$D_3$-breaking I (II)" phases.

\subsection{The ``$D_3$ breaking I" phase}

\subsubsection{The classical solution}
\label{sec:D3I_classical}

We perform a classical analysis in the ``$D_3$ breaking I" phase.
For simplicity, we consider the $\Delta=0$ case with a small negative $J$.
We will use the normalized parameter $\bar{J}=J/\Gamma$.

We take the trial solution given by Eq. (\ref{eq:order_D6I}) and  assume $\lambda_3=\lambda_1$.
Setting $K=\Gamma$ and plugging the trial solution into Eq. (\ref{eq:partial_F1},\ref{eq:partial_F2},\ref{eq:partial_F3}), the saddle point equations reduce to
\bea
(-\lambda_1+J^\prime) x+\Gamma^\prime z+\frac{1}{\sqrt{2}} (K^\prime+J^\prime)&=&0\nn\\
(\Gamma^\prime+J^\prime-\lambda_1)y-\frac{1}{\sqrt{2}} J^\prime&=&0\nn\\
\Gamma^\prime x+(J^\prime -\lambda_1 )z-\frac{1}{\sqrt{2}} \Gamma^\prime&=&0\nn\\
-(K^\prime+J^\prime)x+J^\prime y+\Gamma^\prime z+\frac{1}{\sqrt{2}}\lambda_2&=&0\nn\\
x^2+y^2+z^2-1&=&0.
\label{eq:Saddle_eq_D6I}
\eea
Since there are five variables $x,y,z,\lambda_1,\lambda_2$ and five equations,
generically a solution exists.
Eq. (\ref{eq:Saddle_eq_D6I}) can be solved perturbatively in an expansion over $J$.
The results up to $O(J^3)$ are
\bea
x&=& -\frac{1}{\sqrt{2}}-\frac{1}{6\sqrt{2}}\bar{J}+\frac{5}{72\sqrt{2}} \bar{J}^2-\frac{7}{432\sqrt{2}} \bar{J}^3+O(\bar{J}^4),\nn\\
y&=& \frac{1}{3\sqrt{2}}\bar{J}-\frac{1}{18\sqrt{2}} \bar{J}^2 +\frac{1}{216\sqrt{2}} \bar{J}^3+O(\bar{J}^4),\nn\\
z&=&\frac{1}{\sqrt{2}}-\frac{1}{6\sqrt{2}}\bar{J}-\frac{1}{72\sqrt{2}} \bar{J}^2+\frac{5}{432\sqrt{2}} \bar{J}^3+O(\bar{J}^4),\nn\\
\label{eq:D3I_3rd_order_A}
\eea
and
\bea
\lambda_1&=&\Gamma^\prime (-2+\frac{1}{2}\bar{J}-\frac{1}{24\sqrt{2}} \bar{J}^2-\frac{1}{144} \bar{J}^3)+O(\bar{J}^4),\nn\\
\lambda_2&=&\Gamma^\prime(-2-\bar{J}-\frac{5}{12} \bar{J}^2+\frac{7}{72} \bar{J}^3)+O(\bar{J}^4).
\label{eq:D3I_3rd_order_B}
\eea
Detailed derivations of Eqs. (\ref{eq:D3I_3rd_order_A},\ref{eq:D3I_3rd_order_B}) are included in Appendix \ref{app:D3I}.

Consider the projected Hessian matrix  defined in Eq. (\ref{eq:HFJ}) for $J<0$.
The perturbation Hamiltonian is
\bea
\Delta\mathcal{H}_F(J)&=&P(J) H_F(J) P(J)\nn\\
&&- P(J=0) H_F(J=0) P(J=0).
\eea
Then the first order degenerate perturbation Hamiltonian is given by
\bea
h^{(1)}(J)=\left(\begin{array}{cc}
v_1^T \Delta\mathcal{H}_F(J) v_1& v_1^T \Delta\mathcal{H}_F(J) v_2\\
v_2^T \Delta\mathcal{H}_F(J) v_1 & v_2^T\Delta \mathcal{H}_F(J) v_2
\end{array}
\right),
\eea
in which $v_1,v_2$ are given by Eq. (\ref{eq:v1v2}) where
\bea
\hat{e}_\theta=(-\frac{1}{\sqrt{2}},0,-\frac{1}{\sqrt{2}})^T,~
\hat{e}_\phi=(0,-1,0)^T.
\label{eq:e_thetaphi_D3I}
\eea
Calculations show that 
\bea
h^{(1)}(J)=-JS^2\left(\begin{array}{cc}
\frac{4}{3}&\frac{2\sqrt{2}}{3}\\
\frac{2\sqrt{2}}{3}&\frac{2}{3}
\end{array}\right).
\label{eq:h1J_D3I}
\eea
The two eigenvalues of $h^{(1)}(J)$ are $0$ and $-2JS^2$.
Thus we see that although the first order perturbation already breaks the degeneracy,
one eigenvalue remains zero up to $O(J)$ and higher order perturbation is needed to obtain a nonzero value.
In fact, calculations show that the first nonvanishing term for this eigenvalue appears at $O(J^3)$.
Here we only mention that the result is $-\frac{1}{2}\Gamma S^2\bar{J}^3$,
and detailed derivations are given in Appendix \ref{app:D3I}.

In summary, the two low-lying eigenvalues are
\bea
-\Gamma S^2\bar{J},~-\frac{1}{2}\Gamma S^2\bar{J}^3,
\label{eq:eigenvalues_D6I}
\eea
which are both positive when $J<0$.
This shows that Eqs. (\ref{eq:D3I_3rd_order_A},\ref{eq:D3I_3rd_order_B}) represent a minimum of the free energy.
Numerical calculations provide evidence for Eqs. (\ref{eq:D3I_3rd_order_A},\ref{eq:D3I_3rd_order_B}) to be a global minimum as discussed in Appendix \ref{app:numerics_classical}. 


\subsubsection{Spin wave theory}
\label{sec:sw_D3I}

\begin{figure*}[htbp]
\includegraphics[width=5.0cm]{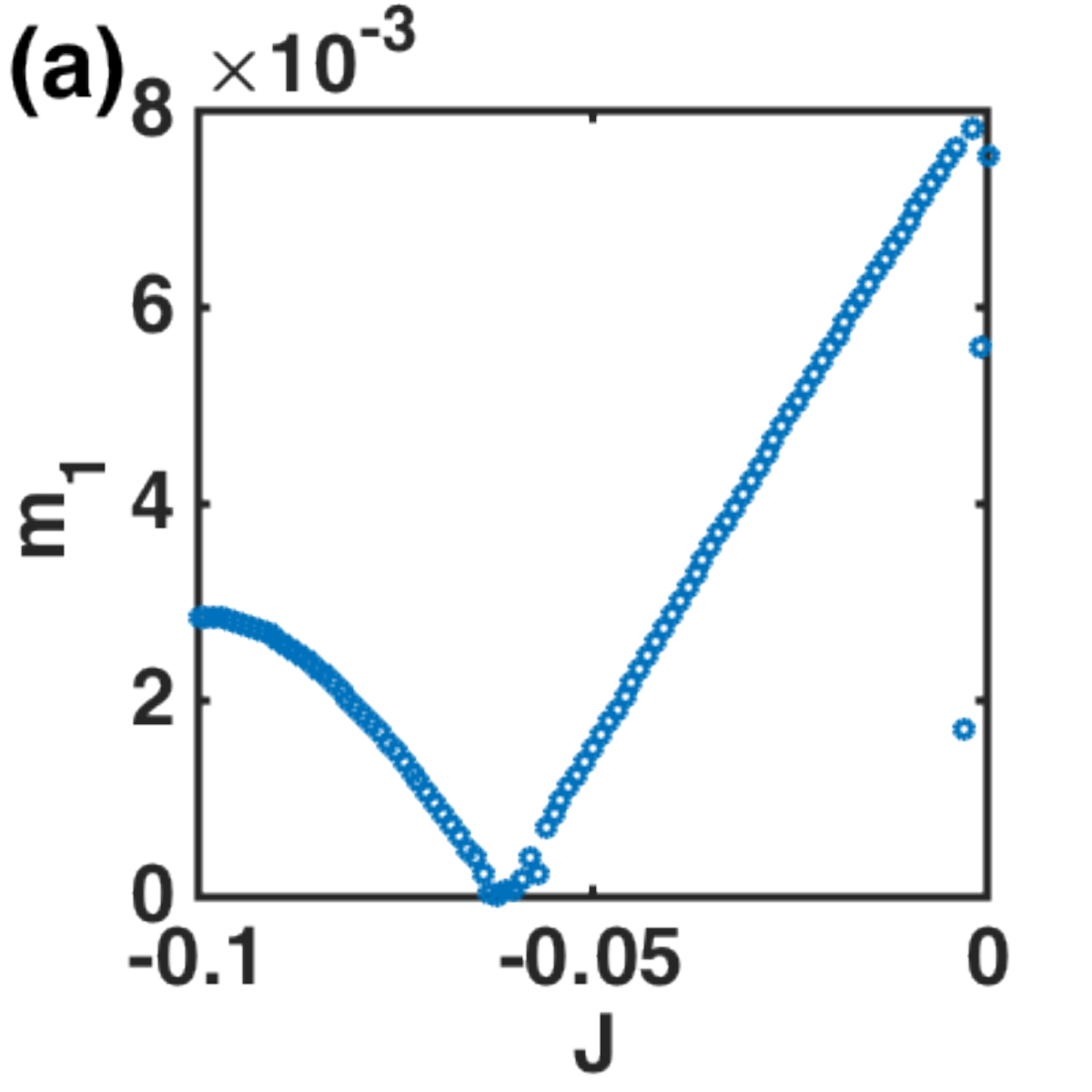}
\includegraphics[width=6.2cm]{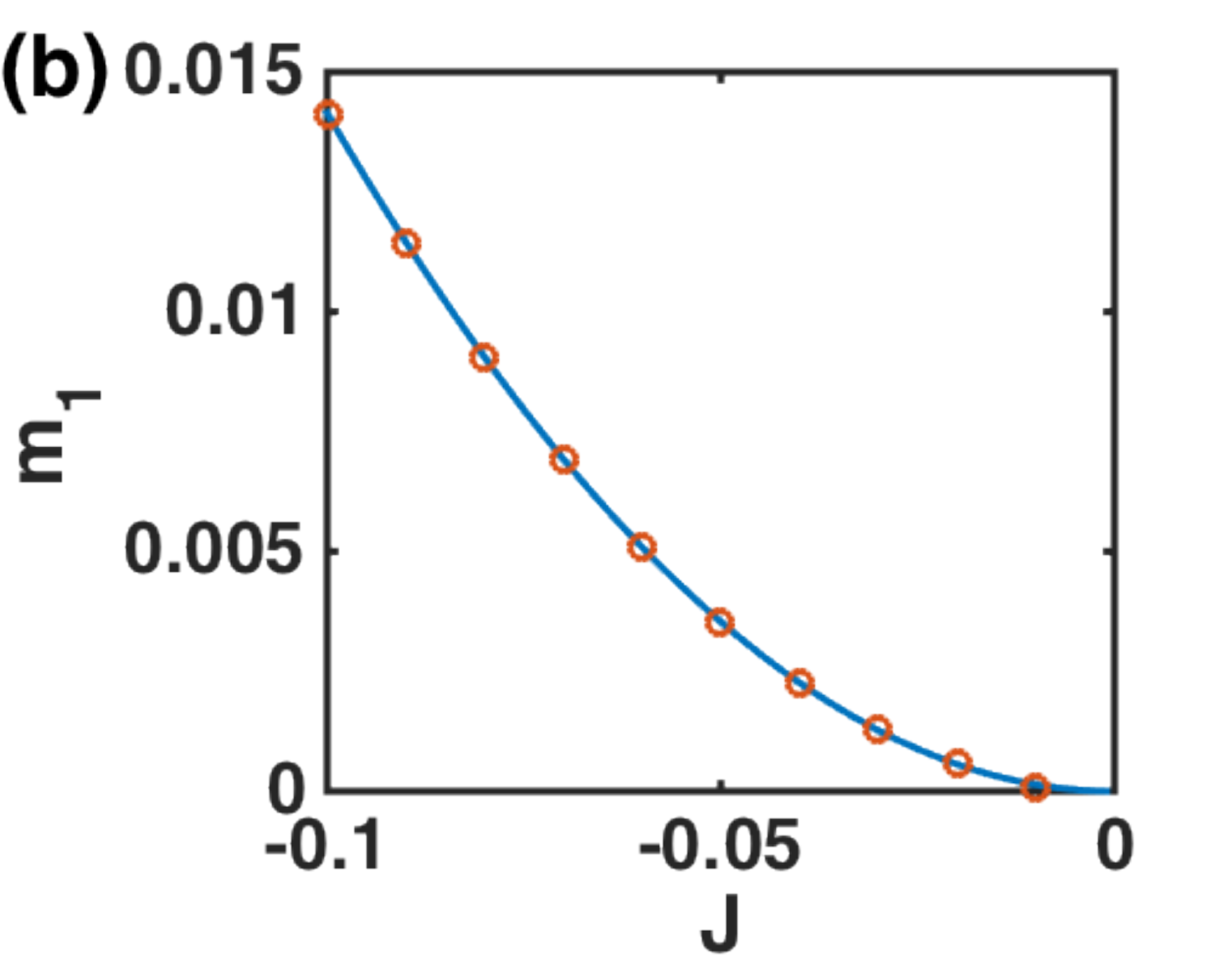}
\includegraphics[width=6cm]{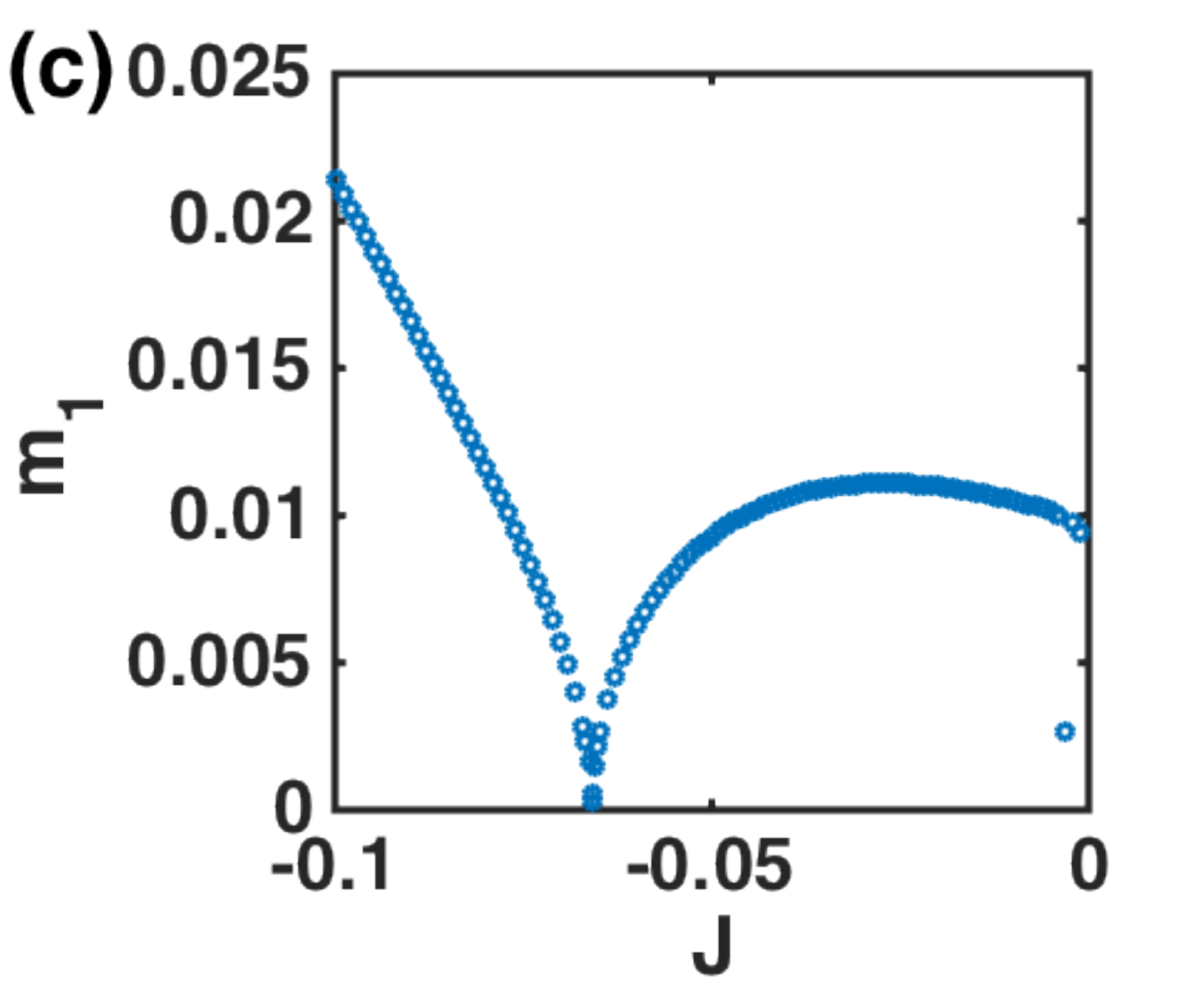}
\caption{Numerically obtained smallest spin wave mass $m_1$ vs. $J$ represented by the hollow circles for (a) $\varphi=0.21\pi$, (b) $\varphi=0.25\pi$,
and (c) $\varphi=0.30\pi$.
In all figures, $m_1$ is in units of $\Gamma S^2$, where $\Gamma=\sin(\varphi)$.
In (b), the solid line represents $\Gamma^\prime \bar{J}^2$ where $\Gamma^\prime=\frac{1}{\sqrt{2}}$.
} 
\label{fig:sw_mass_D3}
\end{figure*}

In this subsection, we  calculate the lowest-lying spin wave mass for the $\Delta=0$ case with a small negative $J$.
Let's first consider the case of a zero wavevector.
Again, we need to diagonalize the Hessian matrix $\mathcal{H}_F(J)$ using symplectic transformations.
As discussed in Sec. \ref{sec:sw_equator_zero_momentum}, the spin wave masses are given by the eigenvalues of $M(J)\mathcal{H}_F(J)$.
We will calculate the smallest spin wave mass up to the leading nonvanishing order of $J$.  

Before proceeding on, notice that the definitions of $v_i$ ($i=1,2$), $w_j$ ($j=1,2,3,4$) are the same as Eq. (\ref{eq:v1v2}) and Eq. (\ref{eq:wi_s}), where $\hat{e}_\theta$ and $\hat{e}_\phi$ should be taken as Eq. (\ref{eq:e_thetaphi_D3I}).
We emphasize that we will use the same notations as Sec. \ref{sec:sw_equator_zero_momentum} for simplicity.
However, the expressions of the quantities are different from those in Eq. (\ref{sec:sw_equator_zero_momentum}),
which are determined by the form of the Hamiltonian and the saddle point solutions.
Let $P_1^{(0)}$ be the projection operation to the subspace spanned by $\{v_1,v_2\}$.
Let $\mathcal{H}^{M,(n)} (J)$ be the order $J^n$ term in the expansion of $M(J)\mathcal{H}_F(J)$ over $J$.
Then the first order degenerate perturbation is given by the following $2\times 2$ matrix,
\bea
h^{M,(1)} (J)=P^{(0)}_1 \mathcal{H}^{M,(1)}_F (J) P^{(0)}_1.
\label{eq:first_D3I_mass}
\eea
According to Eq. (\ref{eq:hM1_equator_relation}), this is simply
\bea
h^{M,(1)} (J)=i\sigma_2 h^{(1)}(J),
\eea
in which $h^{(1)}(J)$ is given by Eq. (\ref{eq:h1J_D3I}), and $i\sigma_2$ is the projection of $M^{(0)}$ to the subspace spanned by $\{v_1,v_2\}$ where $\sigma_\alpha$ ($\alpha=1,2,3$) are the Pauli matrices.
As can be readily checked, since one of the two eigenvalues of $h^{(1)}(J)$  vanishes, the two eigenvalues of $h^{M,(1)} (J)$ are both zero.
Hence, we need to go to second order perturbation.

The second order perturbation is given by the following matrix
\bea
h^{M,(2)}(J)&=&
\left(\begin{array}{c}
v_1^T\\
v_2^T
\end{array}
\right)
\big[\mathcal{H}^{M,(2)}_F+\mathcal{H}^{M,(1)}_F\sum_{i=1}^4 \frac{u_iu_i^\dagger}{E_0-\epsilon_i} \mathcal{H}^{M,(1)}_F\big]\nn\\
&&~~~~~~~~~~~~~~~~~~~~~~~~~~~~~~~~~~~~~\times(v_1\,v_2),
\label{eq:second_D3I_mass}
\eea
in which $v_\pm$ and $u_i$ ($i=1,2,3,4$) are defined in the same way as 
Eq. (\ref{eq:v_pm}) and Eq. (\ref{eq:Def_us});
and the eigenvalues $\epsilon_i$'s ($i=1,2,3,4$) are given by
\bea
\epsilon_1=-3i,~ \epsilon_2=3i,~ \epsilon_3=-3i, ~\epsilon_4=3i.
\eea
Calculations show that 
\bea
h^{M,(2)}(J)=0.
\eea
Hence, the second order perturbation also vanishes, which means that we have to go to the third order perturbation theory.

The third order perturbation matrix $h^{M,(3)}(J)$ is given by
\bea
h^{M,(3)}(J)=\bar{J}^3 \left(\begin{array}{cc}
-\frac{19}{54\sqrt{2}}& \frac{35}{108}\\
-\frac{4}{27}& \frac{19}{54\sqrt{2}}
\end{array}\right).
\label{eq:hM3J}
\eea
Detailed derivation of Eq. (\ref{eq:hM3J}) is included in Appendix \ref{app:D3I_third_order}.

The eigenvalues of $h^{M,(1)}+h^{M,(2)}+h^{M,(3)}$ are $\pm i\Gamma S^2 \bar{J}^2$, which gives
$m_1(\Delta=0,J)=\Gamma S^2 \bar{J}^2$.
In Fig. \ref{fig:sw_mass_D3} (b),
 the hollow circles represent the numerical results for $m_1$ by numerically solving the eigenvalues of $M(J)\mathcal{H}_F(J)$,   
 and the solid line  represents $\Gamma^\prime \bar{J}^2$.
 As can be clearly seen, the numerical results agree well with the obtained perturbative results up to $O(J^2)$.
 
\begin{figure*}[htbp]
\includegraphics[width=15.0cm]{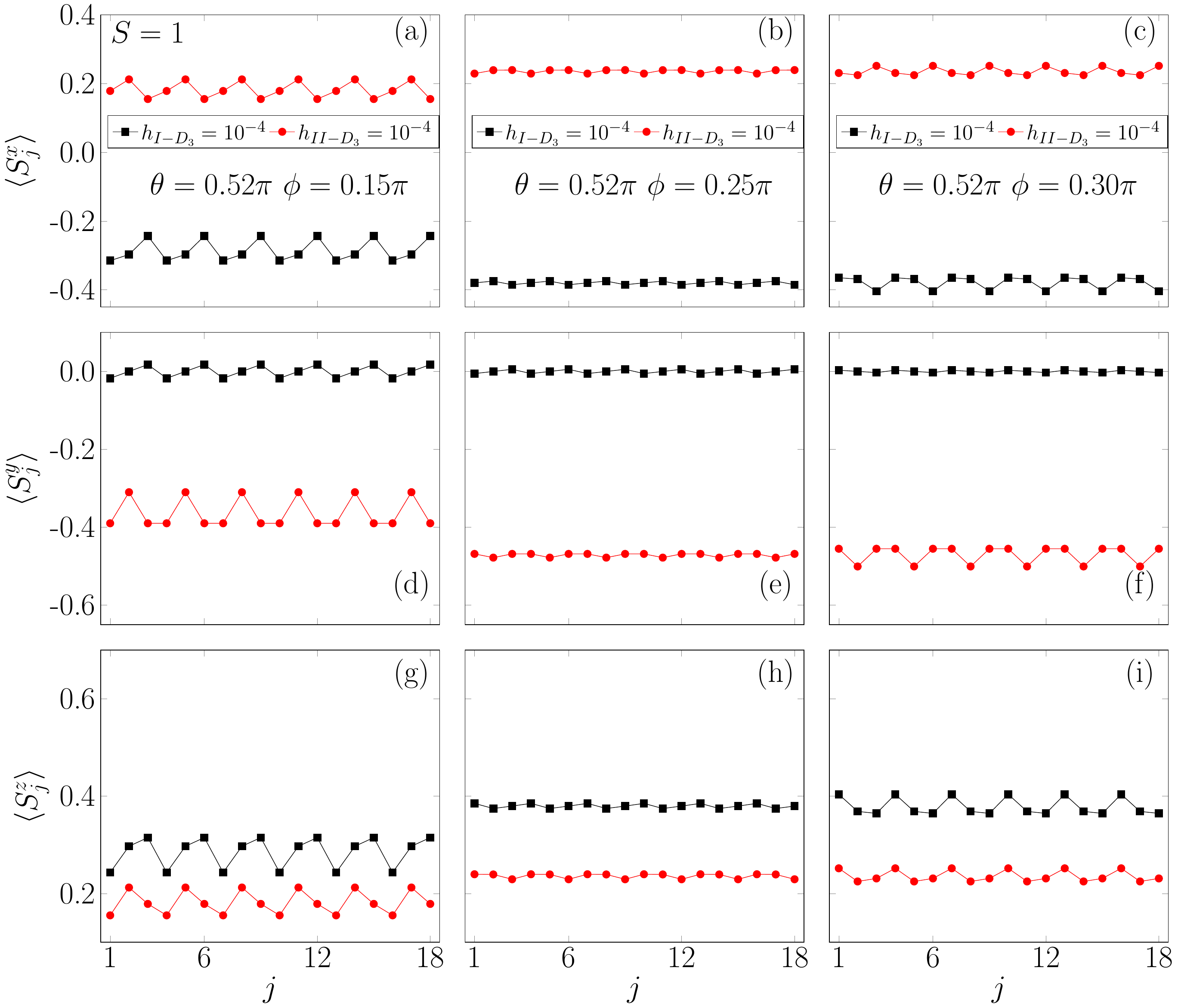}
\caption{(a,b,c) $\langle S_j^x\rangle$, (d,e,f) $\langle S_j^y\rangle$, and (g,h,i) $\langle S_j^z\rangle$ vs $j$ under $h_{\text{I}}$ (black squares) and $h_{\text{II}}$ (red dots) fields for $S=1$ at several different points.
(a,d,g) are for $(\theta=0.52\pi,\phi=0.15\pi)$;
(b,e,h)  for $(\theta=0.52\pi,\phi=0.25\pi)$;
and (c,f,i)  for $(\theta=0.52\pi,\phi=0.30\pi)$.
DMRG numerics are performed on $L=18$ sites with periodic boundary conditions. 
Both $h_{\text{I}}$ and $h_{\text{II}}$ fields are taken to be $10^{-4}$.
} \label{fig:D3_break_Seq1}
\end{figure*}

Based on the above discussions, we are able to obtain the spin wave Hamiltonian for $\Delta=0,|\bar{J}|\ll 1$ as 
\bea
H_{sw}&=& \frac{1}{2}\Gamma S^2\int dx[ (\partial_x \xi_\theta)^2+(\partial_x \xi_\phi)^2]\nn\\
&&+\frac{1}{6}\Gamma S^2\bar{J}^2\int dx (\xi_\theta^2+\xi_\phi^2).
\label{eq:Hsw_D3I}
\eea

\subsection{The ``$D_3$ breaking II" phase}

In this subsection, we discuss the ``$D_3$ breaking II" phase.
To obtain an intuitive understanding, let's start with the case of $\Delta\neq 0, J=0$, and then turn on a small negative $J$.
At $J=0$, the symmetry breaking is $O_h\rightarrow D_3$ and there are eight degenerate ground states.
If $J\neq 0$, since $J$ is planar-like, the two states along $\pm(111)$-direction  in Fig. \ref{fig:spin_orders} will have higher energies than the other six states.
As a result, the two solid light blue circles at the $\pm(111)$ vertices should be removed compared with the $J=0$ case as shown  in Fig. \ref{fig:spin_orders}.
Thus, the ground states now are six-fold degenerate and the spin orientations are slightly distorted away from the $J=0$ case.
This is different from the ``$D_3$ breaking I" phase where the ``center of mass" direction is perpendicular to the $(111)$-direction.
On the other hand, when $J$ is large enough, the ``center of mass" spin orientations will eventually be bent to the plane perpendicular to the $(111)$-direction.
Thus we expect a ``$D_3$ breaking II" to ``$D_3$ breaking I" phase transition classically,
this is indeed the case as shown in Fig. \ref{fig:phase}.

We take the trial solution in Eq. (\ref{eq:order_D6II}) and assume $\lambda_3=\lambda_1$.
Under these assumptions, Eqs. (\ref{eq:partial_F1},\ref{eq:partial_F2},\ref{eq:partial_F3}) reduce to
\bea
-(J^\prime+\lambda_1)x-\Gamma^\prime z-(K^\prime+J^\prime) m&=&0\nn\\
-(K^\prime+J^\prime+\lambda_1)y-J^\prime m-\Gamma^\prime n&=&0\nn\\
-\Gamma^\prime x-(J^\prime+\lambda_1)z-\Gamma^\prime m-J^\prime n&=&0\nn\\
-(K^\prime+J^\prime) x-J^\prime y-\Gamma^\prime z-\lambda_2 m&=&0\nn\\
-2\Gamma^\prime y-2J^\prime z-\lambda_2 n &=&0\nn\\
x^2+y^2+z^2-1&=&0\nn\\
2m^2+n^2-1&=&0.
\label{eq:eq_D6II}
\eea
Since there are seven variables and seven equations, 
generically a solution exists.

Next we try to solve Eq. (\ref{eq:eq_D6II}) in a perturbative expansion over $J$.
However, we find difficulty in carrying out a perturbative expansion. 
The $J=0$ case has been already solved in Sec. \ref{sec:classical_equator},
which is taken as the zeroth order solution.
When $J\neq 0$, up to $O(J)$, the solution is (see Appendix \ref{app:D3II} for details)
\bea
x&=&\frac{1}{\sqrt{3}}\big[x_0+(\frac{3}{\Delta^2}+\frac{7}{6\Delta})\bar{J}\big]+O(\bar{J}^2)\nn\\
y&=&\frac{1}{\sqrt{3}}\big[-x_0+(\frac{6}{\Delta^2}+\frac{1}{3\Delta})\bar{J}\big]+O(\bar{J}^2)\nn\\
z&=&\frac{1}{\sqrt{3}}\big[z_0+(\frac{3}{\Delta^2}+\frac{1}{6\Delta})\bar{J}\big]+O(\bar{J}^2),
\label{eq:sol_D6II_A}
\eea
\bea
m&=&\frac{1}{\sqrt{3}}\big[x_0+(\frac{3}{\Delta^2}-\frac{5}{6\Delta})\bar{J}\big]+O(\bar{J}^2)\nn\\
n&=&\frac{1}{\sqrt{3}}\big[-z_0+(\frac{6}{\Delta^2}+\frac{1}{3\Delta})\bar{J}+O(\bar{J}^2)\big],
\label{eq:sol_D6II_B}
\eea
and
\bea
\lambda_1&=&\lambda_0+\frac{2}{\Delta}\bar{J}+O(\bar{J}^2)\nn\\
\lambda_2&=&\lambda_0-\frac{4}{\Delta}\bar{J}+O(\bar{J}^2),
\label{eq:sol_D6II_C}
\eea
in which the results are obtained up to $O(J)$.
In particular, since the $J$-dependent terms contain negative powers of $\Delta$,
the perturbation is valid only when $|\bar{J}|/\Delta^2\ll 1$.

As usual, the eigenvalues of the Hessian matrix should be calculated to verify that the saddle point solution in Eqs. (\ref{eq:sol_D6II_A},\ref{eq:sol_D6II_B},\ref{eq:sol_D6II_C}) corresponds to a minimum of the free energy. 
However,  the nonanalyticity in $\Delta$ in Eqs. (\ref{eq:sol_D6II_A},\ref{eq:sol_D6II_B},\ref{eq:sol_D6II_C}) complicates the calculation.
As discussed in detail in Appendix \ref{app:D3II},
 one possibly has to go up to at least fifth order perturbation in $\Delta$.
We will not perform such a difficult  fifth order perturbation, 
and in fact, we suspect if a good perturbation exists because of the nonanalytical dependence of the saddle point solution on $\Delta$. 
The smallest eigenvalue is studied by numerics as discussed in Appendix \ref{app:D3II}.

Due to the above mentioned difficulty, the spin wave mass will not be perturbatively calculated.
Instead, we study the spin wave mass numerically by computing the eigenvalues of the matrix $M(\Delta,J)\mathcal{H}_F(\Delta,J)$.
The dependence of $m_1$ on $J$ at three representatively values of $\varphi=0.21\pi$ and $0.30\pi$ are shown in  Fig. \ref{fig:sw_mass_D3} (a) and (c), respectively.
The value of $J=J_c(\varphi)$  where $m_1$ vanishes is the transition point between the ``$D_3$-breaking I" and the ``$D_3$-breaking II" phases.
Notice that for fixed value of $\varphi$,  the ``$D_3$-breaking I (II)" phase occupies the region $J>J_c(\varphi)$ ($J<J_c(\varphi)$).

\subsection{DMRG numerics}

In this subsection, we present the DMRG numerical results which provide numerical evidence for the predicted ``$D_3$-breaking I, II" phases for both $S=1$ and $3/2$.

\begin{figure*}[htbp]
\includegraphics[width=12.0cm]{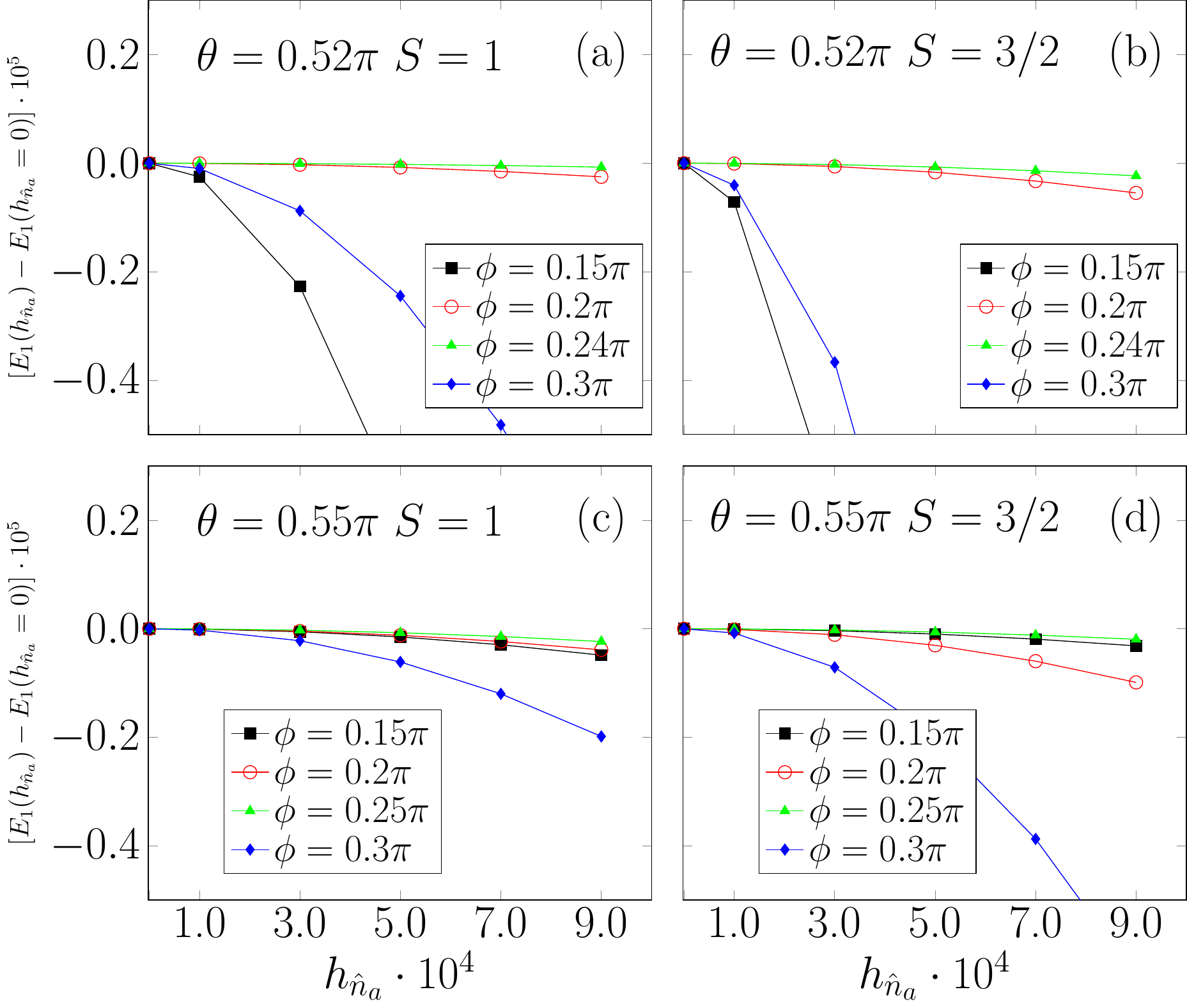}
\caption{$\Delta E$ ($=E(h_{\hat{n}_a})-E(h_{\hat{n}_a}=0)$) vs $h_{\hat{n}_a}$ for (a,c) $S=1$, and (b,d) $S=3/2$ at fixed values of $\theta$ and $\varphi$.
The magnetic field $h_{\hat{n}_a}$ is taken along the $(111)$-direction with a magnitude $h_{\hat{n}_a}=5\times 10^{-4}$.
ED numerics are performed on $L=18$ sites with periodic boundary conditions. 
} \label{fig:energy_vs_field}
\end{figure*}

Before proceeding on, we mention a subtlety in numerical calculations,
which has already been discussed in detail in Ref. \onlinecite{Yang2020b}.
In either the ``$D_3$-breaking I" or ``$D_3$-breaking II" phases, the six symmetry breaking ground states only become exactly degenerate in the thermodynamic limit. 
In a finite size system, the ground state can be some arbitrary linear combination of the six states,
and the coefficients depend on the system size and numerical details.
Because of this, random cancellations occur if the correlation functions $\langle S_i^\alpha S_{i+r}^\beta\rangle$ or the expectation values of the spin operators $\langle\vec{S}_i\rangle$ are directly computed.
To circumvent such difficulty, a small magnetic field has to be applied such that the system is polarized into one of the six degenerate ground states.

For our purpose, we choose the field to be $h_{\text{I}}$ along the $(-1,0,1)$-direction in the ``$D_3$-breaking I" phase, and $h_{\text{II}}$ along the $(1,-1,1)$-direction in the ``$D_3$-breaking II" phase.
According to Fig. \ref{fig:spin_orders},
we expect that the red solid circle located at $(-1,0,1)$ is picked out in the the ``$D_3$-breaking I" phase,
and the solid dark blue circle located at $(1,-1,1)$ is picked out in the the ``$D_3$-breaking II" phase.
Then with the application of such fields, the spins should align according to the pattern given in Eq. (\ref{eq:order_D6I}) (Eq. (\ref{eq:order_D6II})) in the `$D_3$-breaking I (II)" phase. 
However, as discussed in Ref. \onlinecite{Yang2020b},
the ``$D_3$-breaking I (II)" phase responds to the $h_{\text{II}}$- ($h_{\text{(I)}}$-) field as does the ``$D_3$-breaking II (I)" phase.
Therefore, this method is not able to distinguish the two $D_3$-breaking phases.
However, the method is still useful since it can test the existence of either ``$D_3$-breaking I" or ``$D_3$-breaking II" orders.

We have calculated the spin expectation values $\left<S_j^\alpha\right>$ ($\alpha=x,y,z$) at three representative points $(\theta=0.52\pi, \phi=0.15\pi)$, $(\theta=0.52\pi, \phi=0.25\pi)$ and $(\theta=0.52\pi, \phi=0.30\pi)$ under the $h_{\text{I}}$ and $h_{\text{II}}$ fields.
The results for $S=1$ are displayed in Fig. \ref{fig:D3_break_Seq1}.
DMRG numerics are performed on a system of  $L=18$ sites with periodic boundary conditions,
and both $h_{\text{I}}$ and $h_{\text{II}}$ fields are taken to be $10^{-4}$.
As can be clearly seen from Fig. \ref{fig:D3_break_Seq1},
the spin alignments are consistent with the patterns given in Eqs. (\ref{eq:order_D6I},\ref{eq:order_D6II}),
thereby confirming the existence of the ``$D_3$-breaking I, II" phases.
We have also studied the $S=3/2$ case, and the results are included in Appendix \ref{app:more_numerics}.

As discussed in Ref. \onlinecite{Yang2020b},
the two $D_3$-breaking phases can be distinguished by studying the response of the system to a small field $h_{\hat{n}_a}$ along the $(111)$-direction,
since the ``$D_3$-breaking I" phase does not respond to $h_{\hat{n}_a}$,
whereas the ``$D_3$-breaking II" phase does have an response.

\begin{figure*}[htbp]
\includegraphics[width=13.0cm]{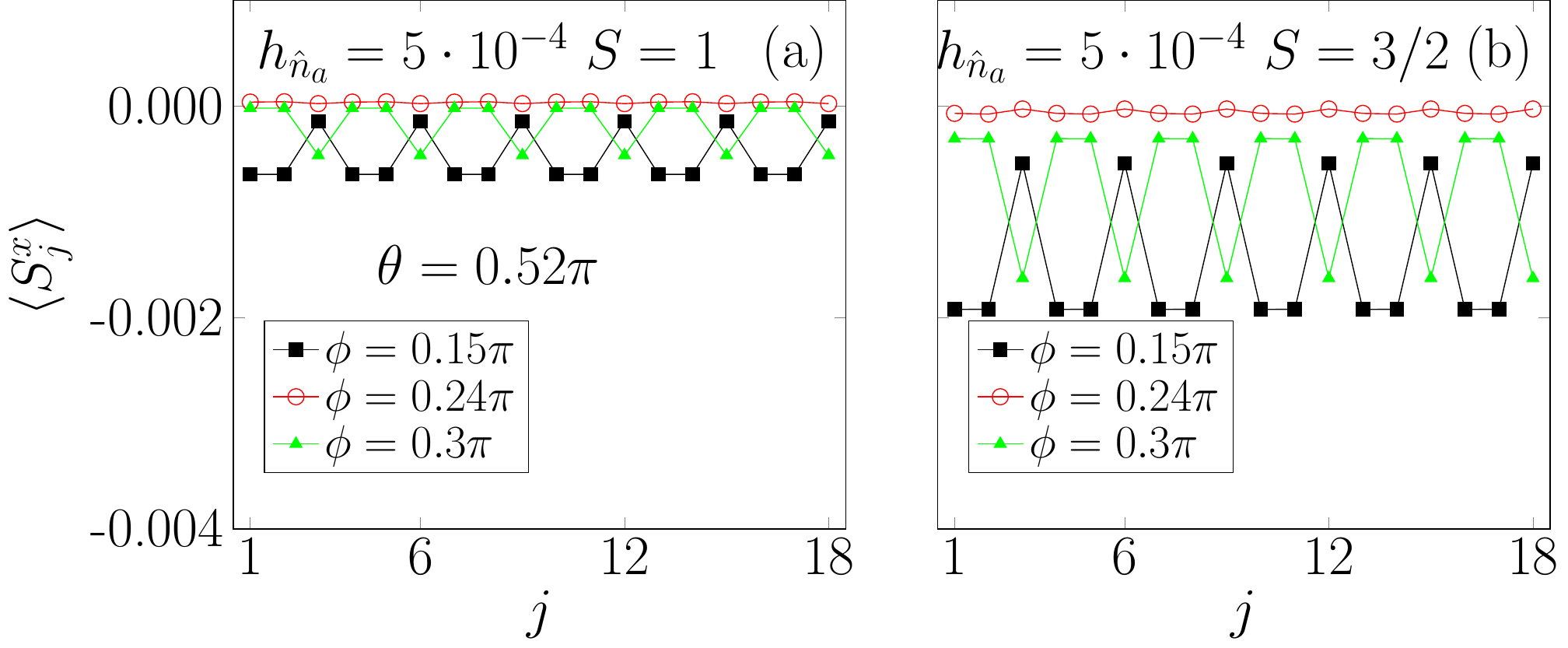}
\caption{$\left< S_j^x \right>$ vs $j$ for (a) $S=1$, and (b) $S=3/2$ at fixed values of $\theta$, $\varphi$.
The magnetic field is taken along the $(111)$-direction with a magnitude $h_{\hat{n}_a}=5\times 10^{-4}$. 
DMRG numerics are performed on $L=18$ sites with periodic boundary conditions. 
} \label{fig:D3_break_111field}
\end{figure*}

Fig. \ref{fig:energy_vs_field} shows the energy change $\Delta E= E(h_{\hat{n}_a})-E(h_{\hat{n}_a}=0)$ as a function of $h_{\hat{n}_a}$ at several representative points in the negative $J$ region for both $S=1$ and $S=3/2$.
Clearly, while the system  has a huge response  at some  points, 
the response nearly vanishes at others. 
Based on the results in Fig. \ref{fig:energy_vs_field}, we arrive at the conclusion that the points 
$(\theta=0.52\pi,\phi=0.2\pi,0.24\pi)$ and $(\theta=0.55\pi,\phi=0.15\pi,0.2\pi,0.25\pi)$ are within the ``$D_3$-breaking I" phase,
whereas the points
$(\theta=0.52\pi,\phi=0.15\pi,0.3\pi)$ and $(\theta=0.55\pi,\phi=0.3\pi)$
are in the ``$D_3$-breaking II" phase.
In particular, as can be seen from Fig. \ref{fig:energy_vs_field}, the range of the ``$D_3$-breaking I" phase expands by increasing $\theta$,
which is consistent with the classical phase diagram as shown in Fig. \ref{fig:phase}.

Fig. \ref{fig:D3_break_111field} displays the response of $\left<S_j^x\right>$ to $h_{\hat{n}_a}=5\times 10^{-4}$ at several different points for both $S=1$ and $S=3/2$.
As can be  seen from Fig. \ref{fig:D3_break_111field},
the response at  the point $(\theta=0.52\pi,\phi=0.24\pi)$ is very small,
hence this point should locate within the ``$D_3$-breaking I" phase.
On the other hand, the response at the points $(\theta=0.52\pi, \phi=0.15\pi,0.3\pi)$
are significant, and they should be within the ``$D_3$-breaking II" phase.

\section{Conclusions}

In conclusion, we have studied the classical phase diagram of the one-dimensional spin-$S$ Kitaev-Heisenberg-Gamma model in the region of an antiferromagnetic Kitaev coupling,
 based on a combination of classical and spin wave analysis.
The  revealed ``N\'eel" and ``$D_3$-breaking I, II" phases are in accordance with the spin-1/2 case as discussed in Ref. \onlinecite{Yang2020b}.
On the other hand, the ``$O_h\rightarrow D_3$" phase in the absence of the Heisenberg term is not the same as the ``$O_h\rightarrow D_4$" phase in the spin-1/2 case.
DMRG numerics provide evidence for the ``$O_h\rightarrow D_3$" symmetry breaking for higher spins including $S=1$ and $3/2$, which are consistent with the classical results.
We have also obtained analytic expressions of the lowest-lying spin wave mass perturbatively in the vicinity of the hidden  SU(2) symmetric ferromagnetic point.

{\it Acknowledgments}
We thank H.-Y. Kee for interesting remarks and helpful discussions.
WY and IA acknowledge support from NSERC Discovery Grant 04033-2016.
AN acknowledges computational resources and services provided by Compute Canada and
Advanced Research Computing at the University of British Columbia.
AN is supported by the Canada First Research Excellence Fund.

\appendix
\begin{widetext}

\section{The Hamiltonians in the six-sublattice rotated frame}
\label{app:Ham}

In this section, we spell out the terms in the Hamiltonians in different frames. 
In general, we write the Hamiltonian $H$ as $H=\sum_{j=1}^L H_{j,j+1}$ where $H_{j,j+1}$ is the term on the bond between the sites $j$ and $j+1$.   
The forms of $H_{j,j+1}$ will be written explicitly. 

In the unrotated frame, the form of $H_{j,j+1}$ has a two-site periodicity. 
We have
\begin{eqnarray}
H_{2n+1,2n+2}&=&K S_{2n+1}^x S_{2n+2}^x +\Gamma (S_{2n+1}^y S_{2n+2}^z+S_{2n+1}^z S_{2n+2}^y)+J\vec{S}_{2n+1}\cdot \vec{S}_{2n+2}, \nn\\
H_{2n+2,2n+3}&=&K S_{2n+2}^y S_{2n+3}^y +\Gamma (S_{2n+2}^z S_{2n+3}^x+S_{2n+2}^x S_{2n+3}^z)+J\vec{S}_{2n+2}\cdot \vec{S}_{2n+3}.
\end{eqnarray}

In the six-sublattice rotated frame, the form of $H^\prime_{j,j+1}$ has a three-site periodicity. 
We have
\begin{eqnarray}
H^\prime_{3n+1,3n+2}&=&-KS_{3n+1}^xS_{3n+2}^x-\Gamma (S_{3n+1}^yS_{3n+2}^y+S_{3n+1}^zS_{3n+2}^z)-J(S_{3n+1}^xS_{3n+2}^x+S_{3n+1}^yS_{3n+2}^z+S_{3n+1}^zS_{3n+2}^y),\nn\\
H^\prime_{3n+2,3n+3}&=&-KS_{3n+2}^zS_{3n+3}^z-\Gamma (S_{3n+2}^xS_{3n+3}^x+S_{3n+2}^yS_{3n+3}^y)-J(S_{3n+2}^zS_{3n+3}^z+S_{3n+2}^xS_{3n+3}^y+S_{3n+2}^yS_{3n+3}^x),\nn\\
H^\prime_{3n+3,3n+4}&=&-KS_{3n+3}^yS_{3n+4}^y-\Gamma (S_{3n+3}^zS_{3n+4}^z+S_{3n+3}^xS_{3n+4}^x)-J(S_{3n+3}^yS_{3n+4}^y+S_{3n+3}^zS_{3n+4}^x+S_{3n+3}^xS_{3n+4}^z).\nn\\
\end{eqnarray}

\section{Numerical minimization of the classical free energy}
\label{app:numerics_classical}

\begin{figure*}[htbp]
\includegraphics[width=7.0cm]{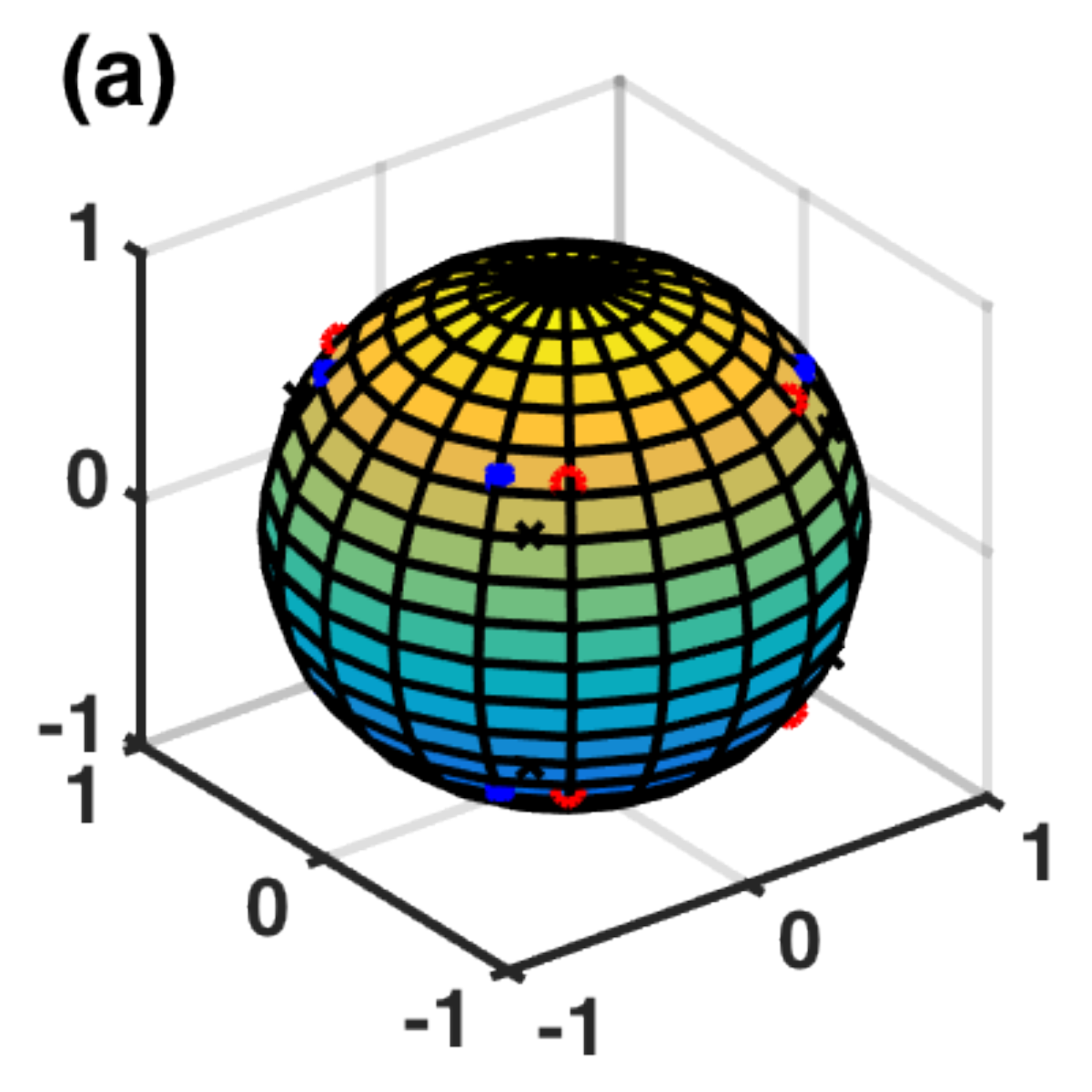}
\includegraphics[width=7.0cm]{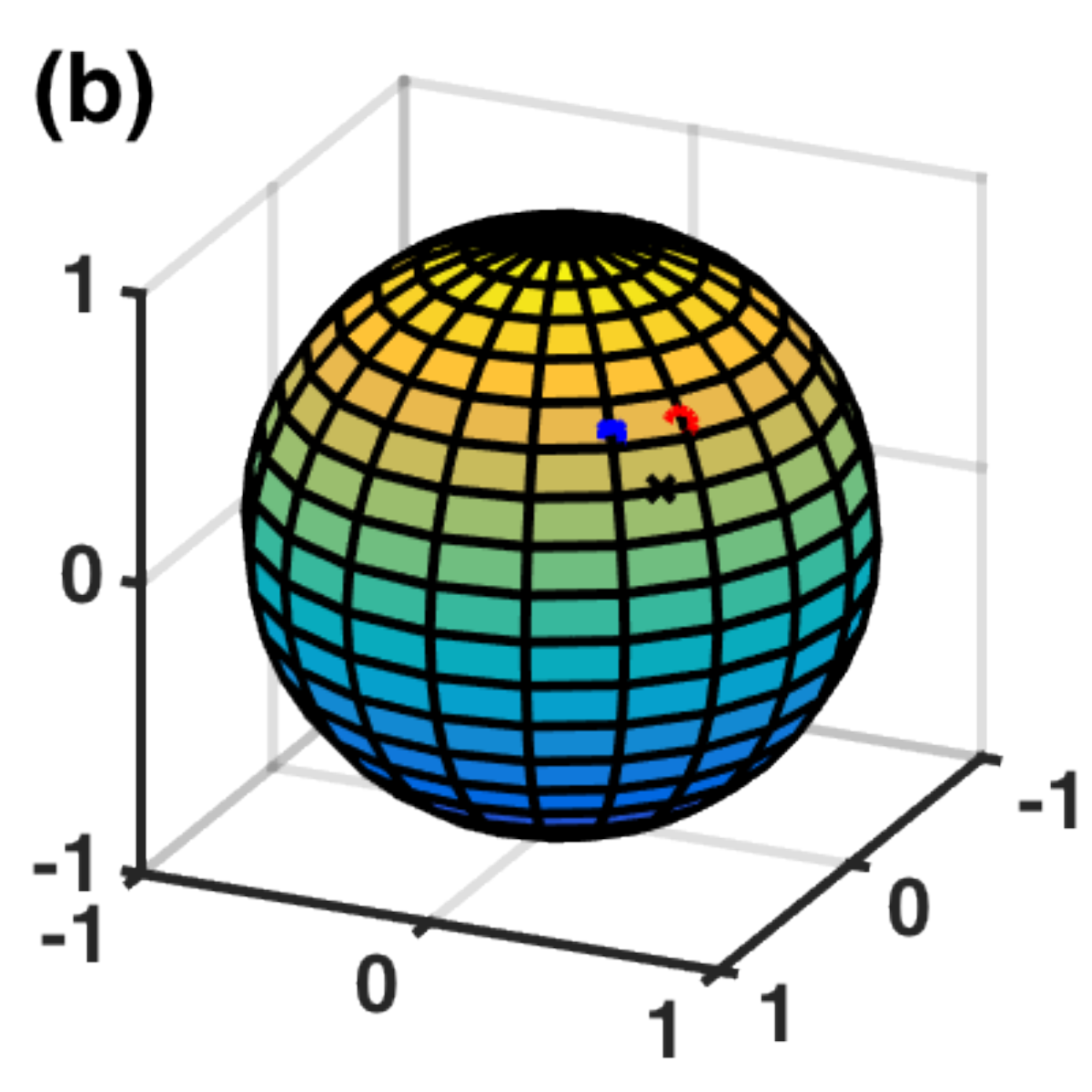}
\includegraphics[width=7.0cm]{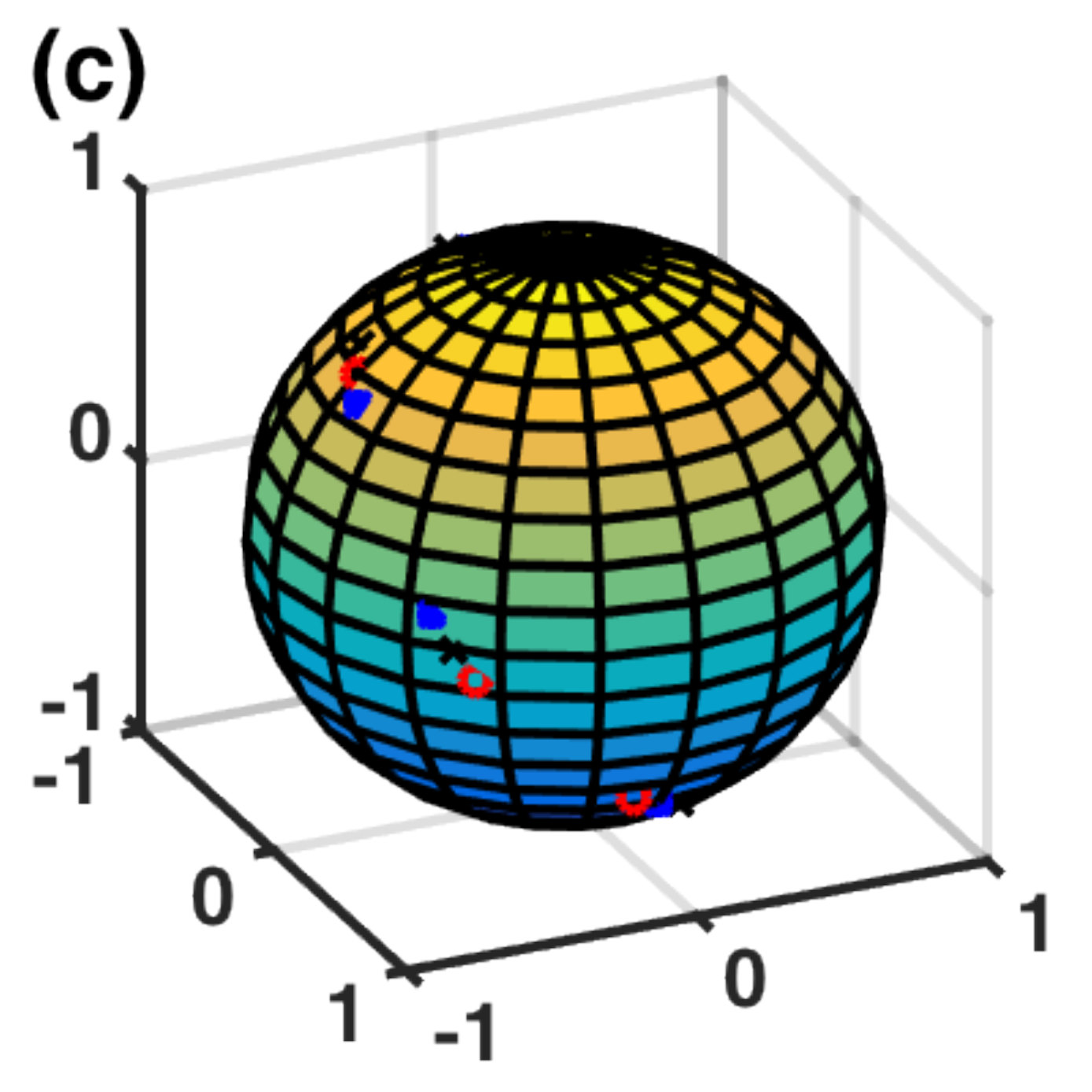}
\includegraphics[width=7.0cm]{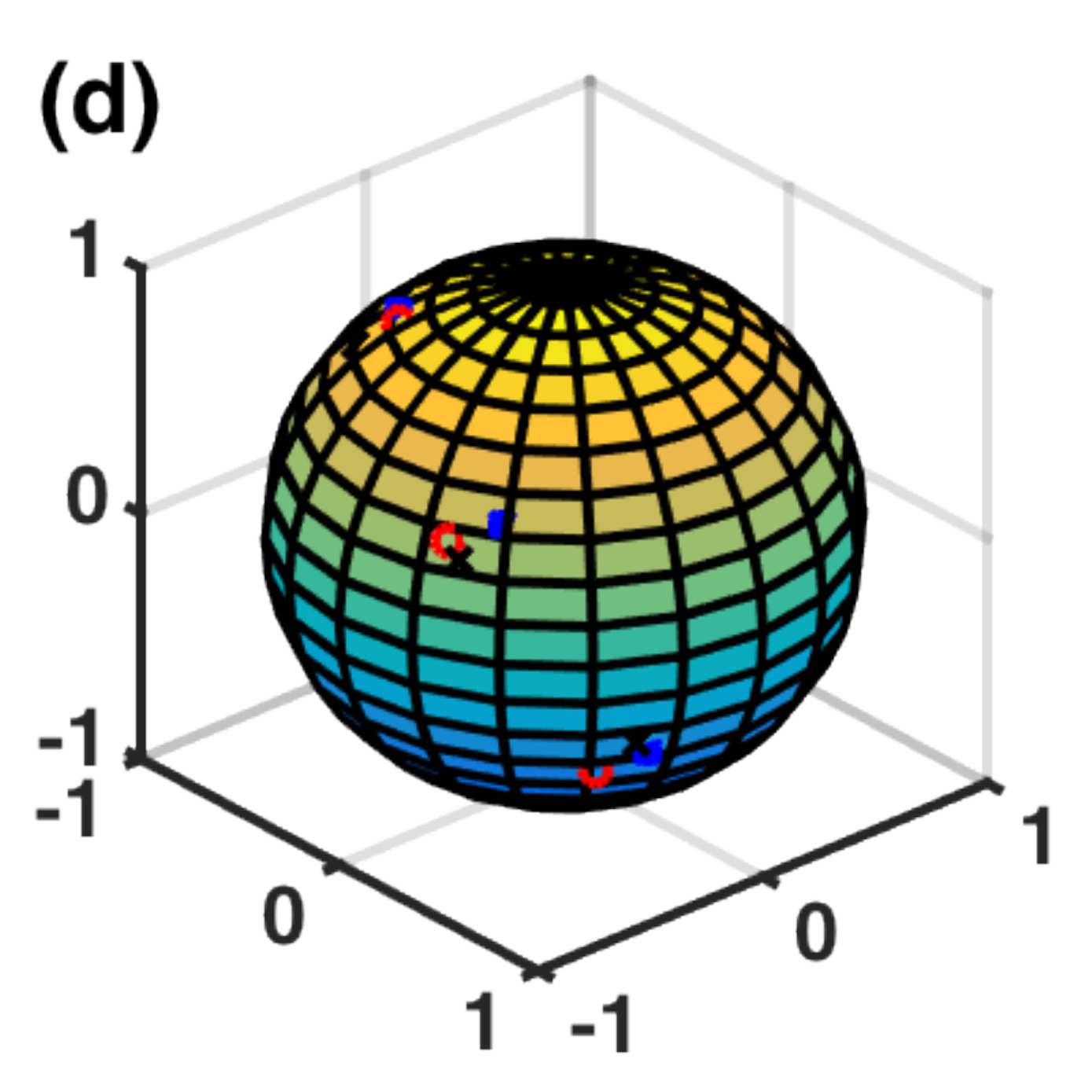}
\caption{Classical minima of the free energy at (a) $\varphi=0.15\pi$, $J=0$; (b) $\varphi=0.15\pi$, $J=0.3$; (c) $\varphi=0.25\pi$, $J=-0.3$; (d) $\varphi=0.15\pi$, $J=-0.1$,
which lie in the ``$O_h\rightarrow D_3$" phase, the ``N\'eel" phase, the ``$D_3$-breaking I" phase and the ``$D_3$-breaking II" phase, respectively.
In (a,b,c,d), the black ``x", blue ``$\times$" and red ``o" symbols represent the spin directions on site 1, 2, 3 within unit cell, respectively.
} 
\label{fig:numerics_classical_orders}
\end{figure*}

\begin{figure*}[htbp]
\includegraphics[width=5.5cm]{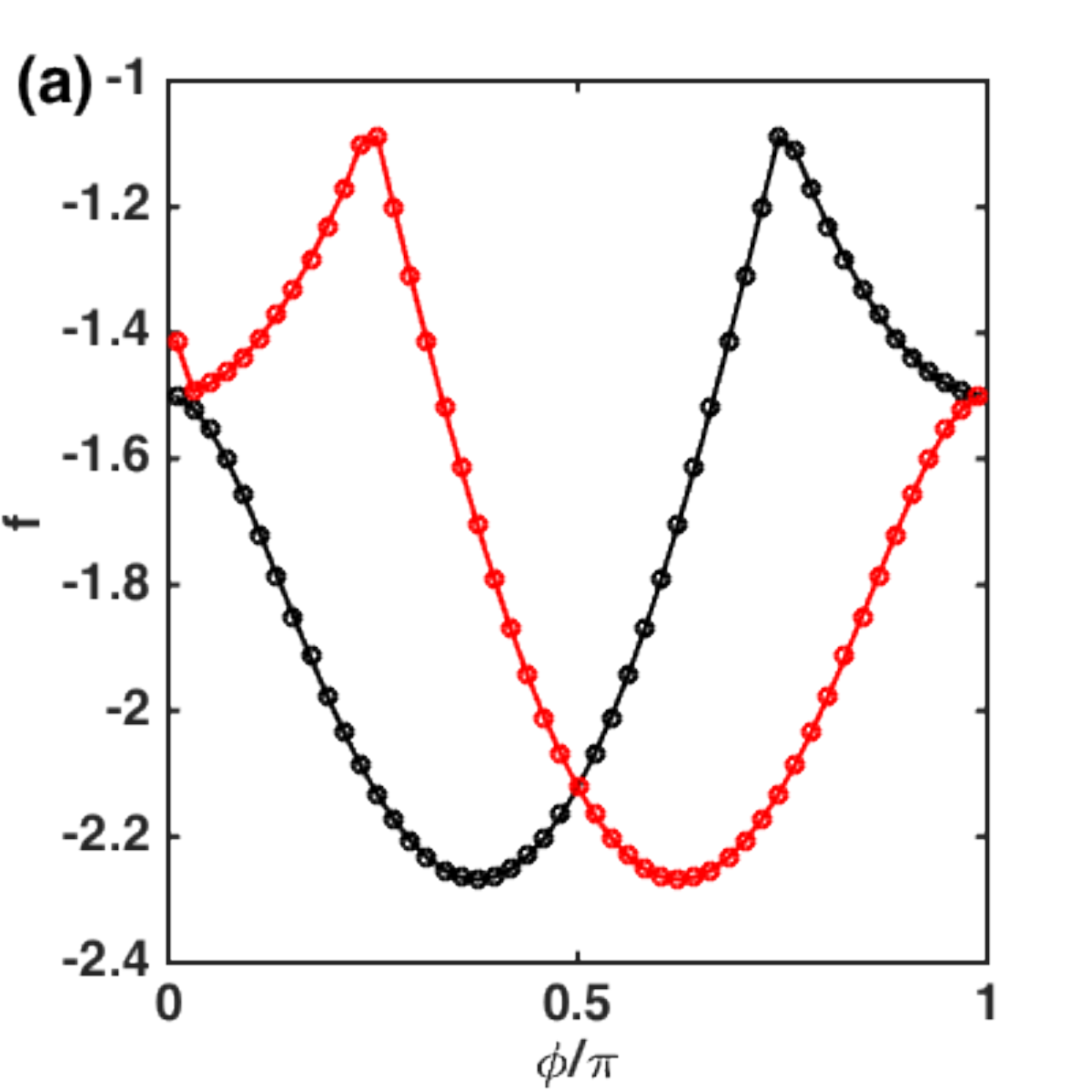}
\includegraphics[width=5.5cm]{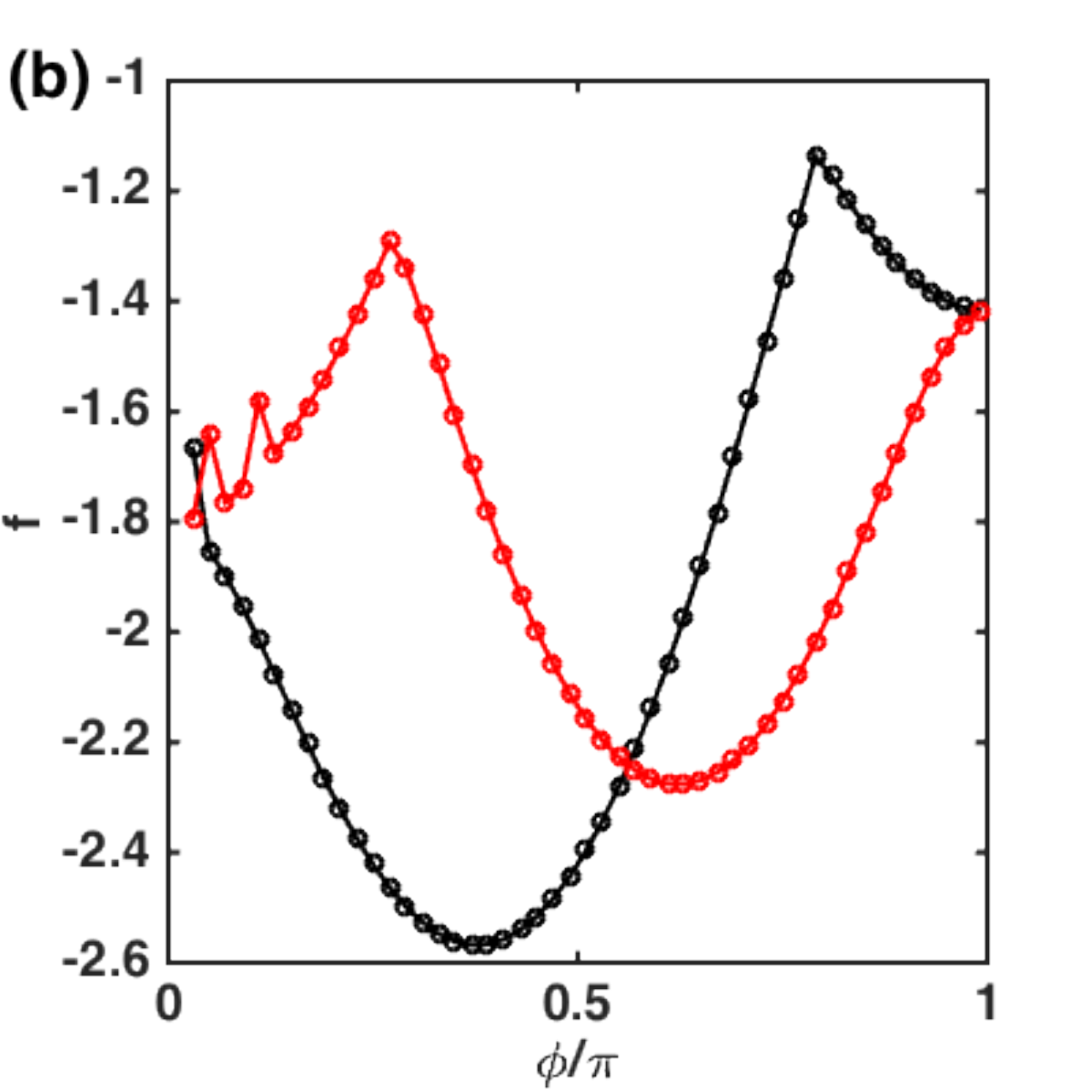}
\includegraphics[width=5.5cm]{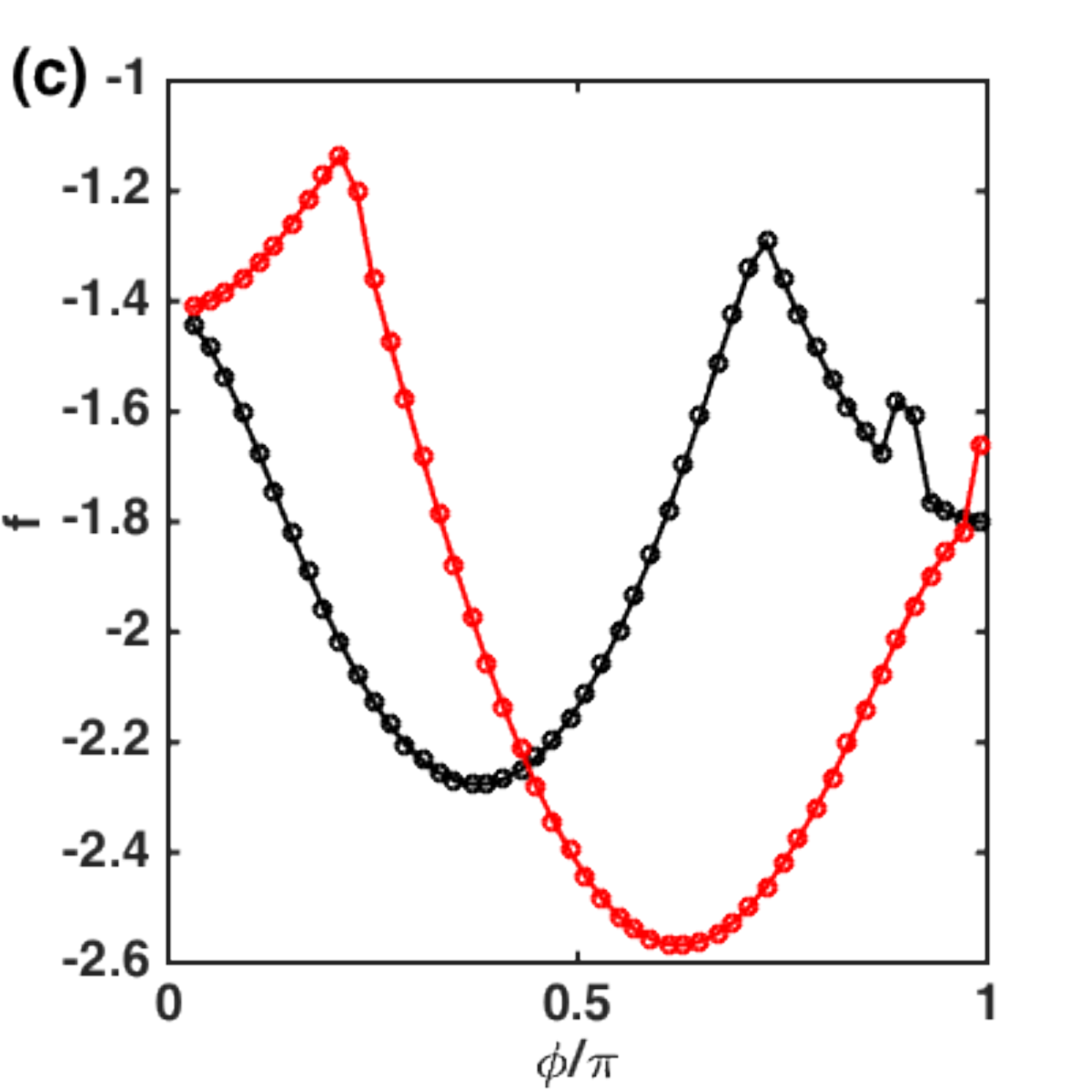}
\caption{Classical energies $f$ of $T_{3a}$-invariant spin configurations (black curve) and $T_{3a}$-staggered spin configurations (red curve) as functions of $\varphi$ at (a) $J=0$, (b) $J=0.1$, and (c) $J=-0.1$.
} 
\label{fig:numerics_classical_transition}
\end{figure*}

In this appendix, we present the numerical results for minimizing the classical free energies.
Throughout this appendix, we work in the six-sublattice rotated frame unless otherwise stated.

In Fig. \ref{fig:numerics_classical_orders}, the classical minima of the free energy are displayed for one representative point in each phase among the ``$O_h\rightarrow D_3$", ``Neel", ``$D_3$-breaking I" and ``$D_3$-breaking II" phases.
They all agree with the patterns of the spin alignments discussed in the main text.

We have also compared the classical energies between FM spin configurations (i.e., invariant under $T_{3a}$) and AFM spin configurations (i.e., staggered under $T_{3a}$) for three representative values of $J$, and the results are displayed in Fig. \ref{fig:numerics_classical_transition}.
(Note: The FM and AFM here refer to spin alignments in the six-sublattice rotated frame.)
As can be seen from Fig. \ref{fig:numerics_classical_transition}, the classical phase transition between FM and AFM  occurs at the $\Gamma$ point.
The transition point is shifted to larger (smaller) $\phi$ when $J>0$ ($J<0$). 
We note that the critical point $\phi_c$ is shifted by quantum fluctuations.
And what is more, the AFM order for $\phi>\phi_c$ may be destroyed by quantum fluctuations.
Indeed, as shown in Ref. \onlinecite{Yang2020a}, $\phi_c$ at $J=0$ is shifted to $0.33\pi$ for $S=1/2$, and the  classical AFM phase does not have any order and the low energy physics is described by the emergent SU(2)$_1$ WZW model.

\section{Proof of degeneracy}
\label{app:proof_degeneracy}

We give an explanation to the question raised at the end of Sec. \ref{sec:classical_equator}; i.e., why the two low-lying eigenvectors are degenerate to all orders in $\Delta$.
Although the Hessian matrix $H_F(\Delta)$ defined in Eq. (\ref{eq:HF_equator}) always has $O_h$ symmetry, the symmetry of the projected Hessian matrix  $\mathcal{H}_F(\Delta)$ in Eq. (\ref{eq:proj_HF_equator}) is reduced to $D_3$ due to the saddle point solutions $r_i(\Delta)$ ($i=1,2,3$) in the definition of the projection matrix $P(\Delta)$.
Thus, the eigenspaces of $\mathcal{H}_F(\Delta)$ form representations of the group $D_3$.
Since the $D_3$ group only has one- and two-dimensional irreducible representations, 
generically, we expect one- and two-fold degenerate eigenvalues of $\mathcal{H}_F(\Delta)$ except accidental degeneracies.
To identify the representations of the eigenspaces (which has to fall into the three irreducible representations of $D_3$, i.e., $A_1$, $A_2$ and $E$), we first consider the $\Delta=0$ case.
As can be easily checked, $\{v_1,v_2\}$ form the $E$ representation of the $D_3$ group, which is two-dimensional.
For a nonzero $\Delta$, this irreducible representation cannot be changed unless there is a level crossing.
Therefore, we conclude that at least for sufficiently small $\Delta$, the degeneracy of the two low-lying eigenvectors should always be two.

\section{Basics of symplectic linear algebra}
\label{app:symplectic}

Let $\mathcal{S}$ be a  symplectic form on a $2n$-dimensional linear space.
Under a suitable basis, $\mathcal{S}$ acquires the form
\bea
\mathcal{S}=\left(\begin{array}{cc}
0&I_n\\
-I_n&0
\end{array}
\right),
\label{eq:define_S}
\eea 
in which $I_n$ is the $n\times n$ identity matrix.
A transformation $V$ is called a symplectic transformation if 
\bea
V \mathcal{S} V^T=\mathcal{S}.
\label{eq:symplectic_def}
\eea

Let $A$ be a $2n\times 2n$ positive-definite real symmetric matrix.
Then: (1) the eigenvalues of $\mathcal{S}A$ are all purely imaginary;  (2) the eigenvalues appear in pairs as $\pm i \lambda_j$ where $\lambda_j\in \mathbb{R}$;
(3) the eigenvectors  satisfy $x^T \mathcal{S} y=0$ if $\lambda_x\neq-\lambda_y$ where $\lambda_x,\lambda_y$ are the eigenvalues of $x,y$ which are eigenvectors of $\mathcal{S}A$.
To see point (1), notice that $A^{1/2}\mathcal{S}A^{1/2}$ has the same eigenvalues as $\mathcal{S}A$ since they differ by a similar transformation $A^{-1/2}(...)A^{1/2}$ which is well-defined because $A$ is assumed to be positive-definite.
Since $A^{1/2}\mathcal{S}A^{1/2}$ is antisymmetric, its eigenvalues have to be purely imaginary.
For point (2), by taking complex conjugate on both sides of the eigenequation $\mathcal{S}Ax=\lambda_x x$, it can be seen that $x^*$ has eigenvalue $\lambda_x^*=-\lambda_x$.
For point (3), notice that on the one hand, $y^T A x=y^T\mathcal{S}^T(\mathcal{S} A x)=-\lambda_x y^T \mathcal{S} x$;
on the other hand,  $y^T A x = (Ay)^T x=(\mathcal{S}^T\mathcal{S}Ay)^Tx=\lambda_y y^T\mathcal{S}x$.
This shows that if $\lambda_x\neq -\lambda_y$, then $y^T\mathcal{S}x=0$.

Next we state the central result for our purpose.
Let $A$ be positive-definite and real as before.
Then there exists a symplectic transformation $V$ (i.e., satisfying Eq. (\ref{eq:symplectic_def})) such that 
\bea
V \mathcal{S}A V^T=\left(
\begin{array}{cc}
\Lambda & 0\\
0& \Lambda
\end{array}
\right),
\eea
where $\Lambda$ is a diagonal matrix.
We will prove this statement based on the previous discussions. 
The eigenvectors of $\mathcal{S}A$ are $e_j\pm i f_j$ ($1 \leq j \leq n$) with eigenvalues $\pm i\lambda_j$,
where $e_j,f_j$ are real vectors.
Using $(e_j^T+i\alpha  f_j^T)\mathcal{S}(e_k+i\beta  f_k)= N_{i\alpha}\delta_{jk}\delta_{\alpha,-\beta}$ ($\alpha,\beta=\pm1$) where $N_{i\alpha}$ is a normalization factor, 
it can be verified that 
\bea
e_j^T \mathcal{S} e_k=f_j^T \mathcal{S} f_k=e_j^T \mathcal{S} f_k=f_j^T \mathcal{S} e_k=0,~j\neq k.
\eea
For $j=k$, the real and imaginary parts of $(e_j^T+i\alpha f_j^T)\mathcal{S} (e_j+i\alpha f_j)$ are $e_j^T\mathcal{S}e_j-f_j^T\mathcal{S}f_j$ and $\alpha (f_j^T\mathcal{S} e_j+e_j^T\mathcal{S} f_j)$, respectively, and both must be vanish according to previous discussions.
Furthermore, since $e_j^T\mathcal{S}e_j=(e_j^T\mathcal{S}e_j)^T=-e_j^T\mathcal{S}e_j$, we have $e_j^T\mathcal{S}e_j=f_j^T\mathcal{S}f_j=0$.
This shows that  the only nonvanishing combinations are $e_j^T\mathcal{S}f_j$ and $f_j^T\mathcal{S}e_j$, which can be normalized to $-1$ and $1$ by a rescaling of $e_j,f_j$.
As a result, 
\bea
(e_1~ ...~ e_n ~f_1~...~f_n)^T\mathcal{S} (e_1~ ...~ e_n ~f_1~...~f_n)=\left(\begin{array}{cc}
0&-I_n\\
I_n&0
\end{array}
\right),
\label{eq:U_sympl}
\eea
in which the right hand side is just $\mathcal{S}^T$.
This means that the matrix $U=(e_1~ ...~ e_n ~f_1~...~f_n)$ is a symplectic transformation.

Now we demonstrate that $U$ is able to diagonalize $A$. 
According to the eigenequations $\mathcal{S}A(e_j\pm i f_j)=\pm i \lambda_j (e_j\pm i f_j)$, we obtain $\mathcal{S}A e_j=-\lambda_j f_j$, and $\mathcal{S}A f_j = \lambda_j e_j$, i.e., 
\bea
\mathcal{S}A U = U \left(
\begin{array}{cc}
0 & -\Lambda\\
\Lambda& 0
\end{array}
\right),
\label{eq:A_eigeneq}
\eea
in which $\Lambda=\text{diag}(\lambda_1,...,\lambda_n)$.
Next multiplying both sides of Eq. (\ref{eq:A_eigeneq}) with $U^T\mathcal{S}^T$, we obtain
\bea
U^T A U = U^T\mathcal{S}^T U\left(
\begin{array}{cc}
0 & -\Lambda\\
\Lambda& 0
\end{array}
\right),
\eea
in which $U^T\mathcal{S}^T U=\mathcal{S}$ according to Eq. (\ref{eq:U_sympl}).
But $\mathcal{S}\left(
\begin{array}{cc}
0 & -\Lambda\\
\Lambda& 0
\end{array}
\right)= \left(
\begin{array}{cc}
\Lambda & 0\\
0& \Lambda
\end{array}
\right)$, thus
\bea
U^T A U= \left(
\begin{array}{cc}
\Lambda & 0\\
0& \Lambda
\end{array}
\right),
\eea
completing the proof.

\section{Proof of Eq. (\ref{eq:hM1_equator_relation})}
\label{app:proof_symplectic_first}

We give a proof of Eq. (\ref{eq:hM1_equator_relation}).
Define $M^{(n)}$ in terms of the power expansions as
\bea
M(\Delta)&=&M^{(0)}+M^{(1)}+M^{(2)}+...,
\eea
where $M^{(n)}$ is proportional to $\Delta^n$.
Writing 
\bea
\mathcal{H}_F^{M,(1)}=M^{(0)}\mathcal{H}_F^{(1)}+M^{(1)}\mathcal{H}_F^{(0)},
\eea
we obtain
\bea
h^{M,(1)}&=&P_1^{(0)}M^{(0)}\mathcal{H}_F^{(1)}P_1^{(0)}+P_1^{(0)}M^{(1)}\mathcal{H}_F^{(0)}P_1^{(0)}.
\label{eq:hM1_1}
\eea
Since $P_1^{(0)}$ commutes with $M^{(0)}$, the first term in Eq. (\ref{eq:hM1_1}) is equal to $M^{(0)}P_1^{(0)}h^{(1)}P_1^{(0)}$.
For the second term in Eq. (\ref{eq:hM1_1}),
since $P_1^{(0)}$ commutes with $\mathcal{H}_F^{(0)}$, 
the second term is equal to the product of $P_1^{(0)}M^{(1)}P_1^{(0)}$ and $P_1^{(0)}\mathcal{H}_F^{(0)}P_1^{(0)}$.
However, $P_1^{(0)}M^{(1)}P_1^{(0)}$ vanishes.
To see this, recall that $M_j(\Delta)$   represents the cross product operation with $\hat{n}_j^{(0)}(\Delta)$.
Denote $T_j(\hat{n}_j^{(0)})$ to be the tangent space of the unit sphere at $\hat{n}_j^{(0)}$,
and $P_{1j}^{(0)}$ to be the projection to $T_j(\hat{n}_j^{(0)})$.
Then $M_j^{(1)}(\Delta)$ corresponds to the cross product with the vector $\delta n_j^{(0)}=n_j^{(0)}(\Delta)-n_j^{(0)}(\Delta=0)$, which lives in $T_j(\hat{n}_j^{(0)})$.
Then clearly, the action of $M^{(1)}_jP_{1j}^{(0)}$ on any vector in the tangent space $T_j(\hat{n}^{(0)})$ is perpendicular to the tangent space,
which means that $P_{1j}^{(0)}M^{(1)}_jP_{1j}^{(0)}=0$.
Hence, $P_{1}^{(0)}M^{(1)}P_{1}^{(0)}=0$.

\section{Equivalence with the Holstein-Primakoff transformation}
\label{app:HP_equiv}

We demonstrate that the calculations in Sec. \ref{sec:SW_equator} based on the path integral formalism are equivalent with the Bogoliubov transformation based on the Holstein-Primakoff transformation.

For site $j$, the coordinate frame in the spin space is set up as $\{\hat{n}_j^{(0)},\hat{e}_\theta^{(0)}(j),\hat{e}_\phi^{(0)}(j)\}$.
Define the spin components $S_j^{\prime\alpha}$ ($\alpha=1,2,3$) as 
\bea
S_j^{\prime3}&=&\hat{n}_j^{(0)}\cdot \vec{S}_j,\nn\\
S_j^{\prime1}&=&\hat{e}_\theta^{(0)}(j)\cdot \vec{S}_j,\nn\\
S_j^{\prime2}&=&\hat{e}_\phi^{(0)}(j)\cdot \vec{S}_j.
\eea
Then by introducing the Holstein-Primakoff boson $\{b_j,b_j^\dagger\}$, the spin operators $S_j^{\prime\alpha}$ can be written as
\bea
S_j^{\prime3}&=&S-b_j^\dagger b_j,\nn\\
S_j^{\prime+}&=&\sqrt{2S-b_j^\dagger b_j}\cdot b_j,\nn\\
S_j^{\prime-}&=&b_j^\dagger\sqrt{2S-b_j^\dagger b_j},\nn\\
\eea
in which $S_j^{\prime\pm}=S_j^{\prime1}\pm iS_j^{\prime2}$.
Within the spin wave approximation, we have
\bea
S_j^{\prime3}&=&S-b_j^\dagger b_j,\nn\\
S_j^{\prime1}&\approx&\sqrt{\frac{S}{2}}\cdot (b_j^\dagger+b_j),\nn\\
S_j^{\prime2}&\approx&i\sqrt{\frac{S}{2}}\cdot (b_j^\dagger-b_j).\nn\\
\label{eq:HP_approx}
\eea

Neglecting the quartic terms in the boson operators, it can be shown that Eq. (\ref{eq:HP_approx}) leads to
\bea
\sum_{\alpha=1,2,3}(S_j^{\prime\alpha})^2=S(S+1),
\eea
which is simply the quantum mechanical value of $\vec{S}_j^2$.
On the other hand, if we take the normal ordered product, then Eq. (\ref{eq:HP_approx}) leads to 
\bea
\sum_{\alpha=1,2,3}:(S_j^{\prime\alpha})^2:=S^2,
\label{eq:S2_normal}
\eea
which coincides with the classical constraints in Eq. (\ref{eq:constraints}).
Therefore, the procedure of plugging Eq. (\ref{eq:HP_approx}) into the Hamiltonian in Eq. (\ref{eq:Ham}) and keeping only the quadratic terms in boson operators is entirely equivalent to expanding the Lagrangian in the path integral into quadratic terms in the coordinates $\{\chi_\theta(j),\chi_\phi(j)\}$, under the following identification
\bea
\chi_\theta(j)&=&\frac{1}{\sqrt{2}}(b_j^\dagger+b_j),\nn\\
\chi_\phi(j)&=&i\frac{1}{\sqrt{2}}(b_j^\dagger-b_j).
\eea
This establishes the equivalence between the two methods.
In particular, it also fixes the operator ordering in Eq. (\ref{eq:Hsw_equator}).
Because of Eq. (\ref{eq:S2_normal}), the operators in Eq. (\ref{eq:Hsw_equator}) should be understood as normal ordered in terms of $\{b_j,b_j^\dagger\}$.

\section{Perturbative calculation in the ``$D_3$-breaking I" phase}
\label{app:D3I}

In this appendix, we calculate the lowest eigenvalue of the Hessian matrix in the ``$D_3$ breaking I" phase via third order perturbation theory.
We consider the $\Delta=0$ case. 
In this appendix, we take $\Gamma^\prime=1$ for simplification of notation.

\subsection{First order perturbation}

Let
\bea
x=-\frac{1}{\sqrt{2}} +x^\prime,~
y=y^\prime,~
z=\frac{1}{2}+z^\prime,~
\lambda_1=-2+\lambda_1^\prime,~
\lambda_2=-2+\lambda_2^\prime.
\eea
Then Eq. (\ref{eq:Saddle_eq_D6I}) becomes
\bea
2x^\prime+z^\prime +\frac{1}{\sqrt{2}} \lambda_1^\prime-|J| x^\prime-\lambda_1^\prime x^\prime &=&0\nn\\
3y^\prime+\frac{1}{\sqrt{2}}|J|-|J|y^\prime-\lambda_1^\prime y^\prime&=&0\nn\\
x^\prime+2z^\prime -\frac{1}{\sqrt{2}} |J|-\frac{1}{\sqrt{2}} \lambda_1^\prime -|J| z^\prime -\lambda_1^\prime z^\prime &=&0\nn\\
-x^\prime+z^\prime+\frac{1}{\sqrt{2}}\lambda_2^\prime-\frac{1}{\sqrt{2}} |J|+|J|x^\prime-|J|y^\prime&=&0\nn\\
-\sqrt{2} x^\prime +\sqrt{2}z^\prime +x^{\prime2}+y^{\prime2}+z^{\prime2}&=&0.
\label{eq:D61_perturbative}
\eea
The quantities $x^\prime,y^\prime,z^\prime,\lambda_1^\prime,\lambda_2^\prime$ can be expanded in a power expansion in $J$, i.e., 
\bea
x^\prime=\sum_{n\geq 1} x^{(n)},~
y^\prime=\sum_{n\geq 1} y^{(n)},~
z^\prime=\sum_{n\geq 1}z^{(n)},~
\lambda_1^\prime=\sum_{n\geq 1}\lambda_1^{(n)},~
\lambda_2^\prime=\sum_{n\geq 1}\lambda_2^{(n)},
\label{eq:Saddle_eq_D6I_1st}
\eea
in which $x^{(n)},y^{(n)},z^{(n)},\lambda_1^{(n)},\lambda_2^{(n)}$ are all proportional to $J^n$.

Plugging Eq. (\ref{eq:Saddle_eq_D6I_1st}) into Eq. (\ref{eq:Saddle_eq_D6I}) and keeping terms only up to $O(J)$,
we obtain
\bea
2x^{(1)}+z^{(1)}+\frac{1}{\sqrt{2}}\lambda_1^{(1)}&=&0\nn\\
3y^{(1)}+\frac{1}{\sqrt{2}}|J|&=&0\nn\\
x^{(1)}+2z^{(1)}-\frac{1}{\sqrt{2}}\lambda_1^{(1)}-\frac{1}{\sqrt{2}}|J|&=&0\nn\\
-x^{(1)}+z^{(1)}+\frac{1}{\sqrt{2}} \lambda_2^{(1)}-\frac{1}{\sqrt{2}}|J|&=&0\nn\\
-\sqrt{2}x^{(1)}+\sqrt{2}z^{(1)}&=&0.
\label{eq:D61_1storder}
\eea
The solution of Eq. (\ref{eq:D61_1storder}) gives 
\bea
\vec{r}_1&=&(
-\frac{1}{\sqrt{2}} +\frac{1}{6\sqrt{2}} |J|,~
-\frac{1}{3\sqrt{2}} |J|,~
\frac{1}{\sqrt{2}} +\frac{1}{6\sqrt{2}}|J|
)^T,\nn\\
\vec{r}_2&=&(-\frac{1}{\sqrt{2}},0,\frac{1}{\sqrt{2}})^T,\nn\\
\vec{r}_3&=&(
-\frac{1}{\sqrt{2}} -\frac{1}{6\sqrt{2}} |J|,~
\frac{1}{3\sqrt{2}} |J|,~
\frac{1}{\sqrt{2}} -\frac{1}{6\sqrt{2}}|J|
)^T,\nn\\
\lambda_1&=&-2-\frac{1}{2}|J|,\nn\\
\lambda_2&=&-2+|J|,
\label{eq:order_D6I_sol_1st}
\eea
in which $\vec{S}_i=S\vec{r}_i$, $i=1,2,3$.

To lowest order in $J$, the above equations reduce to Eq. (\ref{eq:D61_1storder}).

Notice that as discussed in Sec. \ref{sec:sw_D3I},
the smallest spin wave mass vanishes in first order perturbation.

\subsection{Second order perturbation}

Now we expand up to $O(J^2)$,
then Eq. (\ref{eq:D61_perturbative}) becomes
\bea
2x^{(2)}+z^{(2)}+\frac{1}{\sqrt{2}}\lambda_1^{(2)} &=& |J|x^{(1)} +\lambda_1^{(1)}x^{(1)}=\frac{1}{12\sqrt{2}}J^2\nn\\
3y^{(2)}&=&|J|y^{(1)} +\lambda_1^{(1)} y^{(1)}=-\frac{1}{6\sqrt{2}}J^2\nn\\
x^{(2)}+2z^{(2)}-\frac{1}{\sqrt{2}}\lambda_1^{(2)}&=&(|J|+\lambda_1^{(1)}) z^{(1)}=\frac{1}{12\sqrt{2}}J^2\nn\\
-x^{(2)}+z^{(2)}+\frac{1}{\sqrt{2}} \lambda_2^{(2)} &=& |J|(-x^{(1)}+y^{(1)})=-\frac{1}{2\sqrt{2}} J^2\nn\\
\sqrt{2}x^{(2)}-\sqrt{2}z^{(2)}&=& (x^{(1)})^2+(y^{(1)})^2+(z^{(1)})^2=\frac{1}{12}J^2.
\eea
The solutions are
\bea
x^{(2)}=\frac{5}{72\sqrt{2}} J^2,~
y^{(2)}=-\frac{1}{18\sqrt{2}} J^2,~
z^{(2)}=-\frac{1}{72\sqrt{2}} J^2,~
\lambda_1^{(2)}=-\frac{1}{24\sqrt{2}} J^2,~
\lambda_2^{(2)}=-\frac{5}{12\sqrt{2}} J^2.
\eea

From this, we are able to expand $\Delta \mathcal{H}_F(J)$ 
as $\Delta\mathcal{H}_F(J)=\Delta \mathcal{H}^{(1)}_F(J)+\Delta \mathcal{H}^{(2)}_F(J)$.
Let $\Delta \mathcal{H}^{(2)}_{\text{red}}(J)$ be the projection of the following matrix
\bea
\Delta \mathcal{H}^{(2)}_F(J)+\Delta \mathcal{H}^{(1)}_F(J) \sum_{i=1}^4 \frac{w_i^Tw_i}{E_0-E_i}\Delta \mathcal{H}^{(1)}_F(J)
\eea
to the subspace spanned by $v_1,v_2$, in which $w_i$ ($i=1,2,3,4$) are given in Eq. (\ref{eq:wi_s}) where
\bea
\hat{e}_\theta=(-\frac{1}{\sqrt{2}},0,-\frac{1}{\sqrt{2}})^T,~
\hat{e}_\phi=(0,-1,0)^T,
\eea
and $E_i=3$ are the eigenvalues of $w_i$.
Calculations show that $\Delta \mathcal{H}^{(2)}_{\text{red}}(J)=0$, which means that we have to go to third order.

Notice that as discussed in Sec. \ref{sec:sw_D3I},
the smallest spin wave mass still vanishes in second order perturbation.

\subsection{Third order perturbation}
\label{app:D3I_third_order}

The third order expansion of Eq. (\ref{eq:D61_perturbative}) gives
\bea
2x^{(3)} +z^{(3)} +\frac{1}{\sqrt{2}} \lambda_1^{(3)}&=& |J| x^{(2)} +\lambda_1^{(1)} x^{(2)} +\lambda_2^{(2)}x^{(1)}=\frac{1}{36\sqrt{2}}|J|^3\nn\\
3y^{(3)}&=& (|J|+\lambda_1^{(1)}) y^{(2)} +\lambda_1^{(2)} y^{(1)}=-\frac{1}{72\sqrt{2}}|J|^3\nn\\
x^{(3)} +2z^{(3)} -\frac{1}{\sqrt{2}} \lambda^{(1)}&=& |J|z^{(2)} +\lambda_1^{(1)} z^{(2)}+\lambda_1^{(2)}z^{(1)} =-\frac{1}{72\sqrt{2}}|J|^3\nn\\
\lambda_2^{(3)}-\sqrt{2}x^{(3)}+\sqrt{2}z^{(3)}&=&-\sqrt{2} |J| x^{(2)}+\sqrt{2} |J| y^{(2)}=\frac{1}{8}|J|^3\nn\\
x^{(3)}-z^{(3)}&=& \sqrt{2} (x^{(1)} x^{(2)}+y^{(1)} y^{(2)}+z^{(1)} z^{(2)})=\frac{1}{36\sqrt{2}}|J|^3.
\eea
The solution is 
\bea
x^{(3)}=\frac{7}{432\sqrt{2}} |J|^3,~
y^{(3)}=-\frac{1}{216\sqrt{2}} |J|^3,~
z^{(3)}=-\frac{5}{432\sqrt{2}} |J|^3,~
\lambda_1^{(3)}=\frac{1}{144} |J|^3,~
\lambda_2^{(3)}=-\frac{7}{72} |J|^3.
\eea
From this, we are able to obtain $\Delta \mathcal{H}^{(3)}_F(J)$
from the expansion   $\Delta\mathcal{H}_F(J)=\Delta \mathcal{H}^{(1)}_F(J)+\Delta \mathcal{H}^{(2)}_F(J)+\Delta \mathcal{H}^{(3)}_F(J)$.

The third order perturbation matrix is given by
\begin{flalign}
&h^{(3)}(J)=P_1^{(0)}\big[\Delta \mathcal{H}^{(3)}_F(J)+\Delta \mathcal{H}^{(1)}_F(J) \sum_{i=1}^4 \frac{w_i^Tw_i}{E_0-E_i}\Delta \mathcal{H}^{(2)}_F(J)+\Delta \mathcal{H}^{(2)}_F(J) \sum_{i=1}^4 \frac{w_i^Tw_i}{E_0-E_i}\Delta \mathcal{H}^{(1)}_F(J)\nn\\
&+\sum_{1\leq i,j\leq 4}\frac{\Delta \mathcal{H}^{(1)}_F(J)w_i w_i^T \Delta \mathcal{H}^{(1)}_F(J) w_jw_j^T\Delta \mathcal{H}^{(1)}_F(J)}{(E_0-E_i)(E_0-E_j)}-\sum_{1\leq i\leq 2,1\leq j\leq 4}\frac{\Delta \mathcal{H}^{(1)}_F(J)v_iv_i^T \Delta \mathcal{H}^{(1)}_F(J)w_jw_j^T\Delta \mathcal{H}^{(1)}_F(J)}{(E_0-E_j)^2}
\big]P_1^{(0)}.
\label{eq:third_D3I_mass}
\end{flalign}
Recall that the degeneracy has already been broken within first order perturbation theory.
The vector up to zeroth order is
\bea
\psi = -\sqrt{\frac{2}{3}}v_1+\frac{1}{\sqrt{3}}v_2.
\eea
Then the energy correction at $O(J^3)$ can be directly obtained by $\psi^T h^{(3)}(J) \psi$, which is $\frac{1}{2}|J|^3$.

Next, we proceed to calculate the spin wave mass.
The perturbation matrix at $O(J^3)$ is given by 
\begin{flalign}
&h^{M,(3)}(J)=P_1^{(0)}\big[\Delta \mathcal{H}^{M,(3)}_F(J)+\Delta \mathcal{H}^{M,(1)}_F(J) \sum_{i=1}^4 \frac{u_i^\dagger u_i}{E_0-\epsilon_i}\Delta \mathcal{H}^{M,(2)}_F(J)+\Delta \mathcal{H}^{M,(2)}_F(J) \sum_{i=1}^4 \frac{u_i^\dagger u_i}{E_0-\epsilon_i}\Delta \mathcal{H}^{M,(1)}_F(J)\nn\\
&+\sum_{1\leq i,j\leq 4}\frac{\Delta \mathcal{H}^{M,(1)}_F(J)u_i u_i^\dagger \Delta \mathcal{H}^{M,(1)}_F(J) u_ju_j^\dagger\Delta \mathcal{H}^{M,(1)}_F(J)}{(E_0-\epsilon_i)(E_0-\epsilon_j)}-\sum_{1\leq i\leq 2,1\leq j\leq 4}\frac{\Delta \mathcal{H}^{M,(1)}_F(J)v_iv_i^\dagger \Delta \mathcal{H}^{M,(1)}_F(J)u_ju_j^\dagger\Delta \mathcal{H}^{(1)}_F(J)}{(E_0-\epsilon_j)^2}
\big]P_1^{(0)}.
\label{eq:hM_third_order}
\end{flalign}
Evaluation of Eq. (\ref{eq:hM_third_order}) gives Eq. (\ref{eq:hM3J}).
To obtain $m_1$, we need to calculate the eigenvalues of $h^{M,(1)}(J)+h^{M,(3)}(J)$ (recall that $h^{M,(2)}(J)=0$). 
Calculations show that the eigenvalues are $\pm iJ^2$.

We make a comment here. 
When $J=0$ ($\Delta=0$ as before), the null space of $M(\Delta=0,J=0)\mathcal{H}_F(\Delta=0,J=0)$ is five-dimensional, and the corresponding eigenvectors are $v_i$ ($i=1,2$) and $r_j$ ($j=1,2,3$), where $r_j$'s are given by Eq. (\ref{eq:define_ri}) in which $\hat{n}^{(0)}_j=(-\frac{1}{\sqrt{2}},0,\frac{1}{\sqrt{2}})$.
Rigorously, we should perform a degenerate perturbation theory in this five-dimensional space, instead of a perturbation within the two-dimensional space spanned by $\{v_1,v_2\}$ as discussed previously.
However, we demonstrate that in obtaining the two nonzero eigenvalues $\pm iJ^2$, it is enough to work within the two-dimensional space.
The perturbation matrix $h^{M,(1,2,3)}_5(J)$ in the five-dimensional space spanned by $\{v_1,v_2,r_1,r_2,r_3\}$ up to third order
can be obtained by replacing $P_1^{(0)}$ in Eqs. (\ref{eq:first_D3I_mass},\ref{eq:second_D3I_mass},\ref{eq:third_D3I_mass}) with the projection to the five-dimensional space. 
The result is
\bea
h^{M,(1,2,3)}_5(J)=\left(\begin{array}{ccccc}
-\frac{2\sqrt{2}}{3}|J|+\frac{19}{54\sqrt{2}}|J|^3 & -\frac{2}{3}|J|-\frac{35}{108}|J|^3&0&0&0\\
\frac{4}{3}|J|+\frac{4}{27}|J|^3 & \frac{2\sqrt{2}}{3}|J|-\frac{19}{54\sqrt{2}}|J|^3&0&0&0\\
-\frac{1}{3}\sqrt{\frac{2}{3}}|J|^2-\frac{1}{9\sqrt{6}}|J|^3& -\frac{1}{3\sqrt{3}}|J|^2-\frac{1}{18\sqrt{3}}|J|^3&0&0&0\\
0&0&0&0&0\\
\frac{1}{3}\sqrt{\frac{2}{3}}|J|^2+\frac{1}{9\sqrt{6}}|J|^3&\frac{1}{3\sqrt{3}}|J|^2+\frac{1}{18\sqrt{3}}|J|^3&0&0&0
\end{array}\right).
\label{eq:D3I_5dim}
\eea
As can seen from Eq. (\ref{eq:D3I_5dim}), to get the two nonzero eigenvalues, it is enough to consider an eigenvalue problem of the upper-left $2\times 2$ block,
since the vectors $r_j$ ($j=1,2,3$) always lie within the null space of the matrix in Eq. (\ref{eq:D3I_5dim}) regardless of the value of $J$. 
The eigenvalues $\pm iJ^2$ are obtained in this way, i.e., by calculating the eigenvalues of the upper-left $2\times 2$ block. 
We also note that the eigenvectors of the two nonzero eigenvalues $\pm iJ^2$ contain components on $r_j$ ($j=1,2,3$) due to the nonzero matrix elements in the third, fourth, and fifth row (but only within the first and second columns) of $h^{M,(1,2,3)}_5(J)$.

\section{Perturbative calculation in the ``$D_3$-breaking II" phase}
\label{app:D3II}

\begin{figure}[h]
\includegraphics[width=6.0cm]{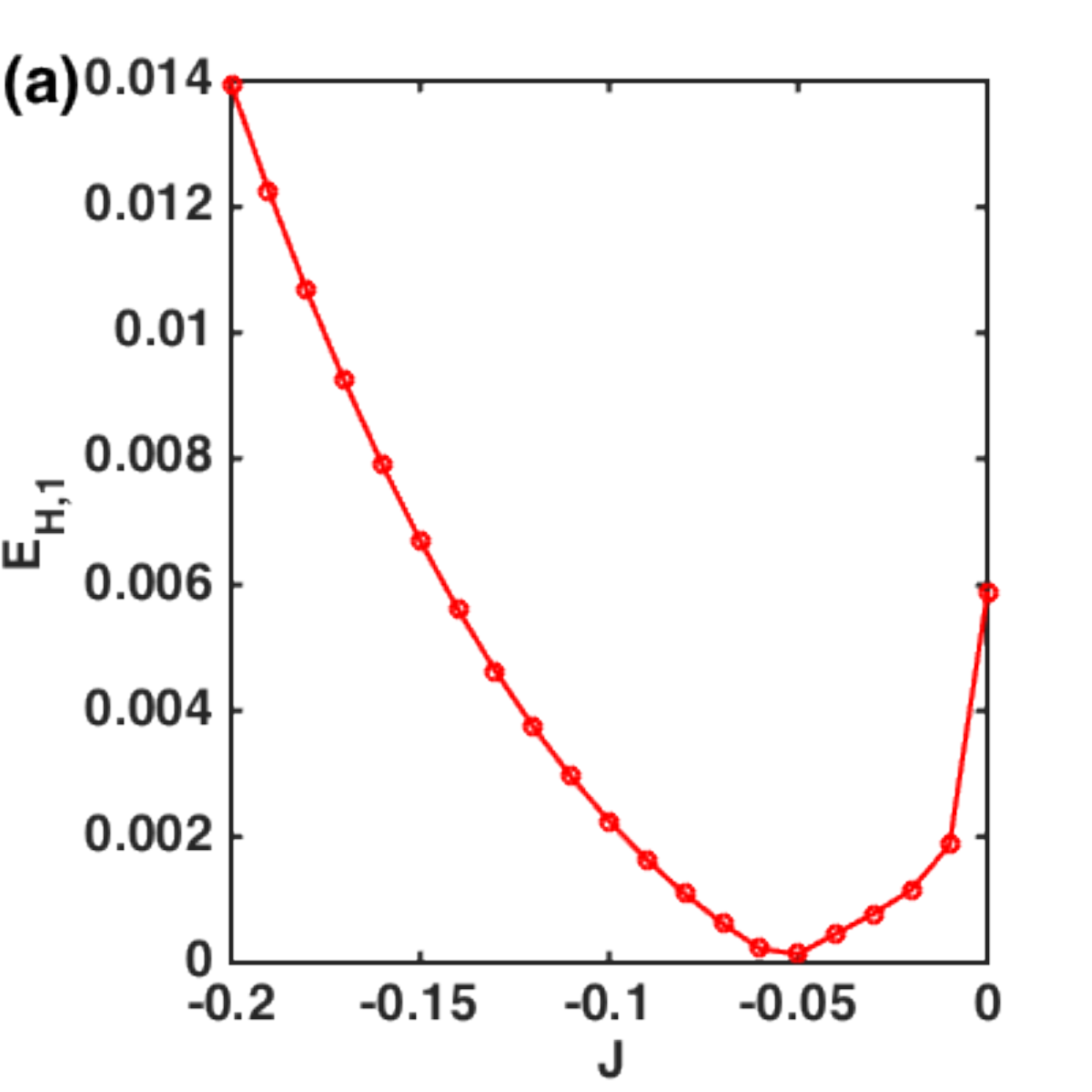}
\includegraphics[width=6.0cm]{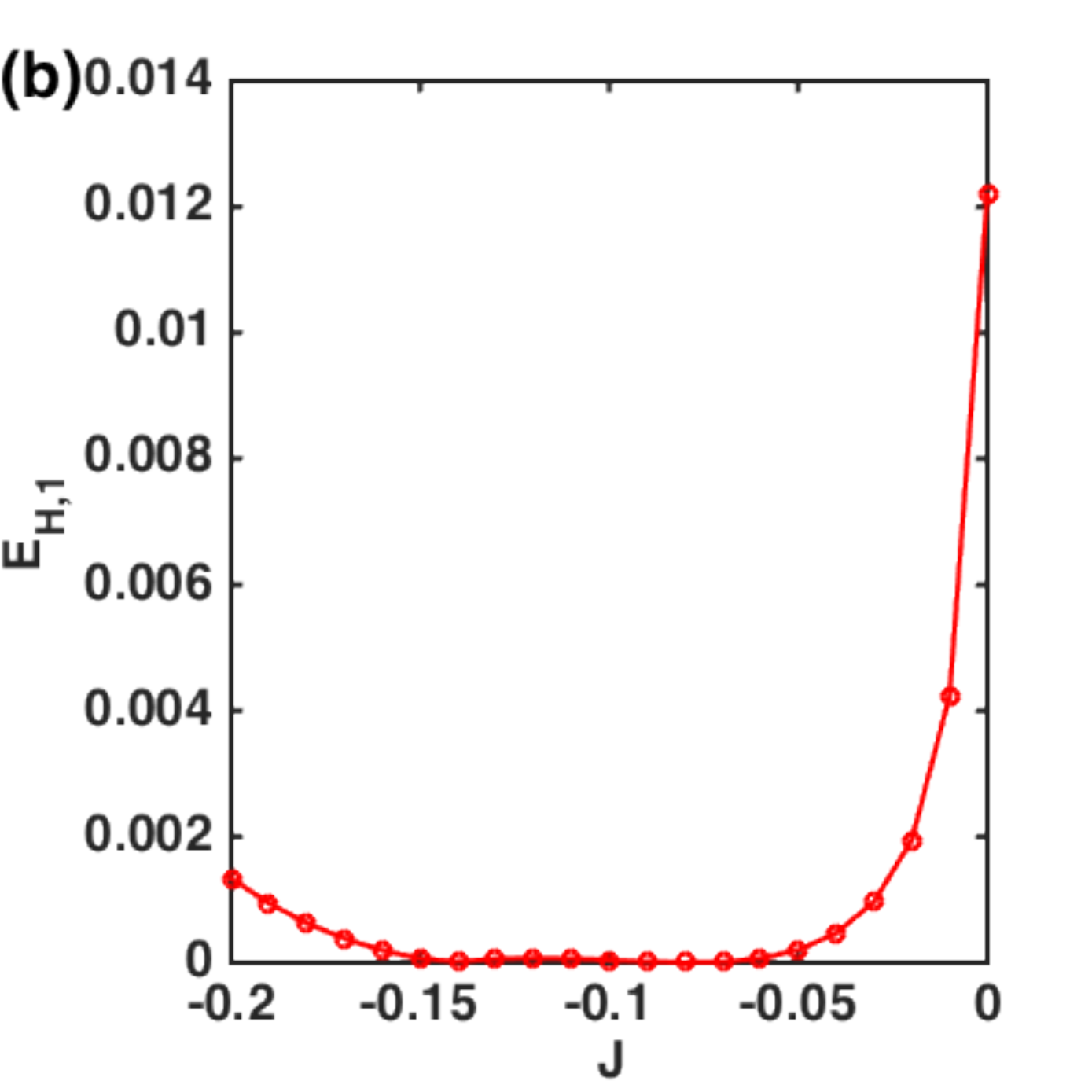}
\caption{Smallest eigenvalue vs. $J$ for (a) $\varphi=0.21\pi$,
and (b) $\varphi=0.30\pi$,
where in accordance with the main text, $\varphi$ is defined through the parametrization $K=\cos(\varphi),\Gamma=\sin(\varphi)$.
Notice that although when $\varphi<0.25\pi$, the dependence of the smallest eigenvalue on $J$ is regular,
the eigenvalue exhibits a rather complicated behavior when $\varphi>0.25\pi$.
On the other hand, as discussed in the main text, the value of $m_1$ is regular even for $\varphi>0.25\pi$.
} 
\label{fig:mass_vs_J}
\end{figure}

Here the zeroth order solution can be taken as the one along the $(1,-1,1)$-direction: 
\begin{flalign}
&\hat{n}^{(0)}_1=\frac{1}{\sqrt{3}}(
x_0,
-x_0,
z_0
)^T,~
\hat{n}^{(0)}_2=\frac{1}{\sqrt{3}}(
x_0,
-z_0,
x_0
)^T,~
\hat{n}^{(0)}_3=\frac{1}{\sqrt{3}}(
z_0,
-x_0,
x_0
)^T,\nn\\
&\lambda_1^{(0)}=\lambda_2^{(0)}=\lambda_3^{(0)}=\lambda_0,
\end{flalign}
in which
\bea
x_0=1+\frac{1}{9}\Delta-\frac{2}{81}\Delta^2,~
z_0=1-\frac{2}{9}\Delta+\frac{1}{81}\Delta^2,~
\lambda_0=-2-\frac{2}{3}\Delta-\frac{2}{27}\Delta^2.
\eea

We solve the saddle point equations perturbatively in an expansion over $J$ starting with a nonzero $\Delta$.
There is some difficulty in calculating the eigenvalues of the Hessian matrix.
Instead of deriving a perturbative result, we study the eigenvalues numerically.
In this appendix, we take $\Gamma^\prime=1$ for simplification of notation.

Let 
\bea
&x=\frac{1}{\sqrt{3}}(x_0+x^\prime),~
y=\frac{1}{\sqrt{3}}(-x_0+y^\prime),~
z=\frac{1}{\sqrt{3}}(z_0+z^\prime),~
m=\frac{1}{\sqrt{3}}(x_0+m^\prime),~
n=\frac{1}{\sqrt{3}}(-z_0+n^\prime),\nn\\
&\lambda_1=\lambda_0+\lambda_1^\prime,~
\lambda_2=\lambda_0+\lambda_2^\prime.
\eea
in which the primed variables are assumed to be $O(J)$.
Plugging these into the saddle point equations and only keeping the $O(J)$ terms, we obtain
\bea
\left(\begin{array}{ccccccc}
-\lambda_0&0&-1&-(1+\Delta)&0&-x_0&0\\
0&-(1+\Delta+\lambda_0)&0&0&-1&x_0&0\\
-1&0&-\lambda_0&-1&0&-z_0&0\\
-(1+\Delta)&0&-1&-\lambda_0&0&0&-x_0\\
0&-2&0&0&-\lambda_0&0&z_0\\
x_0&-x_0&z_0&0&0&0&0\\
0&0&0&2x_0&-z_0&0&0
\end{array}\right)
\left(\begin{array}{c}
x^\prime\\
y^\prime\\
z^\prime\\
m^\prime\\
n^\prime\\
\lambda_1^\prime\\
\lambda_2^\prime
\end{array}\right)
=
\left(\begin{array}{c}
-2x_0|J|\\
0\\
0\\
0\\
-2z_0|J|\\
0\\
0
\end{array}\right).
\eea
The solution gives Eqs. (\ref{eq:sol_D6II_A},\ref{eq:sol_D6II_B},\ref{eq:sol_D6II_C}).

Next we try to proceed as before by defining $\mathcal{H}_F(\Delta,J) = P(\Delta,J)H_F(\Delta,J) P(\Delta,J)$, and 
$\Delta\mathcal{H}_F(\Delta,J) = \mathcal{H}_F(\Delta,J) -\mathcal{H}_F(\Delta,J=0)$.
Consider the first order degenerate perturbation 
\begin{flalign}
h^{(1)}(\Delta,J)=\left(\begin{array}{cc}
v_1^T \Delta\mathcal{H}_F(\Delta,J) v_1& v_1^T \Delta\mathcal{H}_F(\Delta,J) v_2\\
v_2^T \Delta\mathcal{H}_F(\Delta,J) v_1 & v_2^T\Delta \mathcal{H}_F(\Delta,J) v_2
\end{array}
\right).
\label{eq:hM1_D3II}
\end{flalign}
This time, the leading order contribution is $O(\bar{J}/\Delta^2)$.
On the other hand, calculations show that the leading nonvanishing terms in $h^{(1)}(\Delta,J)$ is $O(J)$.
However, if we want to reach $O(J)$, the calculations in Eq. (\ref{eq:hM1_D3II}) are not enough.
Let $v_1(\Delta)$ and $v_2(\Delta)$ be the two lowest spin wave vectors at $J=0$.
Since $\Delta\mathcal{H}_F(\Delta,J)$ contains $O(1/\Delta^2)$ terms, we have to keep $v_i(\Delta)$ ($i=1,2$) to $O(\Delta^2)$ so that $O(J)$ can be reached for $h^{(1)}(\Delta,J)$.
Recall that up to $O(\Delta^2)$, the eigenvalues are still degenerate for $v_i(\Delta)$ ($i=1,2$), both equal to $\frac{4}{27}\Delta^2$.
The best situation is that they split in the third order perturbation, i.e., to $O(\Delta^3)$.
Then we have linear combinations $\alpha_1 v_1+\beta_1 v_2$ and $\alpha_2 v_1+\beta_2 v_2$, with an energy difference $\sim O(\Delta^3)$.
To get an $O(\Delta^2)$ mixture between $\alpha_i v_1+\beta_i v_2$ ($i=1,2$),
we have to go to another two orders of perturbations, i.e., fifth order perturbation in $\Delta$.

The smallest eigenvalue is calculated by numerics  shown in Fig. \ref{fig:mass_vs_J}.
As can be seen from Fig. \ref{fig:mass_vs_J}, the results show a very complicated behavior when $\phi>0.25\pi$.

\section{Numerical results in the ``$D_3$-breaking I, II" phases for  $S=3/2$}
\label{app:more_numerics}

Fig. \ref{fig:D3_break_Seq1p5} shows the results for the spin expectation vales $\left<S_j^\alpha\right>$ ($\alpha=x,y,z$) at three representative points $(\theta=0.52\pi, \phi=0.15\pi)$, $(\theta=0.52\pi, \phi=0.25\pi)$ and $(\theta=0.52\pi, \phi=0.30\pi)$ under the $h_{\text{I}}$ and $h_{\text{II}}$ fields for the $S=3/2$ case.
ED numerics are performed on a system of  $L=18$ sites with periodic boundary conditions,
and both $h_{\text{I}}$ and $h_{\text{II}}$ fields are taken to be $10^{-4}$.
As can be clearly seen from Figs. \ref{fig:D3_break_Seq1},
the spin alignments are consistent with the patterns given in Eqs. (\ref{eq:order_D6I},\ref{eq:order_D6II}),
thereby confirming the existence of the ``$D_3$-breaking I, II" phases for $S=3/2$.

\begin{figure*}[htbp]
\includegraphics[width=15.0cm]{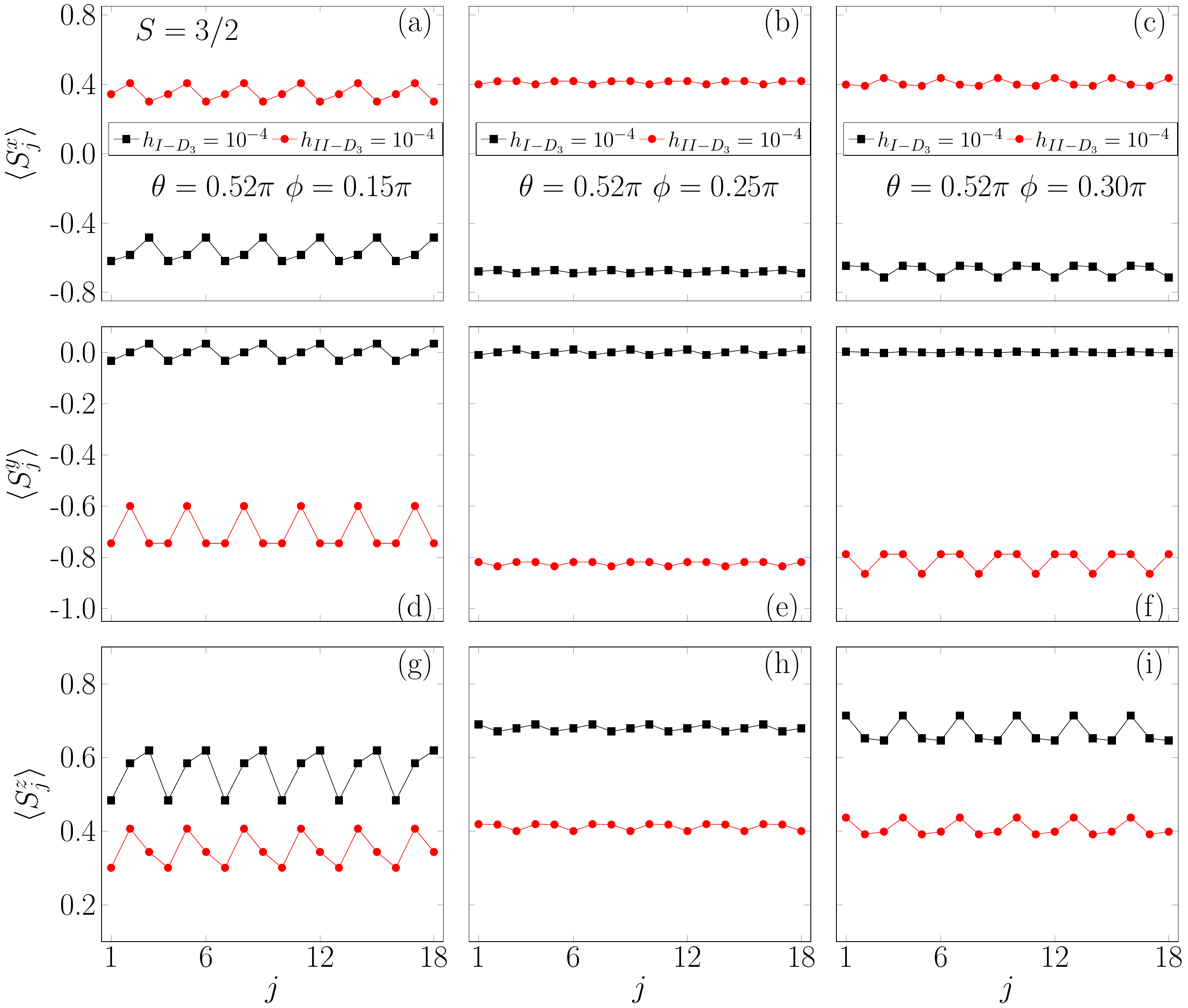}
\caption{(a,b,c) $\langle S_j^x\rangle$, (d,e,f) $\langle S_j^y\rangle$, and (g,h,i) $\langle S_j^z\rangle$ vs $j$ under $h_{\text{I}}$ (black squares) and $h_{\text{II}}$ (red dots) fields for $S=3/2$ at several different points.
(a,d,g) are for $(\theta=0.52\pi,\phi=0.15\pi)$;
(b,e,h) for $(\theta=0.52\pi,\phi=0.25\pi)$;
and (c,f,i)  for $(\theta=0.52\pi,\phi=0.30\pi)$.
DMRG numerics are performed on $L=18$ sites with periodic boundary conditions. 
Both $h_{\text{I}}$ and $h_{\text{II}}$ fields are taken to be $10^{-4}$.
} \label{fig:D3_break_Seq1p5}
\end{figure*}

\end{widetext}


\end{document}